\documentclass[aps,twocolumn,showpacs,preprintnumbers,amsmath,amssymb,nofootinbib,superscriptaddress,showkeys]{revtex4}

\usepackage{epsfig}
\usepackage{graphicx}

\def\slashchar#1{\setbox0=\hbox{$#1$}
   \dimen0=\wd0 \setbox1=\hbox{/} \dimen1=\wd1
   \ifdim\dimen0>\dimen1 \rlap{\hbox to \dimen0{\hfil/\hfil}} #1
   \else  \rlap{\hbox to \dimen1{\hfil$#1$\hfil}} / \fi}

\begin{document}

\title{Renormalization vs Strong Form Factors for One Boson Exchange
  Potentials.}

\author{A. Calle Cord\'on}\email{alvarocalle@ugr.es}
  \affiliation{Departamento de F\'isica At\'omica, Molecular y
  Nuclear, Universidad de Granada, E-18071 Granada, Spain.}
\author{E. Ruiz
  Arriola}\email{earriola@ugr.es}
  \affiliation{Departamento de F\'isica At\'omica, Molecular y Nuclear,
  Universidad de Granada,
  E-18071 Granada, Spain.}
\date{\today}

\begin{abstract} 
\rule{0ex}{3ex} We analyze the One Boson Exchange Potential from the
point of view of Renormalization theory. We show that the
nucleon-meson Lagrangean while predicting the NN force does not
predict the NN scattering matrix nor the deuteron properties
unambiguously due to the appearance of short distance
singularities. While the problem has traditionally been circumvented
by introducing vertex functions via phenomenological strong form
factors, we propose to impose physical renormalization conditions on
the scattering amplitude at low energies. Working in the large $N_c$
approximation with $\pi$,$\sigma$,$\rho$ and $\omega$ mesons we show
that, once these conditions are applied, results for low energy phases
of proton-neutron scattering as well as deuteron properties become
largely insensitive to the form factors and to the vector mesons 
yielding reasonable agreement with the data and for realistic values of
the coupling constants.
\end{abstract}

\maketitle



\section{Introduction} 


The One Boson Exchange (OBE) Potential has been a cornerstone for
Nuclear Physics during many years. It represents the natural
generalization of the One Pion Exchange (OPE) potential proposed by
Yukawa~\cite{Yukawa:1935xg}, and also the scalar meson potential
introduced by Johnson and Teller~\cite{PhysRev.98.783}. With the
advent of vector mesons these degrees of freedom were included as
well~\cite{Bryan:1964zz,Bryan:1967zz,Bryan:1969mp,Partovi:1969wd}.
Actually, Regge theory yields such a potential within a suitable
approximation~\cite{Nagels:1977ze}. The disturbing short distance
divergences were first treated by using a hard core boundary condition
\cite{Bryan:1964zz,Bryan:1967zz,Bryan:1969mp} and it was soon realized
that divergences in the potential could be treated by introducing
phenomenological form factors mimicking the finite nucleon
size~\cite{Ueda:1969er}.  The field theoretical OBE model of the NN
interaction~\cite{Erkelenz:1974uj,Machleidt:1987hj} includes all
mesons with masses below the nucleon mass, i.e., $\pi$, $\eta$,
$\rho(770)$ and $\omega(782)$. We refer to
Ref.~\cite{Machleidt:1989tm,Machleidt:2007ms} for accounts of the many
historical revolves of the problem. An important lesson from these
developments has been that the non-perturbative nature of the NN force
is better handled in terms of quantum mechanical potentials at low
energies where relativistic and nonlocal effects contribute at the few
percent level. Although such a framework has remained a useful,
appealing and accurate phenomenological model after a suitable
introduction of phenomenological strong form
factors~\cite{Machleidt:1987hj,Machleidt:2000ge} it is far from being
a complete description of the intricacies of the nuclear force.  The
highly successful Partial Wave Analysis (PWA) of the Nijmegen
group~\cite{Stoks:1993tb} while providing a spectacular fit with
$\chi^2 /{\rm DOF}< 1$ comprising a large body of pn and pp scattering
data checks mainly OPE and {\it some} contributions from other mesons,
since the interaction below $1.4 {\rm fm}$ is parameterized by an
energy dependent square well potential.

A traditional test to NN forces in general and OBE potentials in
particular has been NN scattering in the elastic region. In such a
situation relative NN de Broglie wavelengths larger than half a fermi
are probed; a factor of two bigger scale than the Compton wavelengths
of the vector and heavier mesons. However, while from this simple
minded argument we might expect those mesons to play a marginal role,
OBE potentials have traditionally been sensitive to short distances
requiring a unnatural fine tuning of the vector meson coupling. As a
consequence there has been some inconsistency between the couplings
required from meson physics, SU(3) or chiral symmetry on the one hand
and those from NN scattering fits on the other hand (see also
\cite{Stoks:1996yj,Furnstahl:1996wv,Papazoglou:1998vr}). Part of the
disagreement could only be overcome after even shorter scales were
explicitly considered~\cite{Janssen:1994kh,Janssen:1996kx}. A more
serious shortcoming stems from the use of strong form factors which
have mainly been phenomenologically motivated and loosely related to
the field theoretical meson-baryon Lagrangean from which the meson
exchange picture is derived. It is therefore not exaggerated to say
that strictly speaking the OBE potentials have not been solved yet.
Of course, this may appear as a mathematically interesting problem
with no relevance to the physics of NN interactions. However, as we
will see, the Meson-Nucleon Lagrangean itself while providing the NN
OBE potential from the Born approximation, {\it does not predict} the
NN S-matrix and the deuteron unambiguously beyond perturbation theory
from the OBE potential. The unspecified information in the Lagrangean
can be advantageously tailored to fit the data in the low
energy region. We will also show that once this is done, the vertex
functions play a minor role, with a fairly satisfactory description of
central waves and the deuteron.

It is notorious that the OBE potentials although exponentially
suppressed with the corresponding meson mass, $\sim e^{-m r}$, are by
themselves large at short distances and mostly even diverge as
$1/r^3$.  For a {\it singular} potential, i.e. a potential fulfilling
$\lim_{r \to 0} 2 \mu |V(r)| r^2 =
\infty$~\cite{Case:1950an,Frank:1971xx}, the Hamiltonian is unbounded
from below preventing the existence of a stable two nucleon bound
state when $\lim_{r \to 0} 2 \mu V(r) r^2 < -1/4$. Of course, the
singularity is unphysical as it corresponds to the interaction of two
point-like static classical particles and not to extended nucleons
with a finite size of about half a fermi~\footnote{Of course, the
  nucleon size depends on the particular electroweak probe. We give
  here a typical number.}.  From a quantum mechanical viewpoint,
however, the relative NN de Broglie wavelength provides the limiting
resolution scale physically operating in the problem. This suggests
that finite nucleon size effects should also play a marginal role in
NN scattering in which case it should be possible to formulate the
problem {\it without} any explicit reference to form factors. In fact,
as we will explicitly demonstrate Renormalization is the natural
mathematical tool to implement the physically desirable decoupling of
short distance components of the interaction at the energies involved
in NN elastic scattering.

Within the NN system the problem of infinities has traditionally been
cured~\cite{Ueda:1969er} by the introduction of phenomenologically or
theoretically motivated strong form factors in each meson-nucleon
vertex, $\Gamma_{mNN}(q^2)$ ($m=\pi,\sigma,\rho,\omega$ etc.) where
the off-shellness of the nucleon legs is usually neglected. This
procedure somewhat mimics the finite nucleon size but strong form
factors are fitted and constrained in practice to NN scattering data
and deuteron properties. This corresponds to the replacement of the
potential $V_m (q) \to V_m (q) \left[ \Gamma_{mNN} (q)\right]^2 $
where typically a monopole form is taken for each separate meson
$\Gamma_{mNN}(q) = (\Lambda_{mNN}^2-m^2)/(\Lambda_{mNN}-q^2)$ and
generally $\Lambda_{mNN} \sim 1-2 {\rm GeV}$~\footnote{For a monopole
  the operating scale is lower, $\Lambda_{mNN}/ \sqrt{2}$, because the
  square of the form factor enters in the modification of the
  potential.}.  Due to the long distance distortion introduced by the
vertex function deuteron properties impose limitations on the lowest
cut-off value $\Lambda_{\pi NN} > 1.3 {\rm GeV}$ still fitting the
result~\cite{Machleidt:1987hj}.

Because of their fundamental character and the crucial role played in
NN calculations there have been countless attempts to evaluate strong
form factors by several means, mainly $\Gamma_{\pi NN}(q^2)$. These
include meson
theory~\cite{1972NuPhA.185..131W,Gari:1984pq,Kaulfuss:1984tw,Flender:1994uh,Schutz:1995dj,Bockmann:1999nu},
Regge models~\cite{Bryan:1979ve}, chiral soliton
models~\cite{Cohen:1986ux,Alberto:1990ru,Christov:1995vm,Holzwarth:1996bc}, QCD sum
rules~\cite{Meissner:1995ra} the Goldberger-Treimann
discrepancy~\cite{Coon:1990fh} or 
lattice QCD~\cite{Liu:1994dr,Alexandrou:2007zz} and quark
models~\cite{Melde:2008dg}. Most calculations yield rather small
values $\Lambda_{\pi NN} \sim 800 {\rm MeV}$ generating the soft form
factor puzzle for the OBE potential for several years since the
cut-off could not be lowered below $\Lambda_{\pi NN}=1.3 {\rm GeV}$
without destroying the quality of the fits and the description of the
deuteron~\cite{Machleidt:1987hj}. The contradiction was solved by
including either $\rho\pi$ exchange~\cite{Janssen:1993nj}, a strongly
coupled excited $\pi'(1300)$ state~\cite{Holinde:1990fe}, two pion
exchange~\cite{Haidenbauer:1994zz} or three pion
exchange~\cite{Ueda:1991ca}. Some of these ways out of the paradox
assume the meson exchange picture seriously to extremely short
distances.  However, as noted in Ref.~\cite{Holzwarth:1996bc} the
contradiction is misleading since a large cut-off is needed {\it just}
to avoid a sizable distortion of the OBE potential in the region $r>
0.5 {\rm fm}$ which can also be achieved by choosing a suitable shape
of the form factor. This point was explicitly illustrated by using the
Skyrme soliton model form factors~\cite{Holzwarth:1996bc}. In fact,
this conclusion is coherent with the early hard core
regularizations~\cite{Bryan:1964zz,Bryan:1967zz,Bryan:1969mp}, recent
lattice calculations~\cite{Alexandrou:2007zz} (where an extremely hard
$\Lambda_{\pi NN} \sim 1.7 {\rm GeV}$ and a rather flat behaviour are
found) and, as we will show, with the renormalization approach we
advocate.

The implementation of vertex meson-nucleon functions has also
notorious side effects, in particular it affects gauge invariance,
chiral symmetry and causality via dispersion relations. As it is
widely accepted, besides the description of NN scattering and the
deuteron, one of the great successes and confirmations of Meson theory
has been the prediction of Meson Exchange Currents (MEC's) for
electroweak processes (see \cite{Ericson:1988gk,Riska:1989bh} for
reviews and references therein). In the case of gauge invariance, the
inclusion of a form-factor introduced by hand, i.e., not computed
consistently within meson theory, implies a kind of non-locality in
the interaction. This can be made gauge invariant by introducing link
operators between two points, thereby generating a path dependence,
and thus an ambiguity is introduced. In the limit of weak non-locality
the ambiguity is just the standard operator ordering problem, for
which no obvious resolution has been found yet. Form factors can also
be in open conflict with dispersion relations, particularly if they
imply that the interaction does not vanish as a power of the momentum
everywhere in the complex plane. We will show that within the
renormalization approach, all singularities fall on the real axes and
spurious deeply bound states are shifted to the real negative. The
extremely interesting issue of analyzing the consequences of
renormalization for electroweak processes is postponed for future research.

In the present paper we approach the NN problem for the OBE potential
from a renormalization viewpoint. We analyze critically the role
played by the customarily used phenomenological form factors. As a
viable alternative we carry out the renormalization program to this
OBE potential to manifestly implement short distance insensitivity as
well as completeness of states by removing the cut-off. In practice,
we use the coordinate space renormalization by means of boundary
conditions~\cite{PavonValderrama:2005gu,Valderrama:2005wv,PavonValderrama:2005uj}. The
equivalence to momentum space renormalization using counterterms for
regular and singular potentials was discussed in
Refs.~\cite{PavonValderrama:2004td,Entem:2007jg}.  In order to
facilitate and simplify the analysis we will use large $N_c$ relations
for meson-nucleon
couplings~\cite{Witten:1979kh,Manohar:1998xv,Jenkins:1998wy} which are
well satisfied phenomenologically and pick the leading tensorial
structures for the OBE potential. In this picture mesons are stable
with their mass scaling as $m \sim N_c^0$, nucleons are heavy with
their mass scaling as $M_N \sim N_c$, the NN potential also scales as
$V_{NN} \sim N_c$.  The OBE component is dominated by the $\pi$,
$\sigma$, $\rho$ and $\omega$
mesons~\cite{Kaplan:1996rk,Banerjee:2001js}.  Further advantages of
using this large $N_c$ approximation have been stressed in in regard
to Wigner and Serber symmetries in
Refs.~\cite{CalleCordon:2008eu,CalleCordon:2008cz,
Cordon:2009ps,Arriola:2009bg}, in particular the fact that
relativistic, spin-orbit and meson widths corrections are suppressed
by a relative $1/N_c^2$ factor suggesting a bold $10\%$ accuracy.
However, we hasten to emphasize that despite the use of this appealing
and simplifying approximation in the OBE potential {\it we do not
claim to undertake a complete large $N_c$ calculation} since multiple
meson exchanges and $\Delta$ intermediate states should also be
implemented~\cite{Banerjee:2001js}. In spite of this, some of our
results fit naturally well within naive expectations of the large
$N_c$ approach. The coordinate space renormalization scheme is not
only convenient and much simpler but it is also particularly suited
within the large $N_c$ framework, where non-localities in the
potential are manifestly suppressed, and an internally consistent
multimeson exchange scheme is possible if energy independent
potentials are used~ \cite{Belitsky:2002ni,Cohen:2002im}.

The paper is organized as follows. In Sect.~\ref{sec:need} we write
down a chiral Lagrangean in order to visualize the calculation of the
OBE potential in the large $N_c$ limit and analyze its singularities.
The standard approach to prevent the singularity has been to include
form factors to represent vertex functions, an issue which is analyzed
critically in Section~\ref{sec:standard} where the alternative between
fine tuning and the appearance of spurious bound states is
highlighted.  In Section~\ref{sec:dispersion} we discuss the physical
conditions under which a description of NN scattering makes sense
within a renormalization point of view. In addition, we discuss some
general features which apply to the solutions of the Schr\"odinger
equation with the local and energy independent large $N_c$
OBE potential on the basis of renormalization. In
Sect.~\ref{sec:singlet} we analyze the $^1S_0$ channel from which the
scalar meson parameters may be fixed. We also discuss the role played
by spurious bound states which appear in this kind of
calculations. The deuteron and the corresponding low energy parameters
as well as the $^3S_1-^3D_1$ phase shifts are analyzed in
Sect.~\ref{sec:singlet}. The marginal influence of form factors in the
renormalization process is shown in Section~ \ref{sec:formfac}.
Finally, in Sect.~\ref{sec:concl} we summarize our main points and
conclusions. In appendix \ref{sec:couplings} we also review current
values for the coupling constants from several sources entering the
potential.

\section{OBE potentials and the need for renormalization}
\label{sec:need}

In this section we briefly sketch the well known process of deriving
the OBE potential from the nucleon-meson Lagrangean. We appeal a
chiral Lagrangean as done in
Refs.~\cite{Stoks:1996yj,Furnstahl:1996wv,Papazoglou:1998vr} and keep
only the leading $N_c$ contributions to the OBE potential due to the
tremendous simplification which proves fair enough to illustrate our
main point, namely the {\it lack of uniqueness} of the S-matrix from the OBE
potential.  In a more ellaborated version the present calculation
should include many other effects such as relativistic corrections,
spin-orbit coupling, meson widths, multi-meson exchange and $\Delta$
intermediate states.

\subsection{Meson-Nucleon Chiral Lagrangian}

We use a relativistic chiral Lagrangean as done in
Refs.~\cite{Stoks:1996yj,Furnstahl:1996wv,Papazoglou:1998vr} as a
convenient starting point. The $\pi-\sigma$ Lagrangean reads 
\begin{eqnarray}
{\cal L}_{\sigma \pi}^{\rm kin} &=& \frac{\sigma^2}{4} \langle
\partial^\mu U^\dagger \partial_\mu U^\dagger \rangle + \frac12
\partial^\mu \sigma \partial_\mu \sigma \nonumber \\ 
&-&  V(\sigma) - \frac{\sigma
m_\pi^2}4 \langle U + U^\dagger \rangle \, , 
\end{eqnarray}
where $U(x)= e^{i \vec \tau \cdot \pi / f_\pi}$ is the non-linearly
transforming pion field and $\langle , \rangle $ represents the trace
in isospin space. The scalar field is invariant under chiral
transformations~\footnote{This is unlike the standard assignment of
  the linear sigma-model where one takes $(\sigma, \vec \pi)$ as
  chiral partners in the $(1/2,1/2) $ representation of the chiral
  $SU(2)_R \otimes SU(2)_L$ group.} and the potential is chosen to
have a minimum at $\sigma = f_\pi $. The sigma mass is then
$m_\sigma^2= V''(\sigma) |_{\sigma = f_\pi}$, so that the physical
scalar field is defined by the fluctuation around the vacuum
expectation value, $\sigma = f_\pi + s $. $f_\pi=92.6~{\rm MeV}$
denotes the pion weak-decay constant, ensuring the proper
normalization condition of the pseudoscalar fields. The vector mesons
kinetic Lagrangeans are represented by Proca fields
\begin{eqnarray}
{\cal L}_\omega^{\rm kin} &=& - \frac{1}{4} (\partial^\mu \omega^\nu
-\partial^\nu \omega^\mu ) (\partial_\mu \omega_\nu -\partial_\nu
\omega_\mu ) + \frac12 m_\omega^2 \omega^\mu \omega_\mu \, , 
\nonumber \\
{\cal L}_\rho^{\rm kin} &=& - \frac{1}{4} (\partial^\mu \rho^\nu
-\partial^\nu \rho^\mu ) (\partial_\mu \rho_\nu -\partial_\nu
\rho_\mu ) + \frac12 m_\rho^2 \rho^\mu \rho_\mu  \, . 
\nonumber \\
\end{eqnarray}
and the kinetic nucleon Lagrangian is
\begin{eqnarray}
{\cal L}_N^{\rm kin} = \bar N  i \slashchar{\partial} N \, . 
\end{eqnarray}
The chirally invariant form of the meson-nucleon Lagrangean can be
looked up in Ref.~\cite{Stoks:1996yj}. From the vacuum expectation
value of the scalar meson we get the nucleon mass $M_N = g_{\sigma NN}
f_\pi $ and the relevant
nucleon-meson interaction vertices can be obtained from a chiral
Lagrangean~\cite{Stoks:1996yj,Furnstahl:1996wv,Papazoglou:1998vr} and
read
\begin{eqnarray}
{\cal L}_{\pi NN} &=& -\frac{g_{\pi NN}}{2 \Lambda_N} \bar N
\gamma_\mu \gamma_5 \tau \cdot \partial^\mu \pi N \, , \nonumber \\
{\cal L}_{\sigma NN} &=& -g_{\sigma NN} \, \sigma \bar N N \, , 
\nonumber 
\\ 
{\cal L}_{\rho NN} &=& -g_{\rho NN} \bar N \tau \cdot
\rho^\mu \gamma_\mu N - \frac{f_{\rho NN}}{2 \Lambda_N} \bar N \sigma_{\mu \nu} \tau \cdot \partial^\mu \rho^\nu N  \, ,  \nonumber \\
{\cal L}_{\omega NN} &=&
-g_{\omega NN} \bar N \gamma_\mu \omega^\mu N 
- \frac{f_{\omega NN}}{2
\Lambda_N} \bar N \sigma_{\mu \nu} \partial^\mu \omega^\nu  N   \, , 
\nonumber \\ 
\end{eqnarray}
Here and $\Lambda_N$ is a mass scale which we take as
$\Lambda_N=3M_N/N_c$ with $N_c$ the number of colours in QCD. An
overview of estimates of couplings from several sources is presented
in appendix \ref{sec:couplings}. In the large $N_c$ limit the
Lagrangean simplifies tremendously since one has the following scaling
relations~\cite{Witten:1979kh}~\footnote{There should be no confussion
in forthcoming sections when we take $N_c=3$ and $\Lambda_N=M_N$ and
the book-keeping becomes less evident.}
\begin{eqnarray}
M_N &\sim& N_c \, , \nonumber \\ 
\Lambda_N &\sim& N_c^0 \, , \nonumber \\ 
g_{\pi NN} & \sim & g_{\sigma NN} \sim g_{\omega NN} \sim f_{\rho NN} \sim \sqrt{N_c} \, , 
\nonumber \\ 
f_{\omega NN} &\sim& g_{\rho NN}  \sim 1/\sqrt{N_c}  \, , 
\nonumber \\  
m_\pi &\sim& m_\sigma \sim m_\rho \sim m_\omega \sim N_c^0 \, , \nonumber \\
\Gamma_\sigma &\sim& \Gamma_\rho \sim 1/N_c \, .
\end{eqnarray} 
The vector/tensor coupling dominance for $\omega$/$\rho$ is well
fulfilled phenomenologically (see appendix \ref{sec:couplings}).
Thus, in the large $N_c$ limit it is convenient to pass to the heavy
baryon formulation by the transformation
\begin{eqnarray}
N (x) = e^{i M_N v \cdot x} B(x) \, . 
\end{eqnarray} 
where $B(x)$ is the heavy iso-doublet baryon field and $v^\mu$ a
four-vector fulfilling $v^2=1$, eliminating the heavy mass
term~\cite{Jenkins:1990jv,Bernard:1992qa}. Choosing $v^\mu = (1,0)$
the meson-nucleon Lagrangean becomes
\begin{eqnarray}
{\cal L} &=& -g_{\sigma NN} s B^\dagger B + g_{\omega NN} \omega^0
B^\dagger B \nonumber \\ &+& g_{\pi NN} B^\dagger \sigma_i \tau_a B
\partial^i \phi^a + \frac{f_{\rho NN}}{2 \Lambda_N} \epsilon^{ijk}
B^\dagger \sigma_i \tau_a B \partial^j \rho^{ka} \, . \nonumber \\
\label{eq:heavy-lag}
\end{eqnarray} 
In the large $N_c$-limit the contracted $SU(4)$ algebra with the
generators given by the total spin $S_i = \sum_A \sigma_i^A /2$, the
total isospin $T_a= \sum_A \tau_a^A /2 $ and the Gamow-Teller $X_{ia}
= \sum_A \sigma_i^A \tau_a^A /4 $ operators is
satisfied~\cite{Manohar:1998xv,Jenkins:1998wy}).  One could, of
course, have started directly from the heavy-baryon Lagrangean,
Eq.~(\ref{eq:heavy-lag}), but the connection with chiral symmetry, in
particular the relativistic mass relation $M_N = g_{\sigma NN} f_\pi $
would be lost.

\subsection{OBE potentials at leading $N_c$}
\label{sec:OBE-largeNc}

From the heavy-baryon Lagrangean, Eq.~(\ref{eq:heavy-lag}), the
calculation of the NN potential in momentum space is
straightforward~\cite{Machleidt:1987hj,Machleidt:2000ge}. However,
passing to coordinate space is somewhat tricky since distributional
contributions proportional to $\delta (\vec x)$ and derivatives may
appear. We discard them but just assuming that $r>r_c$ where $r_c$ is
a short distance radial cut-off~\footnote{As discussed at length in
Ref.~\cite{PavonValderrama:2005wv,Entem:2007jg} these terms are
effectively inessential under renormalization of the corresponding
Schr\"odinger equation via the coordinate boundary condition method,
which will be explained shortly.}. According to their increasing
mass the leading $N_c$ contributions to the OBE potentials read
\begin{eqnarray}
V_\pi (r) &=& \frac1{12} \vec \tau_1 \cdot \vec \tau_2 \frac{g_{\pi
NN}^2}{4 \pi} \frac{m_\pi^2}{\Lambda_N^2} 
\Big[ \vec \sigma_1 \cdot
\vec \sigma_2 \frac{e^{-m_\pi r}}{r}  \nonumber \\  &+& 
 S_{12} \frac{e^{-m_\pi r}}{r}
\left( 1 + \frac{3}{m_\pi r} + \frac{3 }{(m_\pi r)^2}\right) \Big] \, , 
\\ V_\sigma (r) &=& - \frac{g_{\sigma NN}^2}{4 \pi} \frac{e^{-m_\sigma
r}}{r} \, , \\ V_\rho (r) &=& \frac1{12} \vec \tau_1 \cdot \vec \tau_2
\frac{f_{\rho NN}^2}{4\pi}\frac{m_\rho^2}{\Lambda_N^2} \Big[ 2 \vec
\sigma_1 \cdot \vec \sigma_2 \frac{e^{-m_\rho r}}{r} \nonumber \\ 
&-& S_{12}
\frac{e^{-m_\rho r}}{r} \left( 1 + \frac{3}{m_\rho r} + \frac{3
}{(m_\rho r)^2}\right) \Big] \, , \\ V_\omega (r) &=& \frac{g_{\omega
NN}^2}{4 \pi} \frac{e^{-m_\omega r}}{r} \, ,  
\end{eqnarray} 
where the tensor operator $ S_{12} = 3 \sigma_1 \cdot \hat x \sigma_2
\cdot \hat x - \sigma_1 \cdot \sigma_2 $ has been defined. Thus, the
structure of the leading large $N_c$-OBE potential has the general
structure~\cite{Kaplan:1996rk}
\begin{eqnarray}
V (r) = V_C (r) + \tau_1 \cdot \tau_2  \left[ \sigma_1 \cdot \sigma_2 W_S (r)
+ S_{12}  W_T(r) \right] \, .
\label{eq:pot-largeN}
\end{eqnarray} 
Thus, we have as the only non-vanishing components
\begin{eqnarray}
V_C (r) &=& - \frac{g_{\sigma NN}^2}{4 \pi} \frac{e^{-m_\sigma r}}{r}
+ \frac{g_{\omega NN}^2}{4 \pi} \frac{e^{-m_\omega r}}{r} \, , \nonumber 
\\ 
W_S(r)
&=& \frac1{12} \frac{g_{\pi NN}^2}{4\pi} \frac{m_\pi^2}{\Lambda_N^2}
\frac{e^{-m_\pi r}}{r} + \frac1{6} \frac{f_{\rho NN}^2}{4
\pi}\frac{m_\rho^2}{\Lambda_N^2} \frac{e^{-m_\rho r}}{r} \, , \nonumber 
\\ 
W_T(r) &=&
\frac1{12} \frac{g_{\pi NN}^2}{4 \pi}\frac{m_\pi^2}{\Lambda_N^2} \frac{e^{-m_\pi r}}{r} \left[ 1 + \frac{3}{m_\pi r} + \frac{3 }{(m_\pi r)^2}\right] \nonumber
\\ &-& \frac1{12} \frac{f_{\rho NN}^2}{4 \pi}\frac{m_\rho^2}{\Lambda_N^2} \frac{e^{-m_\rho
r}}{r} \left[ 1 + \frac{3}{m_\rho r} + \frac{3 }{(m_\rho r)^2}\right]\, , 
\nonumber \\
\end{eqnarray} 
At short distances we have 
\begin{eqnarray}
V_C (r) & \to & \frac{g_{\omega NN}^2-g_{\sigma NN}^2}{4 \pi} \, 
\frac1{r} \, , \\ W_S(r) &\to & \frac1{12} \frac{ g_{\pi NN}^2 m_\pi^2 + 2
f_{\rho NN}^2 m_\rho^2} {4\pi \Lambda_N^2} \, \frac1{r} \, , \\ W_T(r)
&\to& \frac1{4} \frac{g_{\pi NN}^2- f_{\rho NN}^2}{4 \pi \Lambda_N^2}
\, \frac{1}{r^3} \, .
\label{eq:1/r^3}
\end{eqnarray} 
As we see, the potential is singular at short distances except for the
very special value $f_{\rho NN}= g_{\pi NN} $ (see
Appendix~\ref{sec:exceptional}). While the central $V_C$ and spin
$W_S$ contributions present a mild Coulomb singularity, the tensor
force component $W_T$ is a more serious type of singularity, a
situation appeared already for the simpler OPE
potential~\cite{PavonValderrama:2005gu}.

\subsection{The OBE potential and ambiguities in  the S-matrix}
\label{sec:ambiguities}

We will show next that the S-matrix associated to the OBE potential is
necessarily ambiguous, precisely because of the short distance $1/r^3$
singularity in the non-exceptional situation $g_{\pi NN} \neq f_{\rho
  NN}$. The exceptional case, $g_{\pi NN} = f_{\rho NN}$ will be
treated in Appendix~\ref{sec:exceptional}.  We do so by proving that
the standard regularity conditions for the wave function do not
uniquely determine the solution of the Schr\"odinger
equation. Actually, at short distances, i.e. much smaller than meson
masses, $r \ll 1/m$, the NN problem due to the OBE potential
corresponds to the interaction of two spin-1/2 magnetic dipoles,
namely
\begin{eqnarray}
-\nabla^2 \Psi_k ( \vec x) + U_{dd} (\vec x) \Psi_k (\vec x) = p^2
\Psi_k (\vec x) \, , \qquad r \ll 1/m \, , \nonumber \\ 
\label{eq:sch-E-short}
\end{eqnarray} 
where the reduced dipole-dipole potential\footnote{Note that we are
  {\it not} assuming here this potential at large distances and so the
  standard long range problems of dipole scattering never appear.} is
given by
\begin{eqnarray}
U_{dd} (\vec x) &=& M V_{dd} (\vec x)  = \nonumber \\ &=& 
\pm \frac{R}{r^3} \, \left( 3 \sigma_1 \cdot \hat x \sigma_2 \cdot \hat
  x - \sigma_1 \cdot \sigma_2 \right) \, , 
\end{eqnarray} 
with $R$ is a length scale  and in our particular case 
\begin{eqnarray}
\pm R= \frac{M}{16 \pi \Lambda_N^2} (g_{\pi NN}^2 - f_{\rho NN}^2) \,
,
\end{eqnarray}
the positive or negative sign depends on whether $g_{\pi NN} > f_{\rho NN}$
or $g_{\pi NN} < f_{\rho NN}$ respectively.

\begin{figure*}[ttt]
\begin{center}
\includegraphics[height=5.5cm,width=5cm,angle=270]{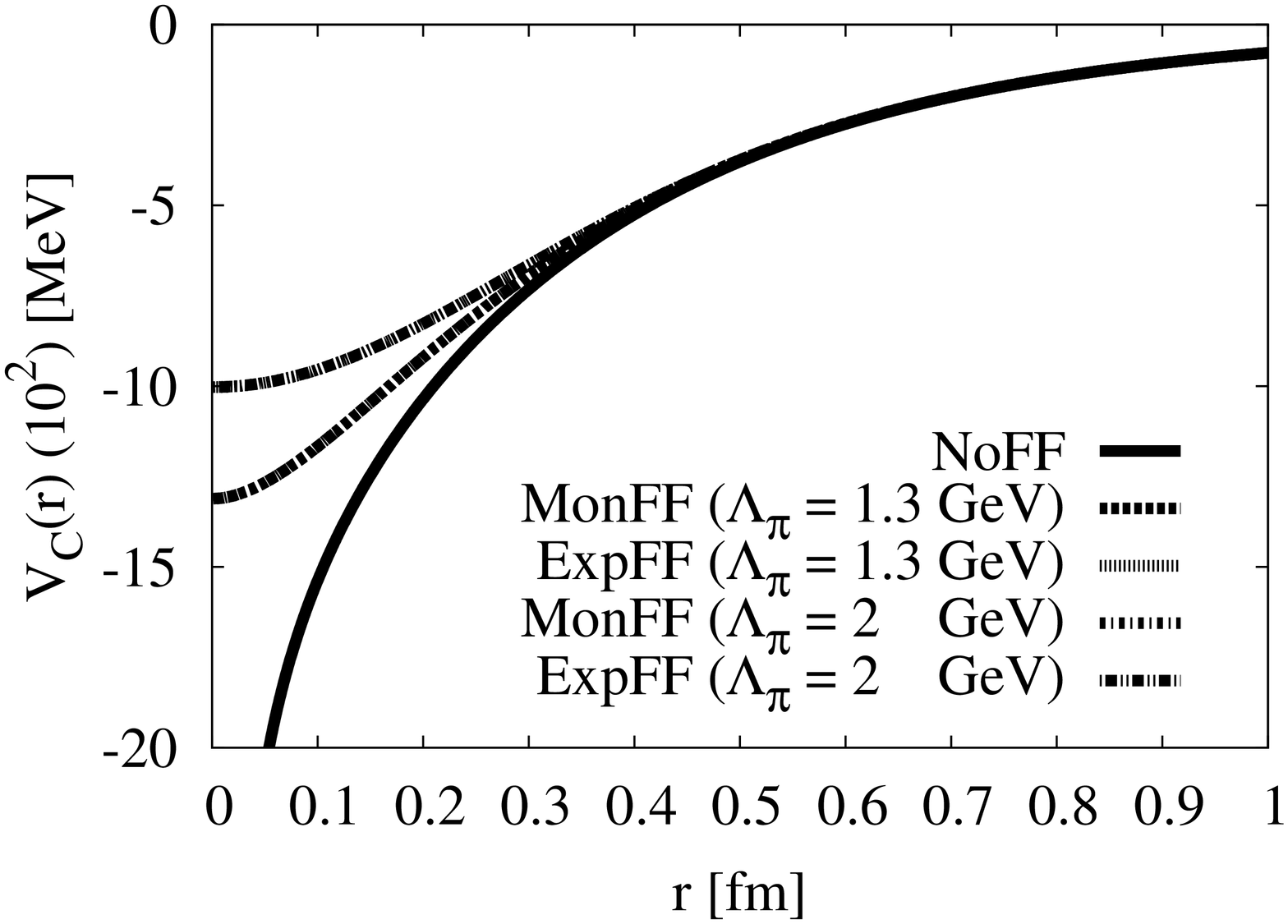}
\includegraphics[height=5.5cm,width=5cm,angle=270]{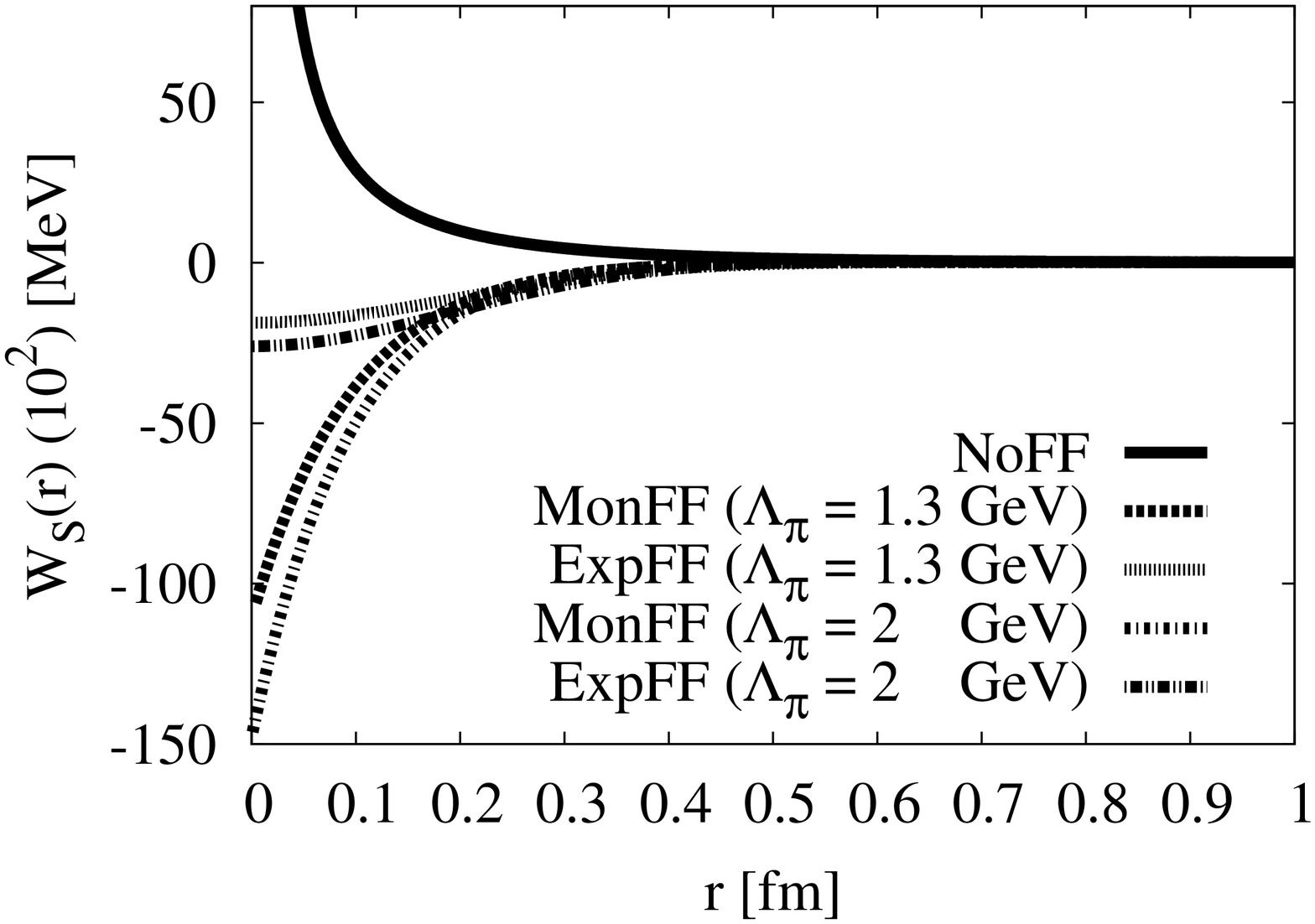}
\includegraphics[height=5.5cm,width=5cm,angle=270]{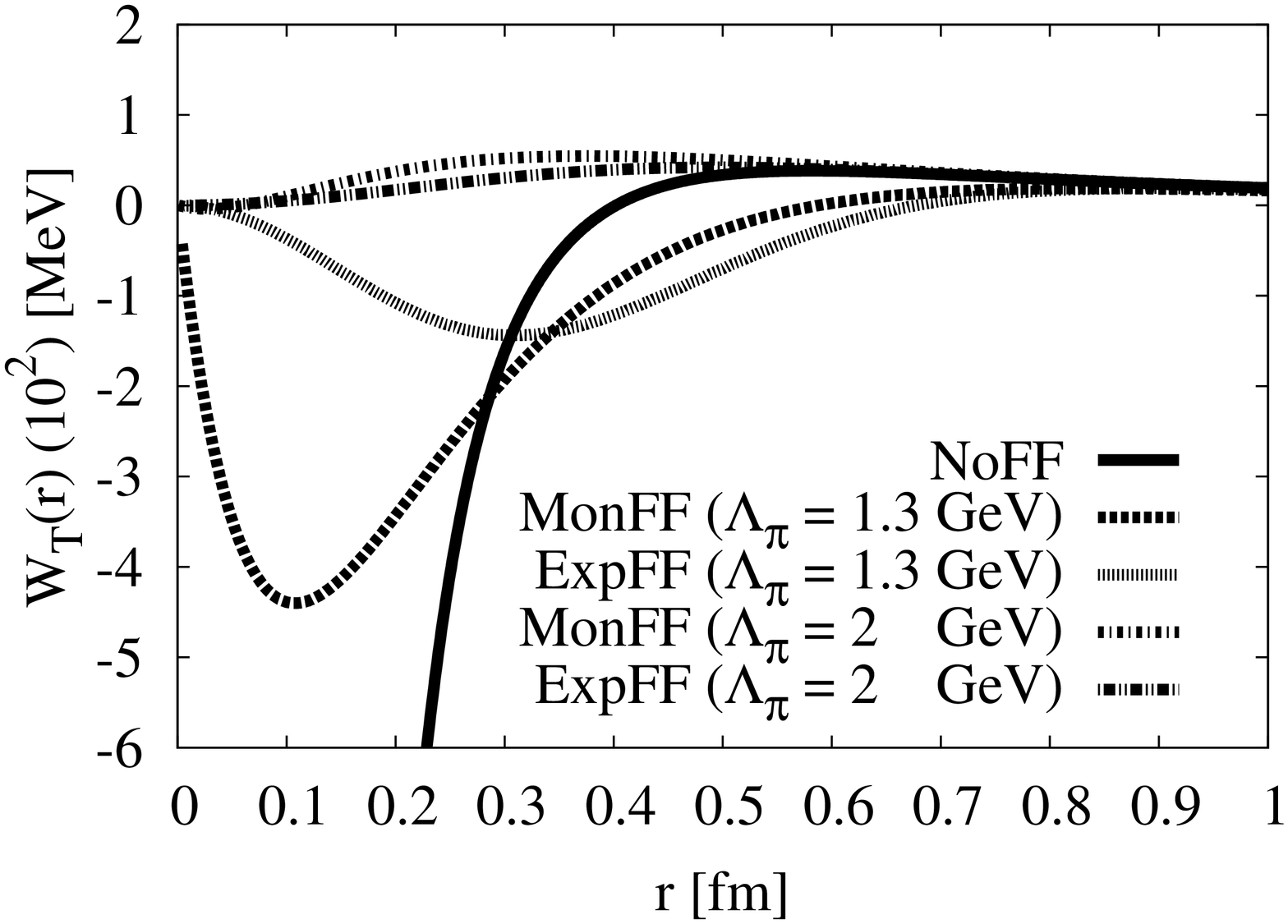}
\end{center}
\caption{The potentials $V_C(r)$, $W_S(r)$ and $W_T(r)$ in ${\rm MeV}$
  as a function of the distance (in fm). We include the effect of both
  exponential, Eq.~(\ref{eq:exponential}), and monopole
  Eq.~(\ref{eq:monopole}) form factors for $\Lambda_{\pi NN}=1300 {\rm
    MeV}$ and $\Lambda_{\pi NN}=2000 {\rm MeV}$. All other cut-offs
  are kept to $\Lambda_{\sigma NN}= \Lambda_{\rho NN}= \Lambda_{\omega
    NN}= 2000 {\rm MeV}$.}
\label{fig:cmp-pots}
\end{figure*}
 
The above potentials become diagonal in the standard total spin $\vec
S^2$, parity $U_P$, isospin $\vec T$ and total angular momentum $\vec
J= \vec L+\vec S$ basis, so the states are labeled by the
spectroscopic notation $^{2S+1}L_J$. We remind that Fermi-Dirac
statistics implies $(-1)^{L+S+T}=-1$. Thus, $\tau_1 \cdot \tau_2 =
2T(T+1)-3$ and $\sigma_1 \cdot \sigma_2 = 2S(S+1)-3$.  For spin
singlet states $S=0$ and $J=L$, the parity is natural $U_P = (-1)^J$
and one has $S_{12}=0$. For uncoupled spin triplet states $S=1$ one
has $J=L$, natural parity $U_P=(-1)^L$ and $S_{12}=2$. For coupled
spin triplet states $S=1$ one has $L=J\pm 1$, unnatural parity
$U_P=(-1)^{J+1}$ and
\begin{eqnarray}
S_{12}=\begin{pmatrix} -\frac{2(J-1)}{2J+1} &
\frac{6\sqrt{J(J+1)}}{2J+1} \\ \frac{6\sqrt{J(J+1)}}{2J+1} &
-\frac{2(J+2)}{2J+1} \end{pmatrix} \, .
\end{eqnarray} 
For the uncoupled spin-triplet channel we have 
\begin{eqnarray}
-v_J '' (r) + \left[ \frac{2 R}{r^3}  + \frac{J(J+1)}{r^2} \right] v_J (r)
&=& p^2 v_J(r) \, .
\end{eqnarray}
At very short distances we may neglect the centrifugal barrier and the
energy yielding
\begin{eqnarray}
-v_J '' (r) \pm \frac{2 R}{r^3}  v_J (r) =0 \, , \qquad r \ll 1/m, R, 1/p \, .
\end{eqnarray}
The general solution can be written in terms of Bessel
functions. Using their asymptotic expansions we may write at short
distances~\footnote{The solutions of $ -y''(x) - y(x)/x^3=0$ are
\begin{eqnarray}
\sqrt{x} J_1 ( 2 /\sqrt{x}) &=& - \frac{x^\frac34}{\sqrt{\pi}} 
\cos(\pi/4+2 /\sqrt{x}) + \dots  
\nonumber \\
\sqrt{x} Y_1 ( 2 /\sqrt{x}) &=& - \frac{x^\frac34}{\sqrt{\pi}} 
\cos(\pi/4-2 /\sqrt{x}) + \dots  \nonumber 
\end{eqnarray}
whereas the  solutions of $ -y''(x) + y(x)/x^3=0$ are
\begin{eqnarray}
\sqrt{x} K_1 ( 2 /\sqrt{x}) &=& \frac12 \sqrt{\pi }x^\frac34 e^{-2/\sqrt{x}}
+ \dots  
\nonumber \\
\sqrt{x} I_1 ( 2 /\sqrt{x}) &=& \frac1{2 \sqrt{\pi }} x^\frac34 e^{2/\sqrt{x}}
+ \dots  
\nonumber \\
\end{eqnarray}
}
\begin{eqnarray}
v_{+,J} (r) &\to & \left(\frac{r}{R}\right)^{3/4} \left[ C_{1R} e^{+ 4
    \sqrt{2} \sqrt{\frac{ R}{r}}} + C_{2R} e^{- 4 \sqrt{2}
    \sqrt{\frac{ R}{r}}} \right] \, , \nonumber \\ 
\label{eq:short_bc_r3}
\\ v_{-,J} (r) &\to
& \left(\frac{r}{R}\right)^{3/4} \left[ C_{1A} e^{- 4 i \sqrt{\frac{
        R}{r}}} + C_{2A} e^{ 4 i\sqrt{\frac{ R}{r}}} \right] \, . \nonumber
\end{eqnarray} 
Clearly in the repulsive case the regularity condition fixes the
coefficient of the diverging exponential to zero, $C_{1R}=0$, whereas
in the attractive case {\it both} linearly independent solutions are
regular and the solution is not unique.  In the case of the triplet
coupled channel, we have for $r \ll 1/m, R, 1/p$, i.e. neglecting
centrifugal barrier and energy, the system of two coupled differential
equations becomes
\begin{eqnarray}
\begin{pmatrix} -u_J''(r) \\ -w_J''(r) \end{pmatrix}
\pm \frac{R}{r^3}\begin{pmatrix} -\frac{2(J-1)}{2J+1} &
  \frac{6\sqrt{J(J+1)}}{2J+1} \\ \frac{6\sqrt{J(J+1)}}{2J+1} &
  -\frac{2(J+2)}{2J+1} \end{pmatrix} \begin{pmatrix} u_J(r)
  \\ w_J(r) \end{pmatrix} =0 \, . \nonumber \\ 
\label{eq:sch_coupled-dip} 
\end{eqnarray}
This system can be diagonalized by going to the rotated basis 
\begin{eqnarray}
\begin{pmatrix} v_{1,J}(r)
  \\ v_{2,J} (r) \end{pmatrix}
 &=& \begin{pmatrix}   
\sqrt{\frac{J}{2J+1}} &  -\sqrt{\frac{J+1}{2J+1}} \\ 
\sqrt{\frac{J+1}{2J+1}} &   \sqrt{\frac{J}{2J+1}} 
\end{pmatrix} \, \begin{pmatrix} u_J(r)
  \\ w_J(r) \end{pmatrix} ,
\end{eqnarray} 
where the new functions satisfy 
\begin{eqnarray}
-v_{1,J} '' (r) \mp \frac{4 R}{r^3} v_{1,J} (r) &=& 0 \, , \\ 
-v_{2,J} '' (r) \pm \frac{8 R}{r^3} v_{2,J} (r) &=& 0 \, .
\end{eqnarray}
Note that here the signs are alternate, i.e. when one of the
short-distance eigenpotentials is attractive the other one is
repulsive and viceversa, and hence the type of solutions in
Eq.~(\ref{eq:short_bc_r3}) can be applied. This means that in general
there will be solutions which {\it are not} necessarily fixed by the
regularity condition at the origin, and thus the OBE potential {\it
  does not} predict the $S-$matrix uniquely. Instead, a complete
parametric family of $S-$matrices will be generated depending on the
particular choice of linearly independent solutions, which are not
dictated by the OBE potential itself.

Thus, some additional information should be given. The traditional way
is to introduce form factors to kill the singularity so that the
regularity condition fixes the solution uniquely as we discuss in
Section~\ref{sec:standard}. Another way, which we discuss in the rest
of the paper, is to fix directly the integration constants from data
with or without form factors. As we will show this way of proceeding
does not make much difference showing a marginal influence of form
factors (see Sect.~\ref{sec:formfac}).

\section{The standard approach to OBE potentials with form factors}
\label{sec:standard}

\subsection{Features of Vertex Functions}

A way out to {\it avoid} the singularities is to implement vertex
functions in the OBE potentials corresponding to the replacement
($q^2= q_0^2 - \vec q^2$ is the 4-momentum)
\begin{eqnarray}
V_{m NN}(q) \to V_{m NN}(q) \left[\Gamma_{mNN}(q^2) \right]^2 \, . 
\end{eqnarray} 
Note that this assumes 1) Off-shell independence and 2) The form
factor is accurately known.  Standard choices are to take form factors
of the monopole~\cite{Machleidt:1987hj} and
exponential~\cite{Nagels:1977ze} parameterizations
\begin{eqnarray}
\Gamma_{mNN}^{\rm mon} (q^2) &=& \frac{\Lambda^2-m^2}{\Lambda^2-q^2} \,
, \label{eq:monopole}  \\ \Gamma_{mNN}^{\rm exp}(q^2) &=&
\exp \left[\frac{q^2-m^2}{\Lambda^2}\right] \, , \label{eq:exponential}
\end{eqnarray} 
fulfilling the normalization condition $\Gamma_{mNN}(m^2)=1$. These
forms are so constructed as to have the {\it same} slope at small
values of $q^2$ in the large cut-off expansion
\begin{eqnarray}
\Gamma_{mNN} (q^2) = 1 + \frac{q^2-m^2}{\Lambda^2} + {\cal O}
(\Lambda^{-4}) \, .
\end{eqnarray} 
so that the meaning for the cut-off is similar. In coordinate space
this can be easily implemented for Yukawa potentials using 
\begin{eqnarray}
Y_\Lambda(r) = \int \frac{d^3 q}{(2\pi)^3} \frac{e^{i q \cdot x }}{q^2+m^2} 
\left[\Gamma_{mNN} (q^2) \right]^2  \, , 
\end{eqnarray} 
yielding 
\begin{eqnarray}
Y_\Lambda^{\rm mon} (r) = \frac{e^{-mr}}{4 \pi r} - \frac{e^{-\Lambda r}}{4 \pi r} 
\left[ 1 + r \frac{\Lambda^2-m^2}{2\Lambda}\right]  \, , 
\end{eqnarray} 
which at short distances becomes finite,  
\begin{eqnarray}
Y_\Lambda^{\rm mon} (r) = \frac1{4\pi} \, \frac{(\Lambda-m)^2}{2 \Lambda} + {\cal O}(r^2)
  \, ,  
\end{eqnarray} 
which diverges linearly for $\Lambda \to \infty$.  The exponentially
regularized Yukawa potential reads
\begin{eqnarray} 
Y_\Lambda^{\rm exp} (r) = \frac{e^{-m r}}{8 \pi r } &+& \frac{e^{-m r}}{8
  \pi r } {\rm Erf} \left( \frac{\Lambda^2 r - 4 m}{2 \sqrt{2} \Lambda}\right) \nonumber \\ 
&-& \frac{e^{m r}}{8 \pi r } {\rm Erfc} \left( \frac{\Lambda^2 r + 4 m}{2
  \sqrt{2} \Lambda}\right) \, , 
\end{eqnarray} 
where ${\rm Erf}$ and ${\rm Erfc}$ are the error function and
complementary error function respectively~\footnote{They are defined
  as
$$ {\rm Erf} (z) = 1 - {\rm Erfc} (z) = \frac{2}{\sqrt{\pi}} \int_0^z
  dt \, e^{-t^2} = 1 - \frac{e^{-z^2}}{\sqrt{\pi} z} \left[ 1 + {\cal
      O} (z^{-1}) \right] $$ }.
For $\Lambda r \ll 1$ we have the finite result 
\begin{eqnarray} 
Y_\Lambda^{\rm exp} (r) = \frac{e^{-2m^2/\Lambda^2}\Lambda}{\sqrt{2 \pi}  4 \pi } - 
\frac{m}{4\pi} {\rm Erfc} \left( \frac{\sqrt{2} m}{
  \Lambda}\right)+ {\cal O} (r^2) \, ,    
\end{eqnarray} 
which diverges linearly for $\Lambda \to \infty$.  In the limit
$\Lambda r \gg 1$ behaves as
\begin{eqnarray} 
Y_\Lambda^{\rm exp} (r) = \frac{e^{-m r}}{4 \pi r } - \frac{e^{-\frac18 \Lambda^2 r^2}
  e^{-2m^2/\Lambda^2}}{\sqrt{2 \pi} \Lambda \pi r^2 } + \dots
\end{eqnarray} 
and the distortion of the original Yukawa potential is much more
suppressed in the exponential than in the case of monopole form
factor.

In any case we note the amazing feature that the form factors have a
radically different effect on different components of the potential.
While $V_C$ and $W_S$ with a mild $\sim 1/r$ short distance behaviour
become finite, the tensor force behaving as $W_T \sim 1/r^3 $ vanishes
at the origin {\it after} due to the form factors, $W_T^{\rm mon} (0)
= W_T^{\rm exp} (0) =0$. This can be seen from the expression
\begin{eqnarray}
&& \lim_{r \to 0} \int \frac{d^3 q}{(2\pi)^3} e^{i q \cdot x } \,
  \frac{\sigma_1 \cdot q \sigma_2 \cdot q }{q^2+m^2}
  \left[\Gamma_{mNN} (q^2) \right]^2 \nonumber \\ &&= \frac13
  \sigma_1 \cdot \sigma_2 \left[ \int \frac{d^3 q}{(2\pi)^3}
    \left[\Gamma_{mNN} (q^2) \right]^2 -m^2 Y_\Lambda (0) \right] \, , \nonumber \\
\label{eq:tensor-ff}
\end{eqnarray} 
which corresponds to take an angular average at short distances. This
feature suggests that the impact of the tensor force at short
distances should be small and looks clearly against the result of the
short distance analysis outlined in Section~\ref{sec:need} where there
is a strong mixing at short distances. As we will show in
Sect.~\ref{sec:triplet}, within the renormalization approach there is
no contradiction; physical observables will naturally display a small
mixing~\footnote{In other words, the counterterm structure is {\it
    not} of the naive form suggested by Eq.~(\ref{eq:tensor-ff}) but a
  more general one {\it including} the tensor operator $S_{12}$. See
  also the discussion in Sect.~\ref{sec:dispersion}}.

We show in Fig.~\ref{fig:cmp-pots} the potentials $V_C(r)$, $W_S(r)$
and $W_T(r)$ in ${\rm MeV}$ as a function of the distance (in fm). We
also include the effect of both exponential,
Eq.~(\ref{eq:exponential}), and monopole Eq.~(\ref{eq:monopole}) form
factors for $\Lambda_{\pi NN}=1.3 {\rm GeV}$ and $\Lambda_{\pi NN}=2
{\rm GeV}$. All other cut-offs are kept to the values
$\Lambda_{\sigma NN}= \Lambda_{\rho NN}= \Lambda_{\omega NN}= 2 {\rm
GeV}$. As we see, the distortion of the tensor component due to the
strong form factor takes place already at $r \sim 1 {\rm fm}$ for
softest cut-off $\Lambda_{\pi NN}=1.3 {\rm GeV}$. The key issue here
is to decide whether this distortion represents a true physical
effect, rather than a mere artifact of the regularization. This boils
down to determine if one can {\it visualize} finite nucleon size
effects when the probing wavelength is not shorter than $0.5 {\rm
fm}\le r \le 1 {\rm fm} $.  The fact that the monopole and exponential
parameterizations agree down to $r \sim 0.5 {\rm fm}$ but differ from
the bare unregularized potential suggests that one could look for a
true physical effect based on model independent distortions in the
region slightly above $0.5 {\rm fm }$. This point will be analyzed
further in Section~\ref{sec:formfac}.

Finally, note that the multiplicative manner in which form factors are
introduced, although it looks quite natural, does build in
correlations which may not reflect the real freedom one has in
general; a dynamical calculation need not comply to this factorization
scheme.

\begin{figure}[ttt]
\begin{center}
\includegraphics[height=6cm,width=5cm,angle=270]{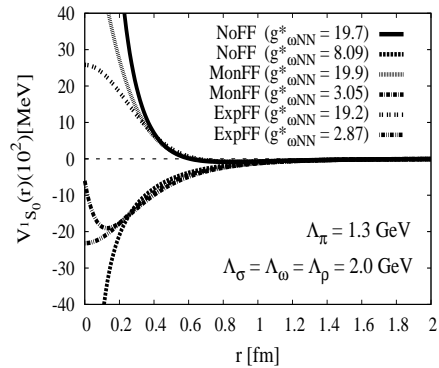}
\end{center}
\caption{The $^1S_0$ potential $V_{^1S_0}(r)$ in ${\rm MeV}$ as a
  function of the distance (in fm) for the different scenarios with
  large and small $\omega-$couplings. We include the effect of both
  exponential, Eq.~(\ref{eq:exponential}), and monopole
  Eq.~(\ref{eq:monopole}) form factors for $\Lambda_{\pi NN}=1300 {\rm
    MeV}$. All other cut-offs are kept to $\Lambda_{\sigma NN}=
  \Lambda_{\rho NN}= \Lambda_{\omega NN}= 2000 {\rm MeV}$.}
\label{fig:cmp-pot-1S0}
\end{figure}

\subsection{The problem of short distance sensitivity vs spurious bound states}
\label{sec:fine-tining}

The advantage of using vertex functions is that they make the OBE
non-singular at short distances. As a consequence, the choice of the
regular solution determines the solution {\it uniquely}. In this
section we analyze critically the use of form factors which are
customarily employed in NN calculations based on the OBE potential. We
will see that for natural choices of meson-nucleon parameters (see
Appendix~\ref{sec:couplings}), the NN potential displays short
distance insensitivity and at the same time spurious deeply bound
states. However, if we {\it insist} on not having spurious bound
states the resulting description is highly short distance sensitive.

As we have mentioned, NN scattering in the elastic region below pion
production threshold involves CM momenta $p < p_{max} = 400$
MeV. Given the fact that $1/m_{\omega} \sim 1/m_{\rho} \sim 0.25
\mathrm{fm} \ll 1/p_{max} = 0.5\mathrm{fm}$ we expect heavier mesons
to be irrelevant, and $\omega$ and $\rho$ to be marginally important,
even in s-waves, which are most sensitive to short distances.  This
desirable property has not been fulfilled in the traditional approach
to OBE forces.  In order
to illustrate this, we consider the $^1S_0$ channel, where the
potential (without form factor) is
\begin{eqnarray}
V_{^1S_0} (r) &=& V_C (r) - 3 W_S (r) \nonumber \\ 
&=&-\frac{ g_{\pi NN}^2 m_{\pi}^2} {16 \pi M_N^2}
\frac{e^{-m_{\pi} r}} {r} - \frac{ g_{\sigma NN} ^2}{4 \pi}\frac{e^{-
m_{\sigma} r}} {r}  
\nonumber \\ 
&+& \frac{g_{\omega NN}^2}{4 \pi}\frac{e^{-m_{\omega}
r}}{r} -
\frac{f_{\rho NN}^2 m_\rho^2}{8 \pi M_N^2 }\frac{e^{-m_{\rho}
r}}{r} \, .  
\label{eq:pot-1S0}
\end{eqnarray}
We take $m_{\pi} = 138$MeV, $M_{N} = 939$MeV, $m_{\rho} = 770$MeV,
$m_{\omega} = 783$MeV and $g_{\pi NN} = 13.1$, which seem firmly
established, and treat $m_{\sigma}$, $g_{\sigma NN }$ and $g_{\omega
  NN}$ and $f_{\rho NN}$ as fitting parameters.  To see the role of
vector mesons we note the redundant combination of coupling constants
$g_{\omega NN}^2 - f_{\rho NN}^2 m_\rho^2 /(2 M_N^2)$ which appears in
the $^1S_0$ potential when we take $m_\rho=m_\omega$, a tolerable
approximation within the present context. To avoid unnecessary strong
correlations we define the effective coupling
\begin{eqnarray} 
g_{\omega NN}^* = \sqrt{g_{\omega NN}^2 - \frac{f_{\rho NN}^2
    m_\rho^2}{2 M_N^2}} \, .
\label{eq:gww*}
\end{eqnarray}
Natural values for the coupling constants $g_{\omega NN} = 9-10.5$ and
$f_{\rho NN} = 14-18$ imply $g_{\omega NN}^* = 0-7$.  In
Fig.~\ref{fig:cmp-pot-1S0} the potential without and with monopole and
exponential vertex functions is depicted for several values of
$g_{\omega NN}^*$. As we see, the differences start below $1 {\rm fm}$
where the standard short distance repulsive core is achieved by large
and unnatural values of $g_{\omega NN}^*$, and not so much depending
on the form factors. On the other hand, if we use the regularized
$^1S_0$ potential at $r=0$ and take the natural values for the
coupling constants $g_{\omega NN} = 9-10.5$ and $f_{\rho NN} = 14-18$
the potential at the origin becomes
\begin{eqnarray}
V_{^1S_0} (0) = -(1000-3000) {\rm MeV} \, . 
\end{eqnarray} 
which is huge and attractive. The number of states is approximately given by the WKB estimate
\begin{eqnarray}
N_B \sim \frac1\pi \int_0^\infty \sqrt{-M V_{^1S_0} (r)} dr \, .  
\end{eqnarray}
which yields numbers around unity. 
In fact the potential accommodates a
deeply bound state, at about 
\begin{eqnarray}
E_B  = -(500-2000) {\rm MeV} 
\end{eqnarray} 
This state does not exist in nature and should clearly be ruled out
from the description on a fundamental level. On the other hand, we do
not expect such state to influence the low energy properties below the
inelastic pion production threshold $E_{\rm CM} =175 {\rm MeV}$ in any
significant manner.

\begin{figure}[ttt]
\begin{center}
\includegraphics[height=6cm,width=5cm,angle=270]{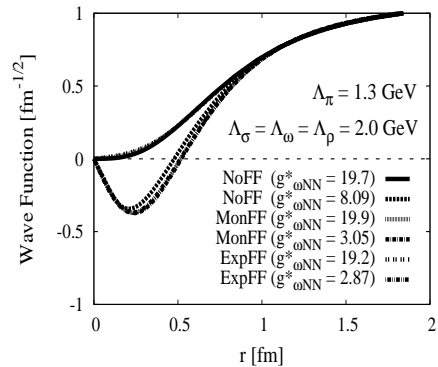}
\hskip1.5cm
\end{center}
\caption{Zero energy wave function for the singlet pn $^1S_0$ channel
  as a function of distance (in fm) and for the different scenarios
  with large and small $\omega-$couplings. We include the effect of both
  exponential, Eq.~(\ref{eq:exponential}), and monopole
  Eq.~(\ref{eq:monopole}) form factors for $\Lambda_{\pi NN}=1300 {\rm
    MeV}$. All other cut-offs
  are kept to $\Lambda_{\sigma NN}= \Lambda_{\rho NN}= \Lambda_{\omega
    NN}= 2000 {\rm MeV}$
This wave function goes
  asymptotically to $u_0(r) \to 1 - r /\alpha_0$ with $\alpha_0 =
  -23.74 {\rm fm}$ the scattering length in this channel. The zero at about 
$r=0.5 {\rm fm}$ signals the existence of a spurious bound state.}
\label{fig:u01S0-reg}
\end{figure}

\begin{table*}
\caption{\label{tab:table_fits_reg} Fits to the $^1S_0$ phase shift of
  the Nijmegen group~\cite{Stoks:1994wp} using the OBE potential
  without or with strong exponential and monopole form factor.  We
  take $m=138.03 {\rm MeV} $, and $g_{\pi NN} =13.1083 $
  \cite{deSwart:1997ep} and $m_\rho=m_\omega=770 {\rm MeV}$ and fit
  $m_\sigma$, $g_{\sigma NN}$ and $g_{\omega NN}^*$. We use
  $\Lambda_{\pi NN}=1300 {\rm MeV}$ and $\Lambda_{\sigma NN}=
  \Lambda_{\rho NN}= \Lambda_{\omega NN}= 2000 {\rm MeV}$. $E_B$
  represents the energy of the (spurious) bound state when it does
  exist.}
\begin{ruledtabular}
\begin{tabular}{|c|c|c|c|c|c|c|c|c|}
 & $r_c ({\rm fm})$ & $m_\sigma ({\rm MeV})$ & $g_{\sigma NN}$ &  $g_{\omega NN}^*$ & $\chi^2 /DOF $ &$\alpha_0 ({\rm fm})$ & $r_0 ({\rm fm})$  & $E_B ({\rm MeV}) $\\ 
\hline 
$\Gamma (q^2)=1$ & 0   & 547.55(4) & 13.559(8) & 19.68(2) & 0.869 & -23.742 & 2.702 & $-$ \\ 
$\Gamma (q^2)=1$ & 0.1 & 500.9 (5) & 9.61(1) & 8.09(2) & 0.484 & -23.742 & 2.504 & -638 \\ 
\hline 
$\Gamma (q^2)= \Gamma^{\rm exp} (q^2) $ & 0 & 552.57(5) & 13.78(2) & 19.21(4) & 0.664 &-23.741 & 2.703 & $-$ \\ 
$\Gamma (q^2)= \Gamma^{\rm exp} (q^2) $ & 0 & 525.1(1)  & 10.41(1) & 2.9(1) & 0.213 & -23.740 & 2.698 &-578 \\ 
\hline 
$\Gamma (q^2)= \Gamma^{\rm mon} (q^2) $ & 0 & 551.7(1) & 13.99(1) & 19.978 (11) & 0.971 & -23.741 & 2.707 & $-$\\ 
$\Gamma (q^2)= \Gamma^{\rm mon} (q^2) $ & 0 & 532.5(2) & 10.81(1) & 3.04(3) & 0.241 & -23.739 & 2.696 &-597 \\  
\end{tabular}
\end{ruledtabular}
\end{table*}

In the standard approach the scattering phase-shift $\delta_0(p)$ is
computed by solving the (s-wave) Schr\"odinger equation in
\textbf{r}-space
\begin{eqnarray}
-u''_p(r) + M_{N}\,V(r)\,u_p(r) &=& p^2\,u_p(r) \label{eq:Scrod-p} \\ 
u_p(r) &\to& \frac{\sin{\left(p r + \delta_0 (p)\right)}}{\sin{\delta_0(p)}}
\label{eq:up-asymp}
\end{eqnarray}
with a regular boundary condition at the origin 
\begin{eqnarray}
u_p(0)=0 
\label{eq:up-bc-reg}
\end{eqnarray}  
This boundary condition obviously implies a knowledge of the potential
in the whole interaction region, and it is equivalent to solve the
Lippmann-Schwinger equation in \textbf{p}-space.  In the usual
approach~\cite{Machleidt:1987hj,Machleidt:2000ge} everything is
obtained from the potential assumed to be valid for $0\leq r <
\infty$. In practice, and as mentioned above, strong form factors are
included mimicking the finite nucleon size and reducing the short
distance repulsion of the potential, but the regular boundary
condition is always kept. One should note, however, that due to the
\textit{unnaturally large} NN $^1S_0$ scattering length ($\alpha_0
\sim -23 {\rm fm}$), any change in the potential $V \to V + \Delta V$
has a dramatic effect on $\alpha_0$, since one obtains 
\begin{eqnarray}
\Delta\alpha_0 = \alpha_0^2 M_N \int_0^{\infty} \Delta V(r)
u_0(r)^2\mathrm{d}r
\label{eq:delta-a0}
\end{eqnarray}
a quadratic effect in the large $\alpha_0$.  This implies that
potential parameters \textit{must be fine tuned}, and in particular
the short distance physics. To illustrate this we make a fit the np
data of Ref.~\cite{Stoks:1994wp}. The results using the OBE potential
without or with strong exponential and monopole form
factor~\footnote{In this particular channel the regularity condition,
Eq.~(\ref{eq:up-bc-reg}) {\it determines} the solution completely
since the potential without vertex functions $V_{^1S_0} (r) \sim 1/r $
is not singular at short distances in the sense that $\lim_{r \to 0} 2
\mu |V(r)| r^2 = \infty$~\cite{Case:1950an,Frank:1971xx}.}  are
presented in Table~\ref{tab:table_fits_reg}. In all cases we have at
least two possible but mutually incompatible scenarios. An extreme
situation corresponds to the case with no form factors~\footnote{We
find strong non-linear and well determined correlations have been
found making a standard error analysis inapplicable. In this situation
we prefer to quote errors by varying independently the fitting
variables $g_{\sigma NN}$, $m_\sigma$ and $g_{\omega NN}^*$ until
$\Delta \chi^2 = 3.53$ as it corresponds to 3 degrees of
freedom.}. The small errors should be noted, in harmony with the fine
tuning displayed by Eq.~(\ref{eq:delta-a0}) and the corresponding
couplings and scalar mass are determined to high accuracy but turn out
to be incompatible. This is just opposite to our expectations and we
may regard these fits, despite their success in describing the data,
as unnatural.  The ambiguity in these solutions are typical of the
inverse scattering problem, and has to do with the number of bound
states allowed by the potential. Actually, this can be seen from
Fig.~\ref{fig:u01S0-reg} where the zero energy wave function is
represented. According to the oscillation theorem, the number of
interior nodes determines the number of bound states. Thus, the larger
values of $g_{\omega NN}^*$ correspond to a situation with no-bound
states since $u_0(r)$ does not vanish, whereas for the smaller
$g_{\omega NN}^*$ values one has a bound state as $u_0(r)$ has a zero,
which energy can be looked up in Table~\ref{tab:table_fits_reg}. Of
course, such a bound state does not exist in nature and it is thus
spurious. On the other hand they always take place at more than twice
the maximum energy probed in NN scattering, $E_{\rm CM}=175 {\rm
MeV}$, and we should not expect any big effect from such an state.
Note that despite the net repulsive $\omega$-vector and attractive
$\rho$-tensor couplings, the total potential would not be repulsive at
short distances with or without form factors in the solution with
natural couplings and a spurious bound state; the net short distance
repulsion comes about only in the solution with unnaturally large
coupling (see Fig.~\ref{fig:cmp-pot-1S0}).

From the table~\ref{tab:table_fits_reg} one can clearly understand the
usually too large values of the $g_{\omega NN}$ coupling constant as
compared to those from SU(3) symmetry $g_{\omega NN} \sim 9$ or from
the radiative decay $\omega \to e^+ e^-$ yielding $g_{\omega NN} =
10.2(4)$. Using the definition of $g_{\omega NN}^*$,
Eq.~(\ref{eq:gww*}), we get for $f_{\rho NN}=14-18$ large values of
$g_{\omega NN}=20-22$ for the case with no bound state, whereas more
natural values $g_{\omega NN}=8.5-10.5$ are obtained for the case with
one (spurious) bound state.

\section{Boundary condition renormalization and ultraviolet completeness}
\label{sec:dispersion}

According to the discussion of Sect.~\ref{sec:ambiguities} the short
distance $1/r^3$ singularity of the OBE potential makes the solution
ambiguous, and thus there is a flagrant need of additional information
not encoded in the potential itself. Of course, once we realize the
freedom of choosing suitable linear combinations of independent
solutions, we may question how general this choice can be, even if the
potential is not singular.  In this Section we derive constraints on
the short distance boundary condition.  As mentioned already, we work
with energy independent potentials.  In this section we show what this
requirement implies for the renormalization program. Using the
potential of Eq.~(\ref{eq:pot-largeN}) we solve the Schr\"odinger
equation,
\begin{eqnarray}
-\frac1 M \nabla^2 \Psi_k ( \vec x) + V (\vec x) \Psi_k (\vec x) = E_k
 \Psi_k (\vec x) \, , 
\label{eq:sch-E}
\end{eqnarray} 
where $\Psi(\vec x)$ is a spin-isospin vector with 4x4=16 components,
which usually satisfies the out-going wave boundary condition, 
\begin{eqnarray}
\Psi_k (\vec x) \to \left[ e^{i \vec k \cdot \vec x} + f(\hat k', \hat
k) \frac{e^{i kr}}{r} \right] \chi_{t,m_t}^{s,m_s} \, , 
\end{eqnarray} 
with $f(\hat k', \hat k)$ the quantum mechanical scattering matrix
amplitude and $\chi_{t,m_t}^{s,m_s}$ a 4x4 total spin-isospin
state. We apply a radial cut-off $r_c$ and consider that the local
potential $V(\vec x)$ is valid for the long distance region $r>
r_c$. The precise form of the interaction for the short distance
region $r < r_c$ is not necessary as the limit $r_c \to 0$ will be
taken at the end. To fix ideas we assume an energy independent
non-local potential, as we expect {\it genuine} energy dependence to
show up as sub-threshold inelastic (e.g. pion production) effects. Any
distributional terms $\sim \delta (\vec x)$ arising from the long
distance potential $V(\vec x)$ are necessarily included in the inner
region, $r < r_c$. The inner wave function $\Phi_k(\vec x)$ satisfies
\begin{eqnarray}
-\frac1 M \nabla^2 \Phi_k ( \vec x) + \int d^3 x' V (\vec x',\vec x)
 \Phi_k (\vec x') = E_k \Phi_k (\vec x) \, ,
\label{eq:sch-non-loc}
\end{eqnarray} 
and will be assumed to vanish at the origin. Using standard
manipulations and the Green identity we get for the inner and outer
regions
\begin{eqnarray}
&& (E_p-E_k) \int_{r < r_c} d^3 x \Phi_k^\dagger (\vec x) \Phi_p (\vec
x) \nonumber \\ &=& \int d\vec S \left[ \vec \nabla \Phi_k^\dagger
(\vec x) \Phi_p (\vec x) - \Phi_k^\dagger (\vec x) \vec \nabla \Phi_p
(\vec x) \right] \Big|_{r=r_c} \, ,
\label{eq:bc-inn}
\end{eqnarray} 
and 
\begin{eqnarray}
&& (E_p-E_k) \int_{r > r_c} d^3 x \Psi_k^\dagger (\vec x) \Psi_p (\vec
x) \nonumber \\ &=& -\int d\vec S \left[ \vec \nabla \Psi_k^\dagger
(\vec x) \Psi_p (\vec x) - \Psi_k^\dagger (\vec x) \vec \nabla \Psi_p
(\vec x) \right] \Big|_{r=r_c} \, ,
\end{eqnarray} 
respectively where the difference in sign form the inner to the outer
integration comes from opposite orientations in the integration
surface.
Clearly, orthogonality of states in the whole space for different energies, 
\begin{eqnarray}
\int_{r < r_c} d^3 x \Phi_k^\dagger (\vec x) \Phi_p (\vec x) + \int_{r
> r_c} d^3 x \Psi_k^\dagger (\vec x) \Psi_p (\vec x)  = 0 \, ,  
\end{eqnarray} 
can be achieved by setting the general and common boundary condition, 
\begin{eqnarray} 
\partial_r \Phi_p (\hat x r_c) &=& L_p (\hat x r_c) \Phi_p (\hat x r_c) \nonumber \\ 
\partial_r \Psi_p (\hat x r_c) &=& L_p (\hat x r_c) \Psi_p (\hat x r_c) \, .
\label{eq:bound-k}
\end{eqnarray} 
Here, $L_p ( \hat x r_c)$ is a self-adjoint matrix which may depend on
energy, and may be chosen to commute with the symmetries of the
potential $V(\vec x)$~\footnote{In practice this would mean taking 
$$ L (\hat x r_c ) = L_C (r_c) + \tau_1 \cdot \tau_2 \left[L_S (r_c)
\sigma_1 \cdot \sigma_2 + L_T (r_c) S_{12} \right]
$$ which implies at most only three counterterms for all partial
waves.  }. Deriving with respect to the energy the inner boundary
condition, Eq.~(\ref{eq:bc-inn}), i.e. taking $E_p = E_k + \Delta E$
and $\Phi_p (\vec x) =\Phi_k (\vec x) + \Delta E \partial\Phi_p (\vec
x) / \partial E$ , we get
\begin{eqnarray}
&&\int d\hat x \, \Phi_p^\dagger (\hat x r_c) \frac{\partial L_p (\hat
    x r_c)}{\partial E} \Phi_p (\hat x r_c) \nonumber \\ &=& \frac1{M
    r_c^2} \int_0^{r_c} r^2 dr\int d \hat x \, \Phi_p^\dagger (\hat x
  r) \Phi_p (\hat x r)\, ,
\end{eqnarray} 
where we see that $ M\partial L_p (\hat x r_c) / \partial E \sim r_c$.
The important issue here is that regardless on the representation at
short distances, the boundary condition must become energy independent
when $r_c \to 0$, namely
\begin{eqnarray}
\lim_{r_c \to 0} \frac{\partial L_p
(\hat x r_c)}{\partial E} =0 \, , 
\end{eqnarray} 
provided one has 
\begin{eqnarray}
\lim_{r_c \to 0} \int_0^{r_c} dr\int d \hat x \Phi_p^\dagger (\hat x
r) \Phi_p (\hat x r) =0 \, .  
\label{eq:prob}
\end{eqnarray}
Thus we may take a fixed energy, e.g. zero energy, as a reference
state.
\begin{eqnarray}
\lim_{r_c \to 0} L_p (\hat x r_c) = \lim_{r_c \to 0} L_0 (\hat x r_c)
 \, , 
\end{eqnarray} 
The condition of Eq.~(\ref{eq:prob}) is the quite natural quantum
mechanical requirement that the contribution to the total probability
in the (generally unknown) short distance region is small. This is the
physical basis of the renormalization program which corresponds to
the mathematical implementation of short distance insensitivity and
which we carry out below for the OBE potential.  It should be noted
that this requirement depends on the potential. The condition of
Eq.~(\ref{eq:prob}) implies that in the limit $r_c \to 0 $ one must
always choose a normalizable outer solution $\Psi_k(\vec x)$ at the
origin and the boundary condition must be chosen independent on
energy. Note that energy dependence would be allowed if the cut-off
was kept finite, and still the requirement of orthogonality in the
{\it whole} space could be fulfilled for an interaction characterized
by a non-local and energy independent potential in the inner
region. This simultaneous disregard of both non-local and energy
dependent effects was advocated long ago by Partovi and
Lomon~\cite{Partovi:1969wd} on physical grounds, and as we see, it is
a natural consequence within the renormalization approach.

The renormalization procedure is then conceptually simple since any
energy state with given quantum numbers can be chosen as a reference
state to determine the rest of the bound state spectrum and scattering
states. For, instance using a bound state (the deuteron) $\Psi_d(\vec
x)$, at long distances (see Sect.~\ref{sec:triplet} for details)
\begin{eqnarray}
\Psi_d ( \vec x) \to \frac{A_S}{\sqrt{4 \pi} r} e^{-\gamma r} \left[
1+ \frac\eta{\sqrt{8}} S_{12} \right] \chi_{pn}^{s m_s} \, , 
\end{eqnarray} 
we integrate in the deuteron equation 
\begin{eqnarray}
-\frac1 M \nabla^2 \Psi_d ( \vec x) + V (\vec x) \Psi_d (\vec x) =
-\frac{\gamma^2}{M} \Psi_d( \vec x) \, ,
\end{eqnarray} 
and determine the short distance boundary condition matrix
$L(\hat x r_c)$ from 
\begin{eqnarray} 
\partial_r \Psi_d (\hat x r_c) = L(\hat x r_c) \Psi_d (\hat x r_c) \, .
\end{eqnarray} 
Then, using the {\it same} boundary condition for the finite
energy state, 
\begin{eqnarray} 
\partial_r \Psi_k (\hat x r_c) = L(\hat x r_c) \Psi_k (\hat x r_c) \, , 
\end{eqnarray} 
we integrate out the finite energy equation (\ref{eq:sch-E}) whence
the scattering amplitude may be obtained. In this manner the deuteron
binding energy defines the appropriate self-adjoint extension spanning
the relevant Hilbert space in the $^3S_1-^3D_1$
channel. Renormalization is achieved by taking the limit $r_c \to 0$
at the end of the calculation. The conditions under which such a
procedure is meaningful will be discussed below for the particular
partial waves under study, but a fairly general discussion can be
found in
Refs.~\cite{PavonValderrama:2005gu,PavonValderrama:2005wv,PavonValderrama:2005uj}. Relevant
cases for chiral potentials where this condition turned out {\it not}
to be true are discussed in Ref.~\cite{PavonValderrama:2005wv}. We will
also encounter below a similar situation in our description of the
deuteron and the $^3S_1-^3D_1$ channel. As a consequence of the
previous limit the completeness relation reads
\begin{eqnarray}
\int \frac{d^3 k}{(2\pi)^3} \Psi_k (\vec x) \Psi_k^\dagger (\vec x') +
\sum_{E_n <0} \Psi_n (\vec x) \Psi_n^\dagger (\vec x') = \delta (\vec
x - \vec x') {\bf 1} \, .\nonumber \\
\end{eqnarray} 
Besides the deuteron, the sum over negative energy states contains
most frequently spurious bound states, and for the singular potential
such as the one we are treating here there are infinitely many. They
show up as oscillations in the wave function at short distances, and
are a consequence of extrapolating the long distance potential to
short distances. On the other hand, from the above decomposition one
may write a dispersion relation for the scattering
amplitude~\footnote{This is done by using the Lippmann-Schwinger equation
in the form $T=V+VGV$ with $G=(E-H)^{-1}$, and normalization $\langle
\vec k | \vec x \rangle = e^{i \vec k \cdot \vec k}$ and $\langle
\Psi_k | \vec x \rangle = \Psi_k( \vec x)$ whence $ f( \hat k' , \hat
k) = -M/(4\pi) \langle \vec k' | T (E) | \vec k \rangle $.} of the form
\begin{eqnarray}
f(\hat k' , \hat k) &=& f_B (\hat k', \hat k) 
-\frac{M}{4\pi} \sum_{E_n < 0} \frac{\langle \vec k' | V | \Psi_n
  \rangle \langle \Psi_n | V | \vec k \rangle}{E-E_n} \nonumber \\ &-&
  \frac{M}{4\pi} \int \frac{d^3 q}{(2\pi)^2} \frac{\langle \vec k' | V
  | \Psi_q \rangle \langle \Psi_q | V | \vec k \rangle}{E- q^2/M}
\end{eqnarray} 
where $f_B (\hat k', \hat k)$ is the Born amplitude and the physical
and spurious bound states occur as poles in the scattering matrix at
negative energies $E=E_n$ and the discontinuity cut along the real and
positive axis is given by the second term only. Clearly, the influence
of these spurious bound states poled is suppressed if their energy
$E_n \ll E_d $. Given the fact that these states do occur in practice
it is mandatory to check their precise location to make sure that they
do not influence significantly the calculations, or else one should
study the dependence of the observables on the short distance cut-off
$r_c$ starting from a situation where it is small but still large
enough as to prevent the occurrence of spurious bound states. It
should be noted, however, that in no case can the spurious states
occur in the first Riemann sheet of the complex energy plane. This
restriction complies to causality, and implies in particular the
fulfillment of Wigner inequalities as was discussed for the $^1S_0$
channel in Ref.~\cite{PavonValderrama:2005wv}~\footnote{Causality
  violations, i.e. poles in the first Riemann sheet of the complex
  energy plane are easy to encounter (see
  e.g. \cite{PavonValderrama:2005wv}), particularly with energy
  dependent boundary conditions. A prominent example is an s-wave
  without potential and having $u_p' (0) / u_p (0) = p \cot
  \delta_0(p) = - 1/\alpha_0 + r_0 p^2 /2 + v_2 p^4$, which for the
  $^1S_0$-channel values of parameters $\alpha_0 = -23.74 {\rm fm}$,
  $r_0 = 2.75 {\rm fm} $ and $v_2=-0.48 {\rm fm}^3$ yields besides the
  well-known virtual state in the second Riemann sheet $E_v=-0.066
  {\rm MeV}$ a spurious bound state at $E_B=-18.37 {\rm MeV}$ and an
  unphysical pole at $E=128.88 \pm i 46.45 {\rm MeV}$. However, finite
  cut-offs {\it and} energy independent boundary conditions are
  guaranteed not to exhibit these problems, while some spurious bound
  states may be removed.}.

\begin{figure*}
\includegraphics[height=6.5cm,width=5.5cm,angle=270]{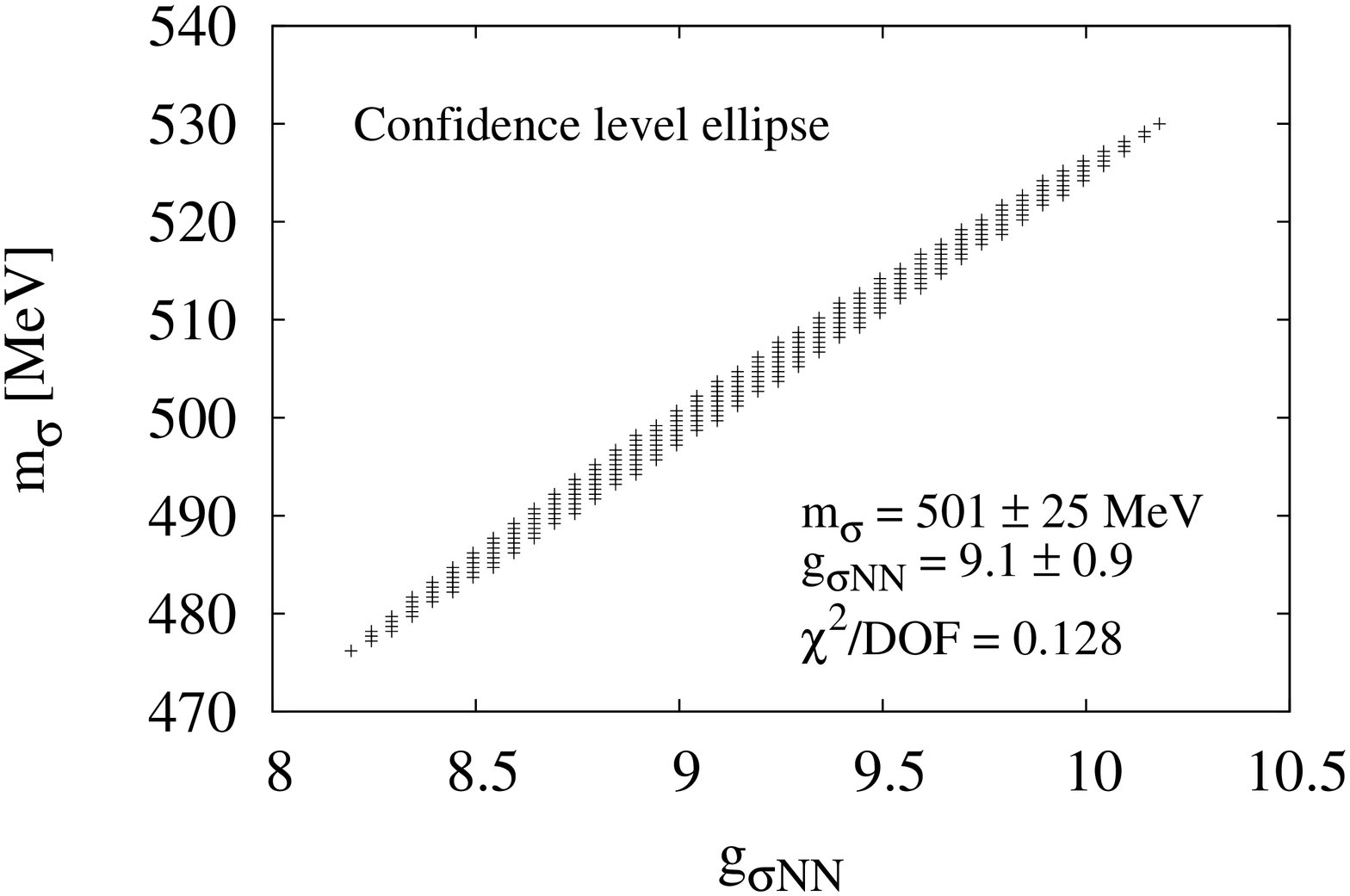}
\includegraphics[height=6.5cm,width=5.5cm,angle=270]{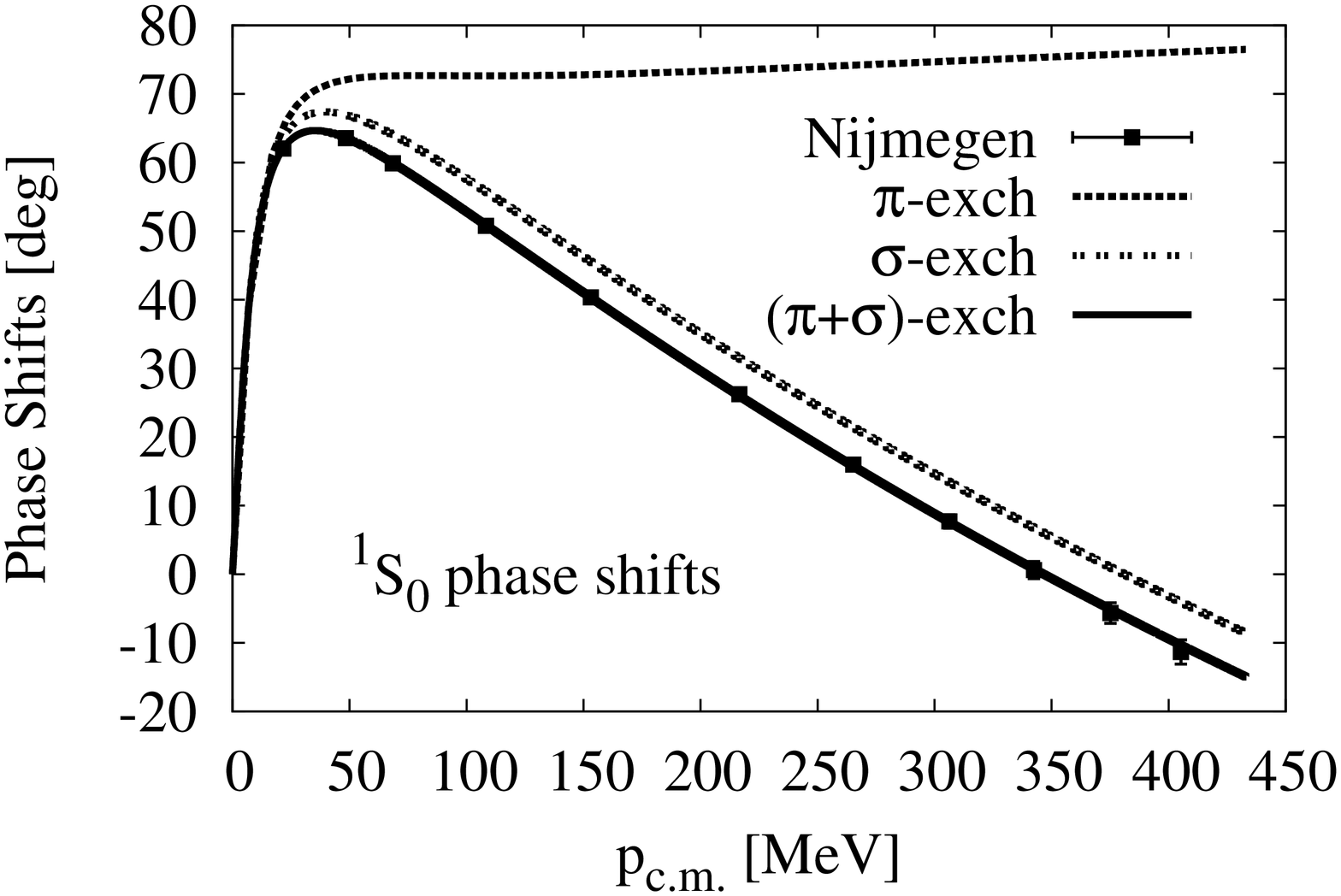}
\caption{ Left: $\Delta \chi^2 = 2.3 $ confidence level ellipse
  (corresponding to $68\%$ for 2 variables) in the $g_{\sigma
    NN}-m_\sigma$ plane without vector mesons $g_{\omega NN}=f_{\rho
    NN}=0$.  Right: Renormalized OBE $^1S_0$ pn phase shifts (in
  degrees) as a function of CM momentum. Data from
  \cite{Stoks:1994wp}.}
\label{fig:elipsefits}
\end{figure*}

\begin{figure*}
\includegraphics[height=5.5cm,width=5.5cm,angle=270]{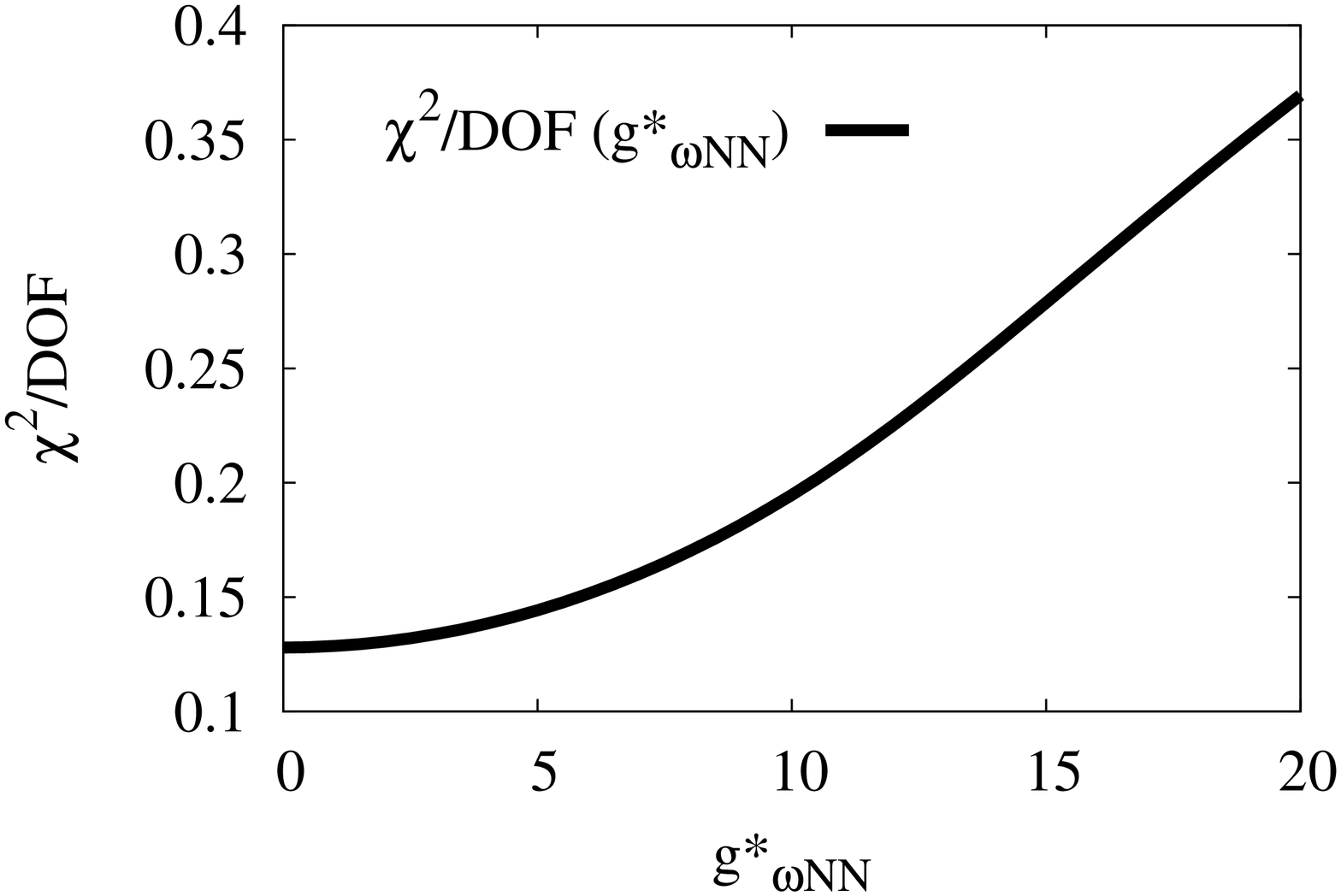}  
\includegraphics[height=5.5cm,width=5.5cm,angle=270]{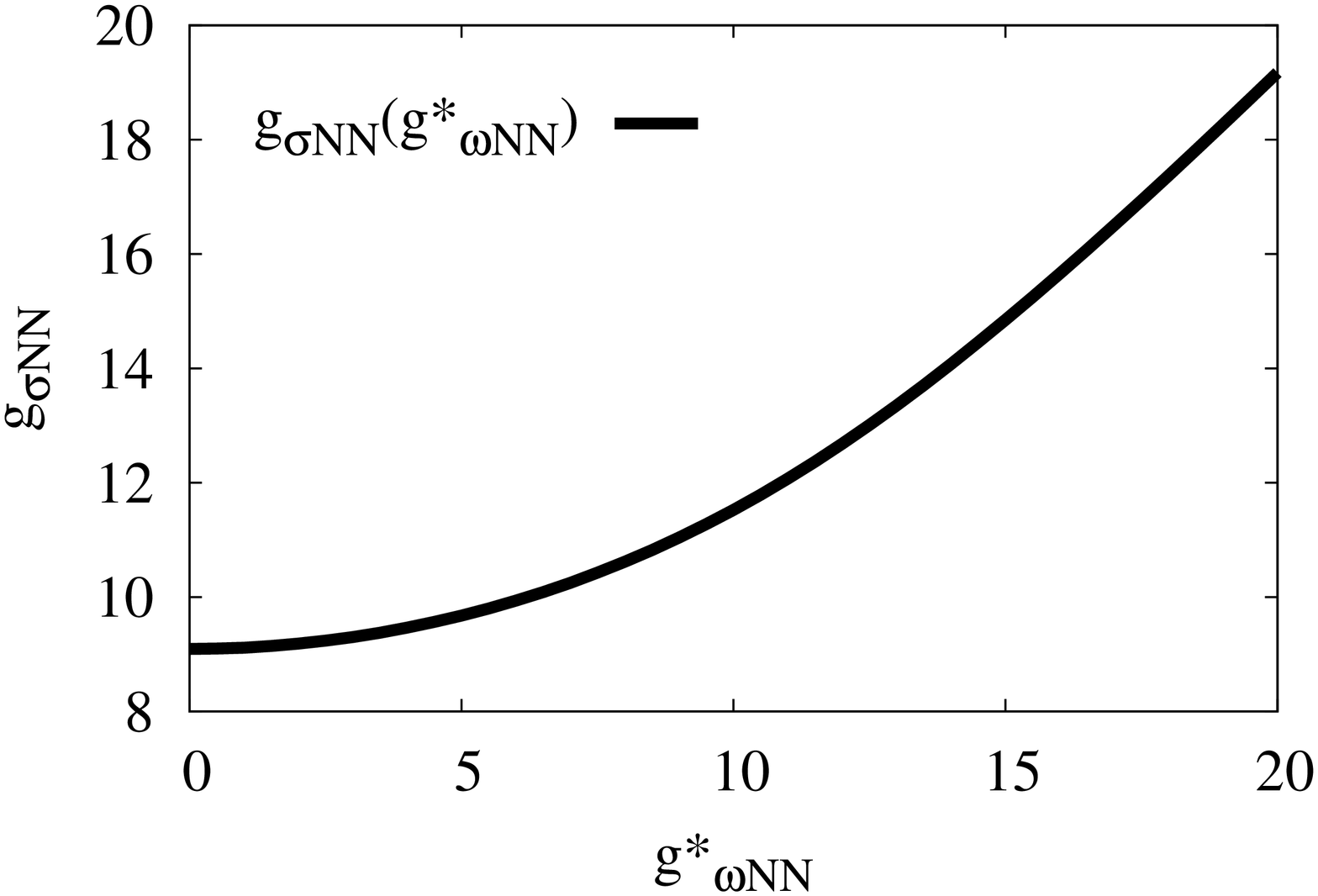}
\includegraphics[height=5.5cm,width=5.5cm,angle=270]{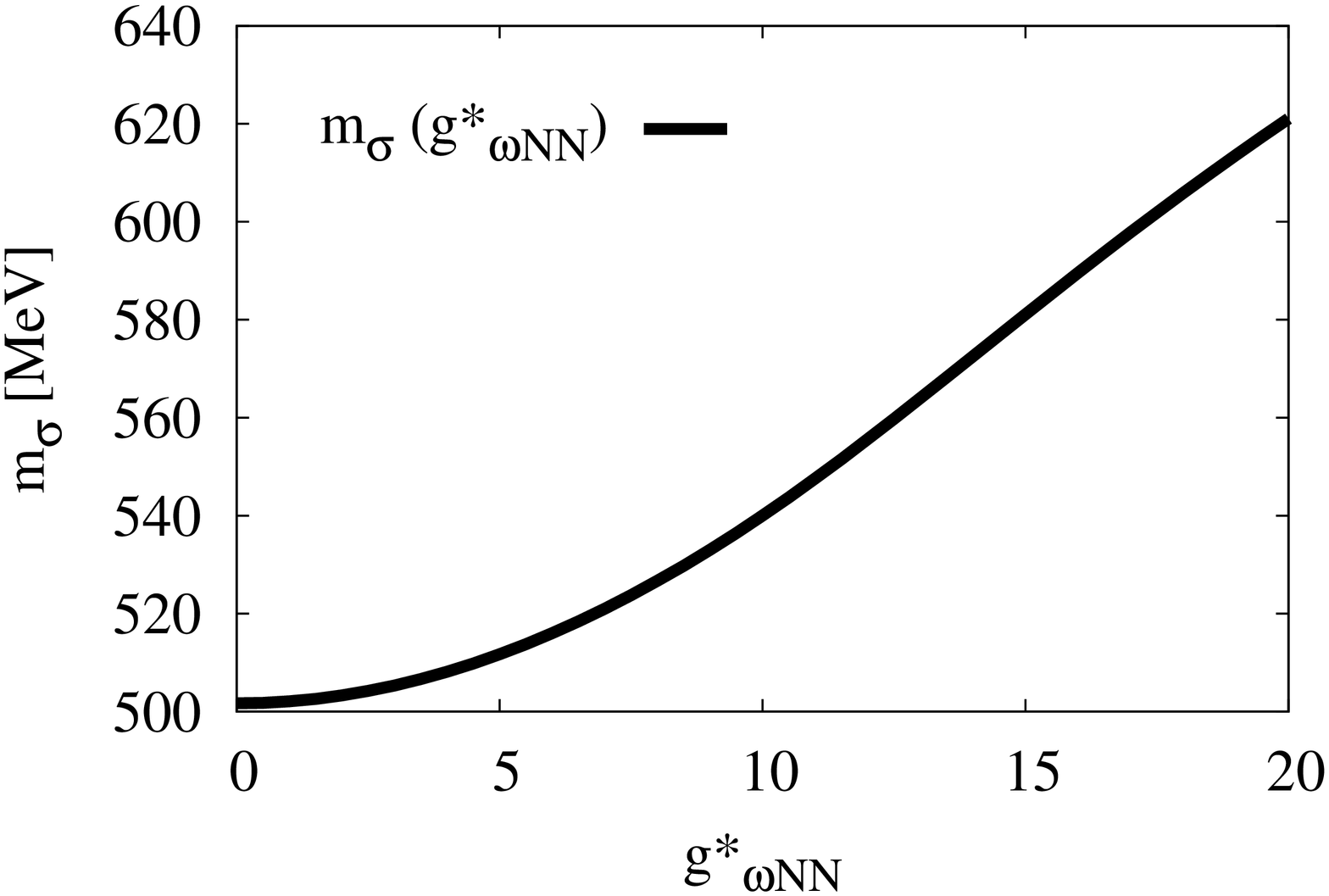}
\caption{$\chi^2 /{\rm DOF}$ (left panel), $g_{\sigma NN}$ (middle
panel) and $ m_\sigma $ (right panel) as a function of the effective
coupling constant $g_{\omega NN}^* = \sqrt{g_{\omega NN}^2- f_{\rho
NN}^2 m_\rho^2 / 2 M_N^2}$ when a fit of the leading $N_c$
contributions to the OBE ($\sigma+\pi+\rho+\omega$) potential is
considered. }
\label{fig:chi2-gwNN}
\end{figure*}

\section{The singlet channel}
\label{sec:singlet}

\subsection{Equations and boundary conditions}

The $^1S_0$ wave function in the pn CM system can be written as
\begin{eqnarray}
\Psi (\vec x) &=& \frac1{\sqrt{4\pi}\, r } \, u(r) \chi_{pn}^{s m_s}
\end{eqnarray} 
with the total spin $s=0$ and $m_s=0 $. The function $u(r)$ is the
reduced S- wave function, satisfying 
\begin{eqnarray}
-u_p '' (r) + U_{^1S_0} (r) u_p (r)  &=& p^2 u_p (r) 
\label{eq:sch_singlet} 
\end{eqnarray}
where one has
\begin{eqnarray}
U_{^1S_0} = M ( V_C -3 W_S  )    
\end{eqnarray} 
At short distances the  OBE potential behaves as a Coulomb type interaction, 
\begin{eqnarray} 
U_{^1 S_0} (r) & \to & \pm \frac1{Rr} 
\label{eq:pot_sing_short}
\end{eqnarray} 
where 
\begin{eqnarray}
\pm \frac1{R} &=& \frac{M}{4\pi} \left[g_{\omega NN}^2 -\frac{f_{\rho NN}^2
    m_\rho^2}{2 M_N^2} - g_{\sigma NN}^2 - f_{\pi NN}^2 \right]
\label{eq:short-R}
\end{eqnarray} 
Here, $f_{\pi NN}= g_{\pi NN} m_\pi /(2 M_N) $. 
The repulsive or attractive character of the interaction depends on a
balance among coupling constants. The short distance solution can be
written as a linear combination of the regular and irregular solution
at the origin
\begin{eqnarray}
u_p (r) &\to & c_1 (p)+ c_2 (p)r/R 
\label{eq:u-short}
\end{eqnarray} 
where in principle the arbitrary constants $c_1 (p)$ and $c_2 (p)$
depend on energy. To fix the undetermined constants we impose orthogonality
for $r > r_c$ between the zero energy state and the state with
momentum $p$ and get
\begin{eqnarray} 
u_p'(r_c) u_k (r_c) &-& u_k'(r_c) u_p(r_c) \nonumber \\ 
&=& ( k^2-p^2) \int_{r_c}^\infty u_k(r)
 u_p(r) dr =0 \, ,
\label{eq:orth_singlet}
\end{eqnarray} 
Taking the limit $r_c \to 0$ implies the following energy independent
combination~\cite{Valderrama:2005wv}
\begin{eqnarray}
\frac{c_1(p)}{c_2(p)} = \frac{c_1(k)}{c_2(k)} = \frac{c_1(0)}{c_2(0)} 
\label{eq:c's-short}
\end{eqnarray} 
leaving one fixed ratio which can be determined from e.g. the zero
energy state or any other reference state.

\subsection{Phase shifts}

For a finite energy scattering state we solve for the OBE potential
with the normalization
\begin{eqnarray}
u_p (r) \to \frac{\sin(pr + \delta_0(p))}{\sin \delta_0(p)} \, ,    
\label{eq:norm}
\end{eqnarray} 
with $\delta_0(p)$ the phase shift. For a potential falling off
exponentially $\sim e^{-m_\pi r}$ at large distances, one has the
effective range expansion at low energies, $|p| < m_\pi/2$,
\begin{eqnarray}
p \cot \delta_0 (p) = - \frac1\alpha_0 + \frac12 r_0 p^2 + v_2 p^4 + \dots 
\end{eqnarray} 
with $\alpha_0$ the scattering length and $r_0$ the effective range.
The phase shift is determined from Eq.~(\ref{eq:norm}).  Thus, for the
zero energy state we solve
\begin{eqnarray}
-u_0 '' (r) + U_{^1S_0} (r) u_0 (r)  &=& 0 \, , 
\label{eq:sch_singlet_0} 
\end{eqnarray}
with the asymptotic normalization at large distances, obtained from 
Eq.~(\ref{eq:norm}),  
\begin{eqnarray} 
u_0 (r) &\to & 1- \frac{r}{\alpha_0}  \, , 
\label{eq:u0-long}
\end{eqnarray}
In this equation $\alpha_0$
is an input, so one integrates in Eq.~(\ref{eq:sch_singlet_0}) from
infinity to the origin. Then, the effective range defined as
\begin{eqnarray} 
r_0 &=& 2 \int_0^\infty dr \left[ \left(1-\frac{r}{\alpha_0} \right)^2-
u_0 (r)^2 \right] 
\label{eq:r0_singlet}
\end{eqnarray} 
can be computed. From the superposition principle of boundary
conditions
\begin{eqnarray} 
u_0 (r) &= & u_{0,c} (r) - \frac1{\alpha_0} u_{0,s} (r)  \, , 
\end{eqnarray} 
where $u_{0,c} (r) \to 1 $ and $ u_{0,s} (r) \to r $ correspond
to cases where the scattering length is either infinity or zero
respectively. Using this decomposition one gets
\begin{eqnarray} 
r_0  &=&  A + \frac{B}{\alpha_0}+ \frac{C}{\alpha_0^2}  \, ,    
\label{eq:r0_univ} 
\end{eqnarray} 
where 
\begin{eqnarray}
A &=& 2 \int_0^\infty dr ( 1 - u_{0,c}^2 ) \, , \\    
B &=& -4 \int_0^\infty dr ( r - u_{0,c} u_{0,s} ) \, , \\    
C &=& 2 \int_0^\infty dr ( r^2 - u_{0,s}^2 )    \, ,  
\end{eqnarray} 
depend on the potential parameters only. The interesting thing is that
all dependence on the scattering length $\alpha_0$ is displayed
explicitly by Eq.~(\ref{eq:r0_univ} ).  To determine the phase shift
$\delta_0(p)$ one proceeds as follows. From Eq.~(\ref{eq:u0-long}) and
integrating in Eq.~(\ref{eq:sch_singlet_0}) one determines $c_1(0)$
and $c_2(0)$ and uses Eq.~(\ref{eq:c's-short}) to determine the ratio
$c_1(p)/c_2(p)$ and integrates out Eq.~(\ref{eq:sch_singlet}) matching
Eq.~(\ref{eq:norm}). This way the phase shift $\delta_0(p)$ is
determined from the potential and the scattering length as {\it
independent} parameters. As it was shown in Ref.~\cite{Entem:2007jg}
this procedure is completely equivalent to renormalize the
Lippmann-Schwinger equation with one counterterm.

\subsection{Fixing of scalar parameters}
\label{sec:fic-scalar}

In this work we will fix our parameters in such a way that the $^1S_0$
phase shift is reproduced. This has the advantage that the scalar
meson parameters are determined for the rest of observables.  Thus,
fixing the scattering length $\alpha_0 = -23.74 {\rm fm}$ and the OPE
potential parameters $g_{\pi NN}=13.1$ and $m_\pi=138.04 {\rm MeV}$ we
fit $g_{\sigma NN}$ and $m_\sigma$ to the $^1S_0$ phase shift of the
Nijmegen group~\cite{Stoks:1994wp}. In the absence of vector meson
contributions, i.e. taking $g_{\omega NN}=f_{\rho NN}=0$ the fit
yields
\begin{eqnarray}
g_{\sigma NN} = 9(1) \qquad m_\sigma = 501(25) {\rm MeV}
\label{eq:sigma-fit}
\end{eqnarray} 
with a $\chi^2 /DOF = 0.13$. As we see from Fig.~\ref{fig:elipsefits},
there is a large, in fact linear, correlation, between the scalar
coupling and mass, while the fit is quite good as we can see.  For
comparison we also show the result with OPE which, despite reproducing
the threshold behaviour does a poor job elsewhere. We quote also the
effective range values from the universal low energy theorem,  
\begin{eqnarray}
r_0 &=& 1.3081 -
\frac{4.5477}{\alpha_0}+\frac{5.1926}{\alpha_0^2} \, \qquad (\pi)
\nonumber 
\\
&=& 1.5089 {\rm fm} \, , 
\nonumber 
\\
r_0&=& 2.4567 - \frac{5.5284}{\alpha_0} + \frac{5.7398}{\alpha_0^2}
\qquad (\pi+\sigma) \nonumber \\
&=& 2.6989 {\rm fm} \, , 
\end{eqnarray} 
where the corresponding numerical values when the experimental $\alpha_0 = 
-23.74 {\rm fm} $ have also been added. 

\begin{figure}
\includegraphics[height=6.5cm,width=5.5cm,angle=270]{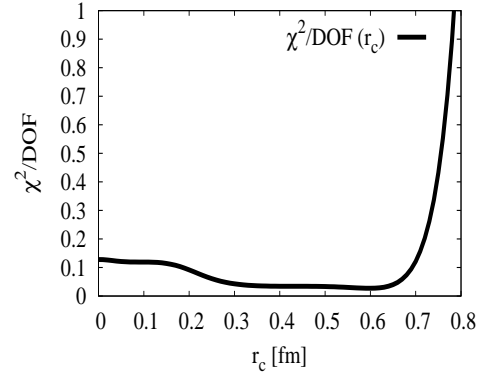} \\ 
\includegraphics[height=6.5cm,width=5.5cm,angle=270]{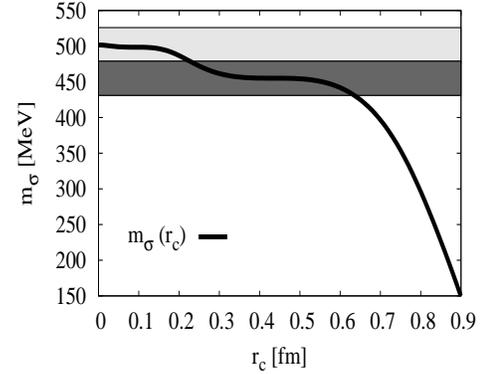}
\caption{Short distance cut-off $r_c$ (in {\rm fm}) dependence of the
  $^1S_0$ phase shift fitting procedure. $\chi^2 /{\rm DOF}$ (upper
  panel) and $m_\sigma$ (lower panel). The bands represent the error
  estimate for two fixed cut-off values: a finite cut-off $r_c=0.4
  {\rm fm}$ and the renormalized case $r_c= 0 {\rm fm}$.}
\label{fig:chi2-rc}
\end{figure}

It is interesting to analyze the dependence of the fitted scalar
parameters on the short distance cut-off radius, $r_c$. A {\it priori}
we should see the $\sigma$ exchange for $r_c \le 1/m_\sigma = 0.4 {\rm
  fm}$. From Fig.~\ref{fig:chi2-rc} we see the masses and the
couplings providing an acceptable fit $\chi^2 /{\rm DOF} < 1$ for
which a reliable error analysis may be undertaken. As we see this
happens for $r_c < 0.6 {\rm fm}$ and two stable plateau regions
yielding two potentially conflicting central $m_\sigma$ values.  An 
error analysis both at a finite cut-off value $r_c=0.4 {\rm fm}$ and
the renormalized cut-off limit $r_c= 0 {\rm fm}$.  gives two
overlapping and hence compatible bands. This shows that in this case
the data do not discriminate below $r_c=0.5 {\rm fm}$. Much above
that scale, the $\sigma$ meson becomes nearly irrelevant, as the
coupling becomes rather small.

Alternatively, we may treat the cut-off itself as a fitting parameter.
To avoid the large $m_\sigma-g_{\sigma NN}$ correlations displayed in
Fig.~\ref{fig:elipsefits} we fix the coupling constant to its central
value $g_{\sigma NN}=9.1$ and get then $r_c=0.10^{+0.13}_{-0.07} {\rm
  fm}$ and $m_\sigma=500(3) {\rm MeV}$. This shows that removing the
cut-off is not only a nice theoretical requirement, but also a
preferred phenomenological choice.

To analyze now the role of vector mesons we note, as already discussed
in Section \ref{sec:fine-tining}, the redundant combination of
coupling constants $g_{\omega NN}^2 - f_{\rho NN}^2 m_\rho^2 /(2
M_N^2)$ which appears in the $^1S_0$ potential when we take
$m_\rho=m_\omega$.  We thus define the effective coupling $g_{\omega
  NN}^* $ defined in Eq.~(\ref{eq:gww*}).  This combination is
responsible for the repulsive contribution to the potential in the
$^1S_0$ channel. From typical values of the couplings $g_{\omega
  NN}=9-10.5$ and $f_{\rho NN} =15-17$ we expect $g_{\omega NN}^*$ to
be effectively small. We show in Fig.~(\ref{fig:chi2-gwNN}) the
corresponding $\chi^2 /DOF $ as well as the readjusted scalar mass
$m_\sigma$ and coupling $g_{\sigma NN}$ as a function of the effective
combination of coupling constants, $g_{\omega NN}^*$. As we see, the
fit is rather insensitive but actually slightly worse than without
vector mesons when their contribution is repulsive. Thus, we will fix
this effective coupling to zero which corresponds to take
\begin{eqnarray}
g_{\omega NN}^2 = \frac{f_{\rho NN}^2 m_\rho^2}{2 M_N^2} 
\label{eq:gwNN-frNN}
\end{eqnarray} 
This choice has the practical advantage of fixing $g_{\sigma NN}$ and
$m_\sigma$ to the values provided in Eq.~(\ref{eq:sigma-fit}) also
when the leading $N_c$ vector meson contributions are
included. Moreover, it is also phenomenologically satisfactory as we
have discussed above. In Sect.~\ref{sec:triplet} we will also see that
deuteron or triplet $^3S_1-^3D_1$ do not fix the deviations from the
relation given by Eq.~(\ref{eq:gwNN-frNN}).

\subsection{Discussion}

The linear $g_{\sigma NN}-m_\sigma$ correlation can be established
solely by requiring that the effective range of the Nijmegen group
$r_0 = 2.67 {\rm fm}$ or any other be
reproduced~\cite{RuizArriola:2007wm}.  Actually,
Eq.~(\ref{eq:sigma-fit}), yields the combination $C_\sigma = g_{\sigma
  NN}^2 / m_\sigma^2 = 331 (50) {\rm GeV}^{-2}$ which is fixed by the
effective range and not by the scattering length. This is in contrast
with the resonance saturation viewpoint adopted in
Ref.~\cite{Epelbaum:2001fm} where this combination fixes the
scattering length.

Furthermore our calculation shows that an accurate fit without
explicit contribution of the vector mesons is possible. In particular,
our potential does not exhibit any repulsive region. This is in
apparent contradiction with the traditional view point that the
$\omega$-meson is responsible for the short range repulsion of the
nuclear force.

\begin{figure}[ttt]
\begin{center}
\includegraphics[height=6cm,width=5cm,angle=270]{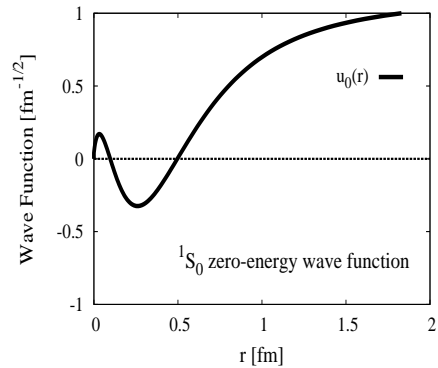}
\hskip1.5cm
\end{center}
\caption{Zero energy wave function for the singlet pn $^1S_0$ channel as
a function of distance (in fm). This wave function goes asymptotically
to $u_0(r) \to 1 - r /\alpha_0$ with $\alpha_0 = -23.74 {\rm fm}$ the
scattering length in this channel.}
\label{fig:u01S0}
\end{figure}

To understand this issue we plot in Fig.~\ref{fig:u01S0} the zero
energy wave function obtained by integrating in with the physical
scattering length $\alpha_0$. As we see, there appear two zeros
indicating, according to the oscillation theorem, the existence of two
negative energy spurious bound states. To compute such a state we
solve Eq.~(\ref{eq:sch_singlet}) with negative energy $E_B=-
\gamma_B^2 /M$, for an exponentially decaying wave function, $u_B(r)
\to A_B e^{-\gamma_B r}$ (normalized to one), and impose orthogonality
to the zero energy state, namely
\begin{eqnarray}
u_0 u_B' - u_0' u_B \Big|_{r=r_c} = 0 \, , \nonumber \\
\end{eqnarray} 
from which $\gamma_B$ can be determined. A direct calculation yields
$E_{B1} = - 777 {\rm MeV}$ and $A_{B1}=15.64 {\rm fm}^{-1/2}$ and
$E_{B2} = - 11077 {\rm MeV}$ and $A_{B2}=27.43 {\rm fm}^{-1/2}$. If we
regard the scattering amplitude as a function of energy in the complex
plane, these spurious bound state energies are beyond well the maximum
CM energy we want to describe in elastic NN scattering, $ E_{\rm CM}
\le 175 {\rm MeV}$, and so have no practical effect on the scattering
region. The appearance of spurious bound states in EFT approaches are
commonplace; one must check that they are beyond the considered energy
range.

In order to discuss this point further we may try several ways of
removing the unwanted poles and to quantify the effect on the results.
Unitarity implies the usual relation between the partial wave
amplitude and the phase shift
\begin{eqnarray}
\left[ f_0(p) \right]^{-1} = p \cot \delta_0(p)-ip
\end{eqnarray} 
Actually, the contribution of a negative energy state to the
s-wave scattering amplitude is a pole contribution
\begin{eqnarray}
 f_0(p)|_B=  -\frac{A_B^2}{M} \frac1{E+|E_B|}= -\frac{A_B^2}{p^2+\gamma_B^2}
\end{eqnarray} 
A simple way of subtracting such a bound state without spoiling
unitarity and preserving the value of the amplitude at threshold
$f_0(0) = F_0(0)=-\alpha_0$ is to modify the real part of the inverse
amplitude as follows, 
\begin{eqnarray}
 \frac1{F_0(p)}=  \frac1{f_0(p)} -\frac{p^2}{A_B^2} \, ,   
\end{eqnarray} 
which has no pole at $E=-|E_B|$, since $F_0 ( i \gamma_B) =
A_B^2/\gamma_B^2$. Using the relation between amplitude and phase
shift $F_0(p) = 1/(p \cot
\Delta_0(p)-ip)$ we get the modified phase shift, 
\begin{eqnarray}
p \cot \Delta_0 (p) =  p \cot \delta_0 (p)  + \frac{p^2}{A_B^2} \, , 
\end{eqnarray} 
which corresponds to a change in the effective range  
\begin{eqnarray}
\Delta r_0|_B = \frac{2}{A_B^2}
\label{eq:delta-r0} 
\end{eqnarray} 
For the values of the two spurious bound states we get $\Delta
r_0|_B=0.008, 0.002 {\rm fm}$, a tiny amount. The change in the phase
shift never exceeds $0.1^0$. Of course, this is not the only procedure
to remove spurious bound states, but the result indicates that the
effect should be small.

Another practical way to verify this issue is to study the influence
of changing the cut-off $r_c$ from the lowest value not generating any
spurious bound state and the origin, corresponding to look for $u_0
(a)=0$. This point is clearly identified as the outer zero of the wave
function, which takes place at about $a = 0.5 {\rm fm}$. Thus, if we
choose $r_c=a$, there will not be any bound state. For this particular
point, the orthogonality of states, Eq.~(\ref{eq:orth_singlet}),
implies that $u_p (a)=0$, resembling the standard hard core picture,
if we assume $u_p(r)=0$ for $r \le a$. Thus, at this $r_c$ our method
would correspond to infinite repulsion below that scale.  In other
words, the boundary condition {\it does} incorporate some effective
repulsion which need not be necessarily visualized as a potential. The
advantage of using a boundary condition is that we need not require
modelling nor deep understanding on the inaccessible and unknown short
distance physics.

The contribution to the effective range from the origin to the ``hard
core'' radius $a$ is $r_0^{\rm in} \sim 0.04 {\rm fm}$, while the
change in the phase shift at the maximum energy due to the inner
region $0 \le r \le a $ is $\Delta \delta_0 = 6^0$ to be compared with
the error estimate $ \Delta \delta_0=0.7^0$ from the PWA analysis of
the Nijmegen group~\cite{Stoks:1993tb} or the $ \Delta \delta_0=2^0$
from the corresponding high quality potentials~\cite{Stoks:1994wp}. If
we identify this hard core radius to the breakdown scale of the
potential, these differences might be interpreted as a systematic
error of the renormalization approach for our OBE potential and, as we
see, they turn out to be rather reasonable.

\section{The triplet channel} 
\label{sec:triplet} 

\subsection{Equations and boundary conditions}

The $^3S_1-^3D_1$ wave function in the pn CM system can be written as
\begin{eqnarray}
\Psi (\vec x) &=& \frac1{\sqrt{4\pi}\, r } \, \Big[ u(r) \sigma_p \cdot
\sigma_n \nonumber \\ &+& \frac{w(r)}{\sqrt{8}} \left( 3 \sigma_p
\cdot \hat x \, \sigma_n \cdot \hat x - \sigma_p \cdot \sigma_n \right)
\Big] \chi_{pn}^{s m_s}
\end{eqnarray} 
with the total spin $s=1$ and $m_s=0,\pm 1$ and $\sigma_p$ and
$\sigma_n$ the Pauli matrices for the proton and the neutron
respectively. The functions $u(r)$ and $w(r)$ are the reduced S- and
D-wave components of the relative wave function respectively. They 
satisfy the coupled set of equations in the $^3S_1 - ^3D_1 $ channel
\begin{eqnarray}
-u '' (r) + U_{^3S_1} (r) u (r) + U_{E_1} (r) w (r) &=& M E u
 (r) \, ,\nonumber \\ -w '' (r) + U_{E_1} (r) u (r) + \left[U_{^3D_1}
 (r) + \frac{6}{r^2} \right] w (r) &=& M E  w (r) \, , \nonumber \\ 
\label{eq:sch_coupled} 
\end{eqnarray}
with $ U_{^3S_1} (r)$, $U_{E_1} (r) $ and $U_{^3D_1} (r)$ the
corresponding matrix elements of the coupled channel potential
\begin{eqnarray}
U_{^3S_1} &=&  M(V_C   -3 W_S) \, , \nonumber   
\\ 
U_{E_1} &=&  -6 \sqrt{2} M W_T  \, ,   \nonumber 
\\ 
U_{^3D_1} &=&  M( V_C    -3 W_S  + 6W_T )\, .    
\end{eqnarray} 
At short distances one has the leading singularity
\begin{eqnarray}
U_{^3S_1} &= &  {\cal O} (r^{-1}) \, , \nonumber   
\\ 
U_{E_1} &=&  -\frac{4 \sqrt{2}R}{r^3} + {\cal O} (r^{-1})  \, ,   \nonumber 
\\ 
U_{^3D_1} &=& -\frac{12R}{r^3} + {\cal O} (r^{-1})  \, .
\end{eqnarray} 
where 
\begin{eqnarray}
\pm R = \frac{g_{\pi NN}^2 - f_{\rho NN}^2}{32\pi M_N}
\label{eq:R-pm}
\end{eqnarray}
This is very similar to the pure OPE case treated in
Ref.~\cite{PavonValderrama:2005gu} but with the important technical
difference that for $f_{\rho NN} < g_{\pi NN}$ and $f_{\rho NN} >
g_{\pi NN}$ there is a turn-over of repulsive-attractive eigenchannels
since the effective short distance scale $R$ changes sign.  Thus, we
must distinguish two different cases~\footnote{The exceptional case,
  $g_{\pi NN} = f_{\rho NN}$ corresponds to a regular potential and
  will be treated in Appendix~\ref{sec:exceptional}}. At short
distances we have for $g_{\pi NN} > f_{\rho NN}$ the plus sign in
Eq.~(\ref{eq:R-pm}) yielding
\begin{eqnarray}
u_A (r) &=& \sqrt{\frac23} u (r)  + \frac{1}{\sqrt{3}}w (r) \, , \nonumber  \\
u_R (r) &=& -\frac1{\sqrt{3}}u (r) + \sqrt{\frac23} w (r) \, , 
\label{eq:eigenvectors1}
\end{eqnarray} 
whereas for $g_{\pi NN} < f_{\rho NN}$ the minus sign in
Eq.~(\ref{eq:R-pm}) is taken and the solutions are interchanged
\begin{eqnarray}
u_R (r) &=& \sqrt{\frac23} u (r)  + \frac{1}{\sqrt{3}}w (r) \, , \nonumber  \\
u_A (r) &=& -\frac1{\sqrt{3}}u (r) + \sqrt{\frac23} w (r) \, , 
\label{eq:eigenvectors2}
\end{eqnarray} 
yielding an attractive singular potential $U_A \to -4 R/r^3 $ for
$u_A$ and $U_R \to 8 R/r^3 $ for $u_R$, which solutions are 
\begin{eqnarray}
u_R (r) &\to & \left(\frac{r}{R}\right)^{3/4} \left[ C_{1R} e^{+ 4
\sqrt{2} \sqrt{\frac{ R}{r}}} + C_{2R} e^{- 4 \sqrt{2} \sqrt{\frac{
R}{r}}} \right] \, , \nonumber \\ \\ u_A (r) &\to &
\left(\frac{r}{R}\right)^{3/4} \left[ C_{1A} e^{- 4 i \sqrt{\frac{
R}{r}}} + C_{2A} e^{ 4 i\sqrt{\frac{ R}{r}}} \right] \, . \nonumber
\label{eq:short_bc}
\end{eqnarray} 
The constants $C_{1R}$, $C_{2R}$, $C_{1A}$ and $C_{2A}$ depend on both
$\gamma $ and $\eta$ and the OBE potential parameters.  As it was
discussed in Ref.~\cite{PavonValderrama:2005gu} we must define a
common domain of wave functions to define a complete solution of the
Hilbert space in this $^3S_1-^3D_1$ channel.  This is achieved taking
\begin{eqnarray}
u_R (r) &\to & C_{R} (\gamma) \left(\frac{r}{R}\right)^{3/4} e^{- 4
\sqrt{2} \sqrt{\frac{ R}{r}}}  \, ,  \nonumber \\ 
u_A (r) &\to & C_{A}(\gamma) \left(\frac{r}{R}\right)^{3/4} \sin \left[ 4
\sqrt{\frac{ R}{r}} + \varphi \right] \, . 
\label{eq:short_bc_reg}
\end{eqnarray} 
Here, the short distance phase $\varphi$ is energy independent. This
can be done by matching the numerical solutions to the short distance
expanded ones, a cumbersome procedure in
practice~\cite{PavonValderrama:2005gu}. It is far more convenient to
use an equivalent short distance cut-off method with a boundary
condition. Thus, at the cut-off boundary, $r=r_c$ we can impose a
suitable regularity condition depending on the sign of $g_{\pi NN}^2 -
f_{\rho NN}^2$. A set of possible auxiliary boundary conditions was
discussed in Ref.~\cite{PavonValderrama:2005gu}, showing that the rate
of convergence was depending on the particular choice. Actually, there
are infinitely many auxiliary boundary conditions which converge
towards the same renormalized value, as we discuss below.

\subsection{The deuteron}

In this case we have a negative energy state
\begin{eqnarray}
E= -\frac{\gamma^2}M \, ,  
\end{eqnarray} 
and we look for normalized solutions of the coupled
equations~(\ref{eq:sch_coupled}) 
and normalized to unity,
\begin{eqnarray}
\int_0^\infty dr \left[ u(r)^2 + w(r)^2 \right] = 1  \, ,  
\label{eq:normalization} 
\end{eqnarray}
which asymptotically behave as
\begin{eqnarray}
u_\gamma (r) &\to & A_S e^{-\gamma r} \, , \nonumber \\ w_\gamma (r)
&\to & A_S \eta e^{-\gamma r} \left( 1 + \frac{3}{\gamma r} +
\frac{3}{(\gamma r)^2} \right) \,
\label{eq:deut-asymptotic},
\end{eqnarray} 
where $A_S$ is the asymptotic wave function normalization and $\eta$
is the asymptotic D/S ratio. To solve this problem we introduce, as
suggested in \cite{PavonValderrama:2005gu}, the auxiliary problems
\begin{eqnarray}
\begin{pmatrix} u_S \\ w_S \end{pmatrix} &\to& 
\begin{pmatrix}  1 \\ 0 \end{pmatrix} e^{-\gamma r} \, ,  \\ 
\begin{pmatrix} u_D \\ w_D \end{pmatrix} &\to& 
\begin{pmatrix} 0 \\  1  \end{pmatrix} e^{-\gamma r} \left( 1 + \frac{3}{\gamma r} + \frac{3}{(\gamma r)^2} \right) \, , 
\end{eqnarray} 
which solutions depend on the deuteron binding energy through $\gamma$
and the OBE potential. Further, we can use the superposition principle
of boundary conditions to write
\begin{eqnarray}
u (r) &=& u_S (r) + \eta \, u_D (r) \, , \nonumber \\ w (r) &=& w_S
(r) + \eta\, w_D (r) \, .
\label{eq:sup_bound}
\end{eqnarray}
The short distance regularity conditions (see below) must be imposed
an a cut-off radius $r_c$ in order to determine the value of $\eta
(r_c)$. Then, for a given solution we compute several properties as a
function of the cut-off radius, $r_c$. From the normalization
condition, Eq.~(\ref{eq:normalization}), in $r_c \le r \le \infty$ we
get $A_S (r_c)$. In this paper we also compute the matter radius,
\begin{eqnarray}
r_m^2 = \frac{\langle r^2 \rangle}{4} = \frac14 \int_{r_c}^\infty r^2
( u(r)^2 + w(r)^2 ) dr \, ,
\end{eqnarray} 
the quadrupole moment (without meson
exchange currents) 
\begin{eqnarray}
Q_d  = \frac1{20} \int_{r_c}^\infty r^2 w(r) ( 2\sqrt{2} u(r)-w(r) ) dr  \, , 
\end{eqnarray} 
the $D$-state probability   
\begin{eqnarray}
P_D = \int_{rc}^\infty w(r)^2  dr \, ,  
\end{eqnarray} 
which in the impulse aproximation and without meson exchange currents
can be related to the deuteron magnetic moment. Finally, we also
compute the inverse moment
\begin{eqnarray}
\langle r^{-1} \rangle = \int_{r_c}^\infty r^{-1} ( u(r)^2 + w(r)^2 ) dr
\, ,
\end{eqnarray} 
which appears, e.g., in the multiple expansion of the $\pi$-deuteron
scattering length. 

\begin{figure}[tbc]
\begin{center}
\includegraphics[height=6.5cm,width=5cm,angle=270]{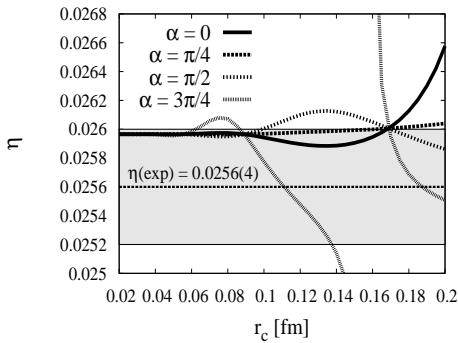}
\end{center}
\caption{Short distance cut-off dependence of the asymptotic
  $D/S$-ratio for the case with $\pi+\sigma+\rho+\omega$.  We show the
  dependence of the asymptotic D/S normalization $\eta$ for several
  choices of the arbitrary and auxiliary short distance condition $\sin \alpha u(r_c) + \cos \alpha w(r_c)=0 $ for several values of $\alpha$.}
\label{fig:finite_cutoff-alpha}
\end{figure}

As mentioned, there are infinitely many possible auxiliary
conditions. This is an important point which we wish to
illustrate. For instance, we could take
\begin{eqnarray}
\sin \alpha u(r_c) + \cos \alpha w(r_c)=0  \, ,  
\label{eq:bc-alpha}
\end{eqnarray}
where we may choose the parameter $\alpha$ arbitrarily~\footnote{This
  arbitrariness is not exclusive to this boundary condition, it is
  also present when the standard from factor regularization is
  introduced.  The exponential, Eq.~(\ref{eq:exponential}), and
  monopole Eq.~(\ref{eq:monopole}) form factors are {\it just} two
  possible choices which do not cover the most general form which
  might allow a theoretical estimate on the systematic error.}. This
is illustrated in Fig.~(\ref{fig:finite_cutoff-alpha}). Note that
despite possible wild behaviour all choices converge to the same
value, although at a quite different rate. This is indeed another
reason for removing the cut-off although it may be appealing and less
demanding to choose one particular scheme where stability is found at
the largest possible distances.

Here we will take the smoothest auxiliary
condition (labeled as BC6 in Ref.~\cite{PavonValderrama:2005gu})
\begin{eqnarray}
u'(r_c) - \sqrt{2} w'(r_c) &=& 0 \, ,\,  \,  g_{\pi NN}^2 - f_{\rho
NN}^2 > 0 \, , \nonumber \\
\label{eq:bc6} \\ 
\sqrt{2} u'(r_c) + w'(r_c) &=& 0 \, ,\,  \,  g_{\pi NN}^2 - f_{\rho
NN}^2 < 0 \, . \nonumber 
\end{eqnarray} 
Clearly, for the values that we will be using the convergence is
determined by the size of the short distance scale characterizing the
most singular component of the potential. As we see from
Eq.~(\ref{eq:R-pm}) it depends strongly on the combination $g_{\pi
  NN}^2- f_{\rho NN}^2$. This is an important point since the short
distance cut-offs, $r_c$, for which convergence is achieved may change
by orders of magnitude~\footnote{An extreme example is given by the
  exceptional case $f_{\rho NN}= g_{\pi NN}$ since the $1/r^3$
  singularity turns into a slowly and logarithmically converging
  Coulomb singularity. This case is treated specifically in
  Appendix~\ref{sec:exceptional}.}. An additional numerical problem
arises due to undesired amplification of the short distance growing
exponential, setting some limitations to the numerics due to roundoff
errors. In all our calculations we have payed particular attention to
these delicate issues.

\begin{figure*}[ttt]
\begin{center}
\includegraphics[height=5cm,width=5cm,angle=270]{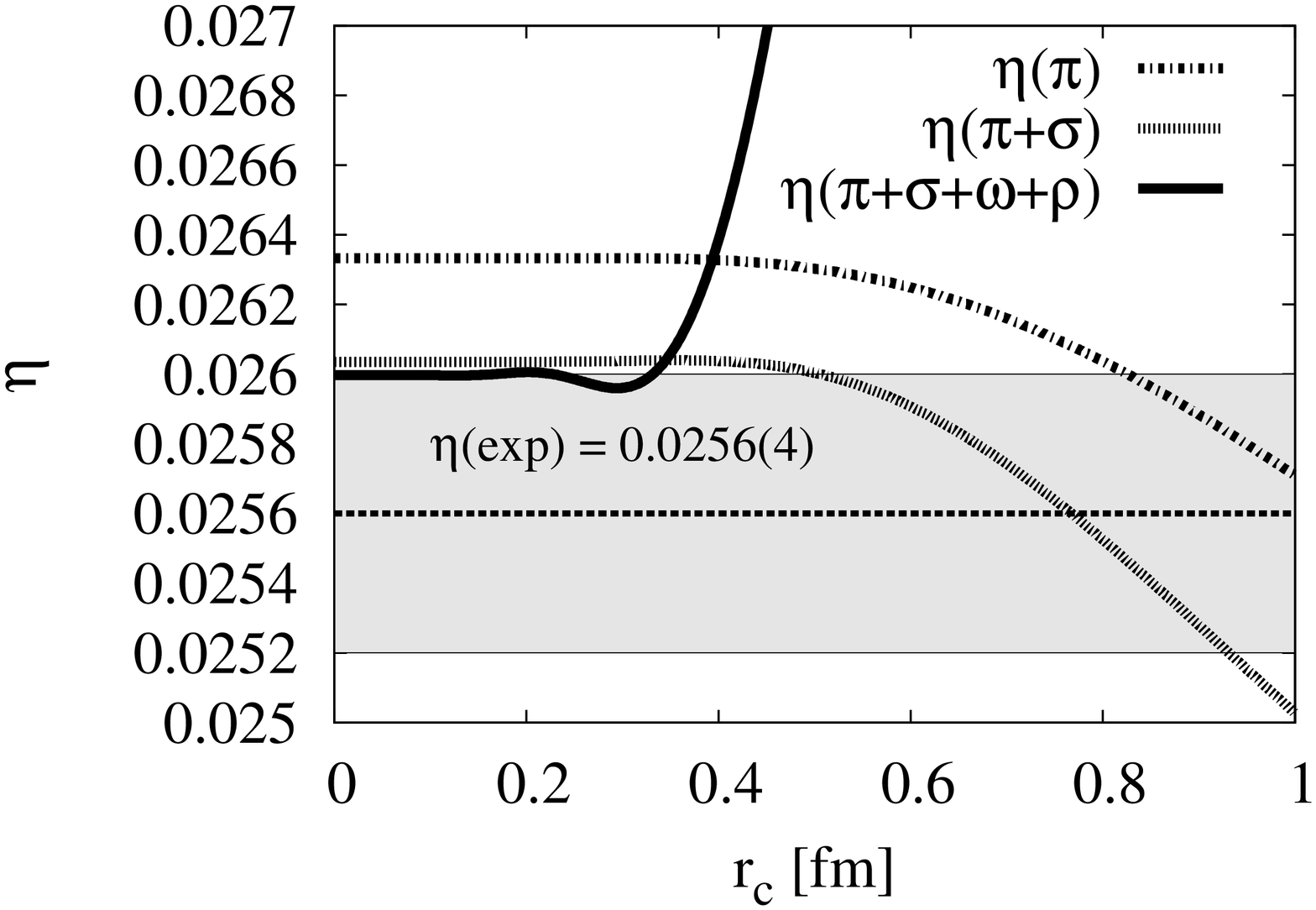}
\includegraphics[height=5cm,width=5cm,angle=270]{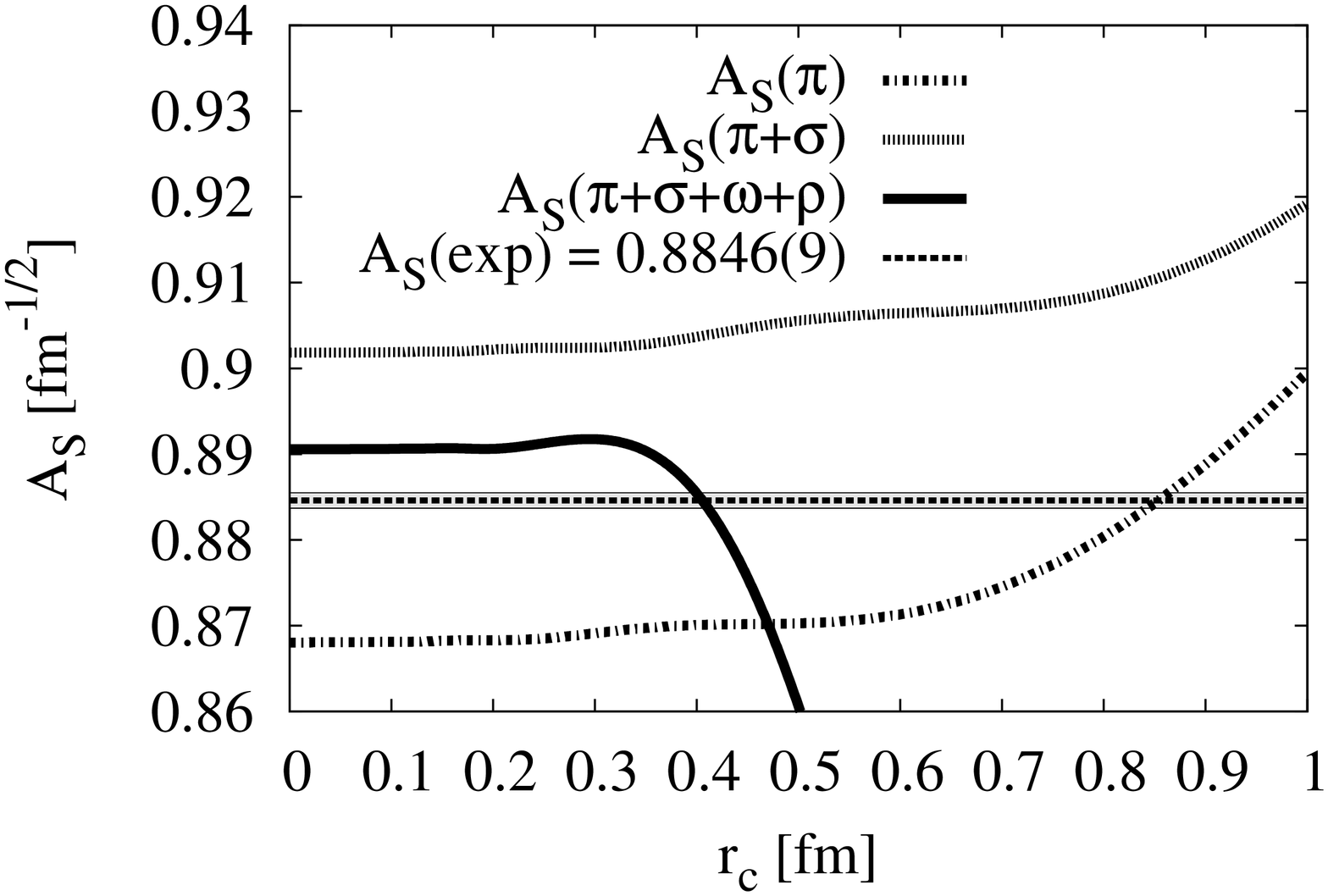}
\includegraphics[height=5cm,width=5cm,angle=270]{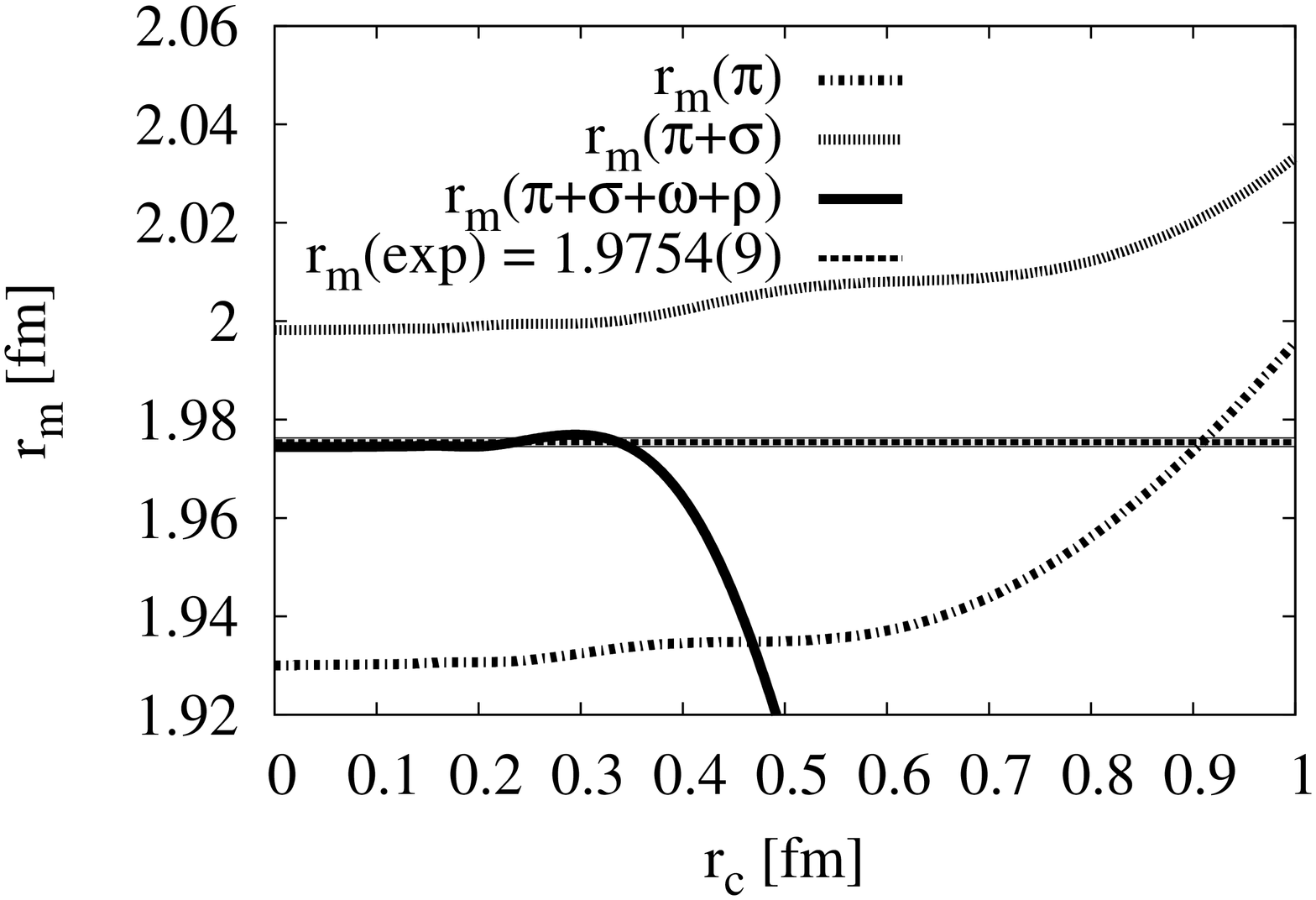}\\ 
\includegraphics[height=5cm,width=5cm,angle=270]{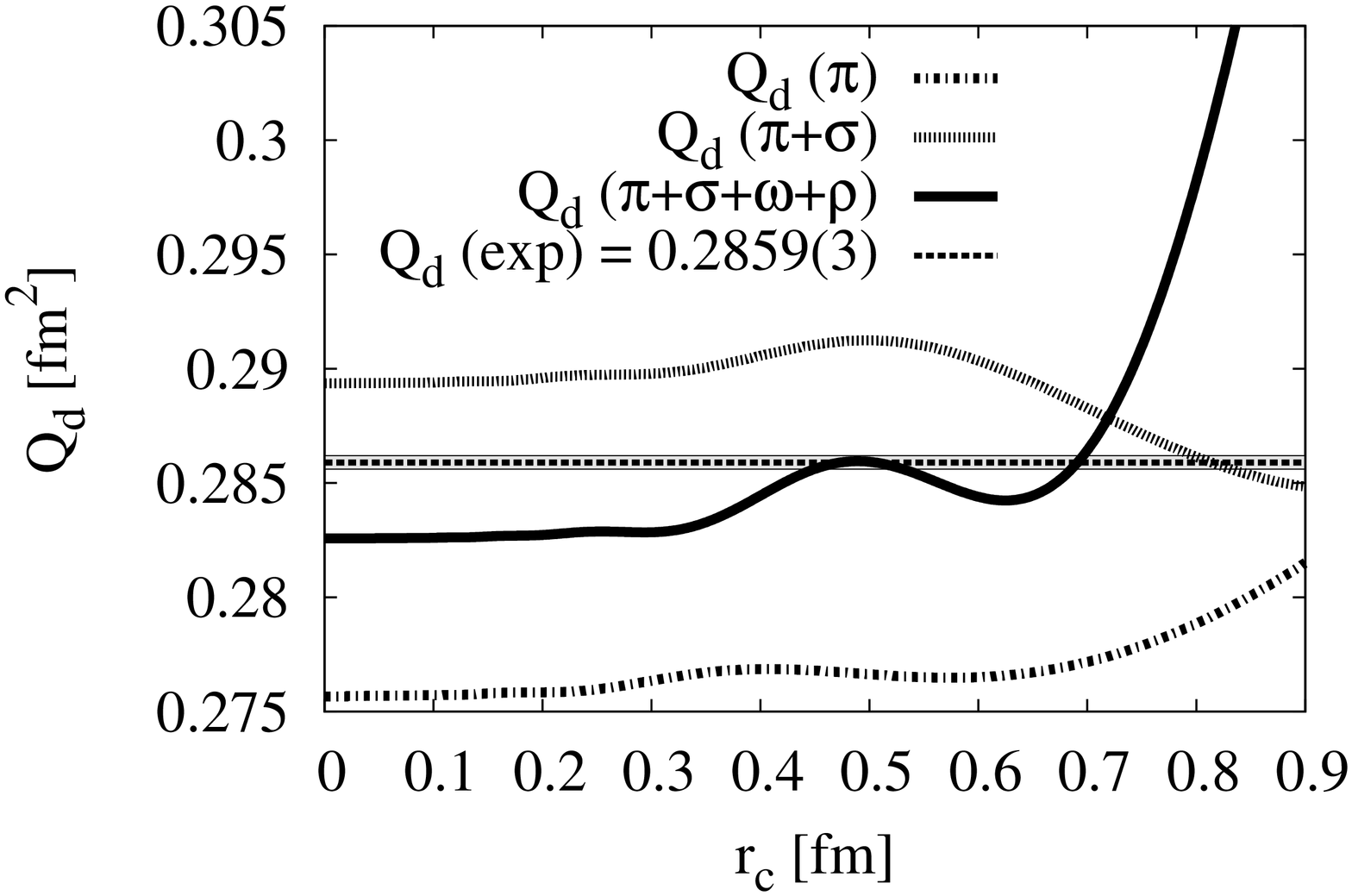}
\includegraphics[height=5cm,width=5cm,angle=270]{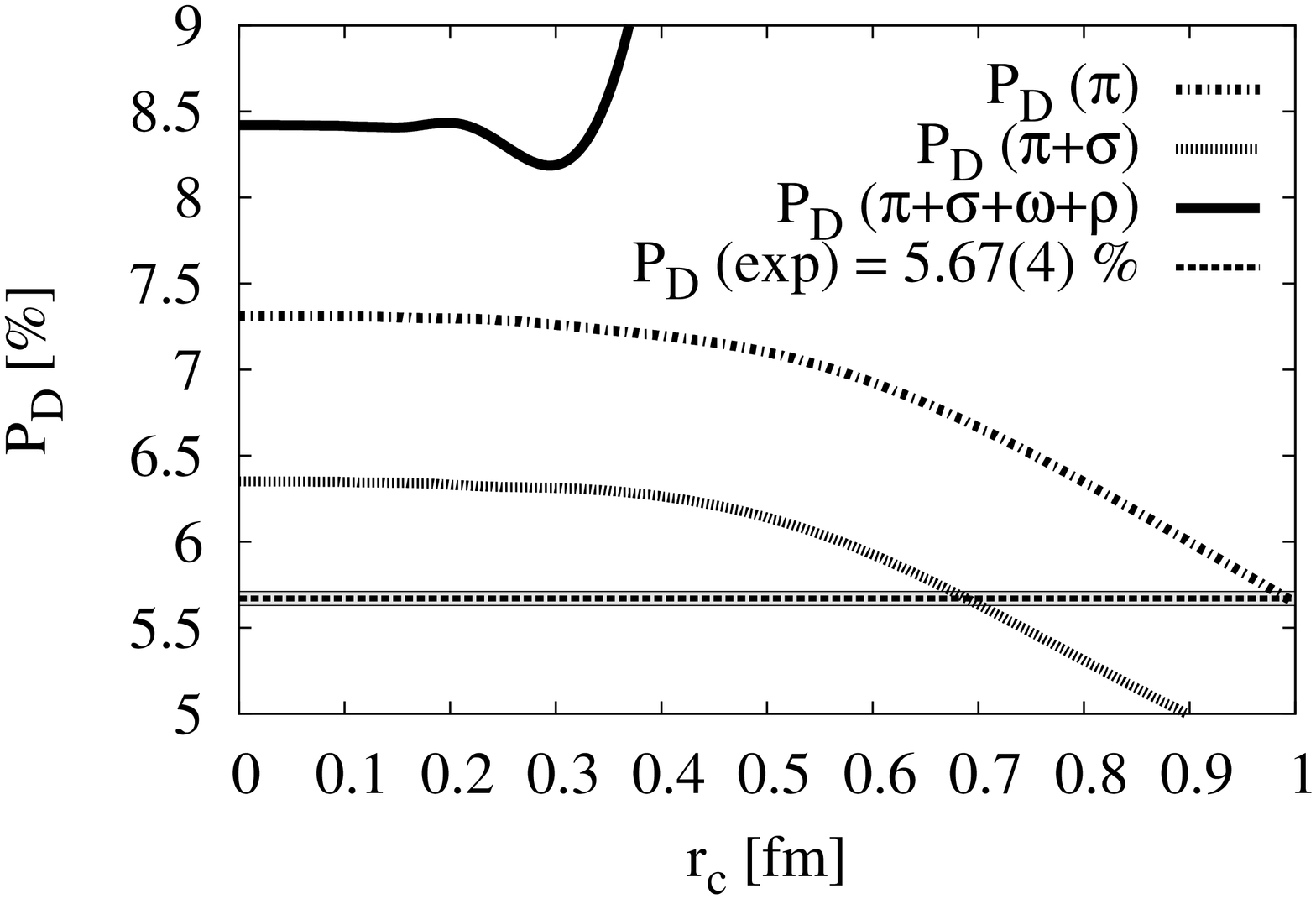}
\includegraphics[height=5cm,width=5cm,angle=270]{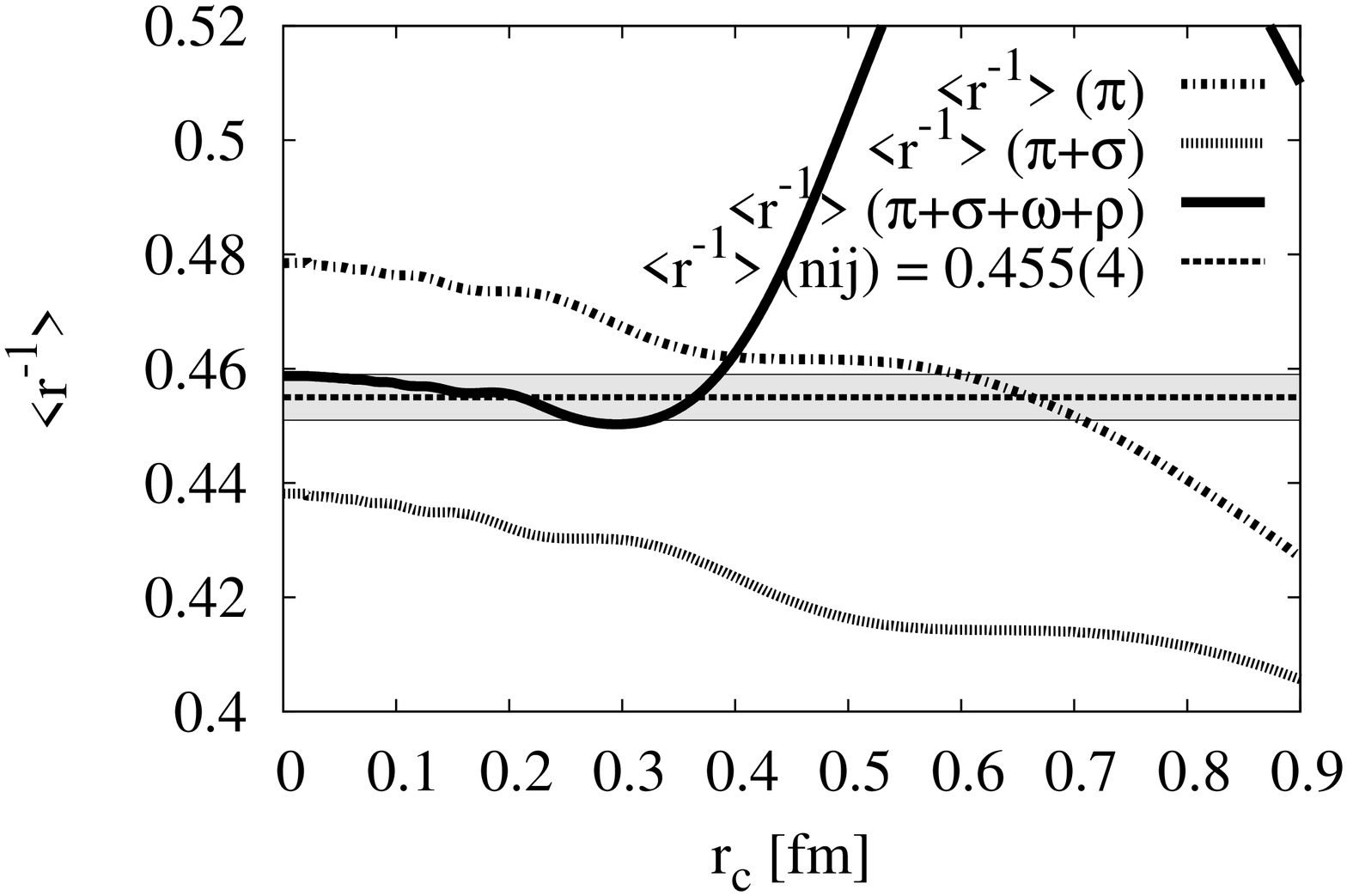}
\end{center}
\caption{Short distance cut-off dependence of deuteron properties for
the cases with $\pi$, $\pi+\sigma$ and $\pi+\sigma+\rho+\omega$.  We
show the dependence of the asymptotic D/S normalization $\eta$ (upper
left panel), the S-wave normalization $A_S$ (in ${\rm fm}^{-1/2} $,
upper middle panel), the matter radius $r_m$ (in ${\rm fm} $, upper
right panel), the quadrupole moment $ Q_d $ (in ${\rm fm}^2 $, lower
left panel), the $D$-state probability (lower middle panel) and the
inverse radius $\langle r^{-1} \rangle $ (in ${\rm fm}^{-1}$ lower
right panel).
Experimental or recommended
  values can be traced from Ref.~\cite{deSwart:1995ui}.}
\label{fig:finite_cutoff}
\end{figure*}

The cut-off dependence of these observables is
shown in Fig.~\ref{fig:finite_cutoff}, for the case of $\pi$ only
(Ref.~\cite{PavonValderrama:2005gu}), $\pi+\sigma$ and
$\pi+\sigma+\rho+\omega$ and as we see good convergence can be
achieved as $r_c \to 0$. As already mentioned, the rate of convergence
depends on the scale of the singularity.

\begin{figure*}[ttt]
\begin{center}
\includegraphics[height=.3\textheight,angle=270]{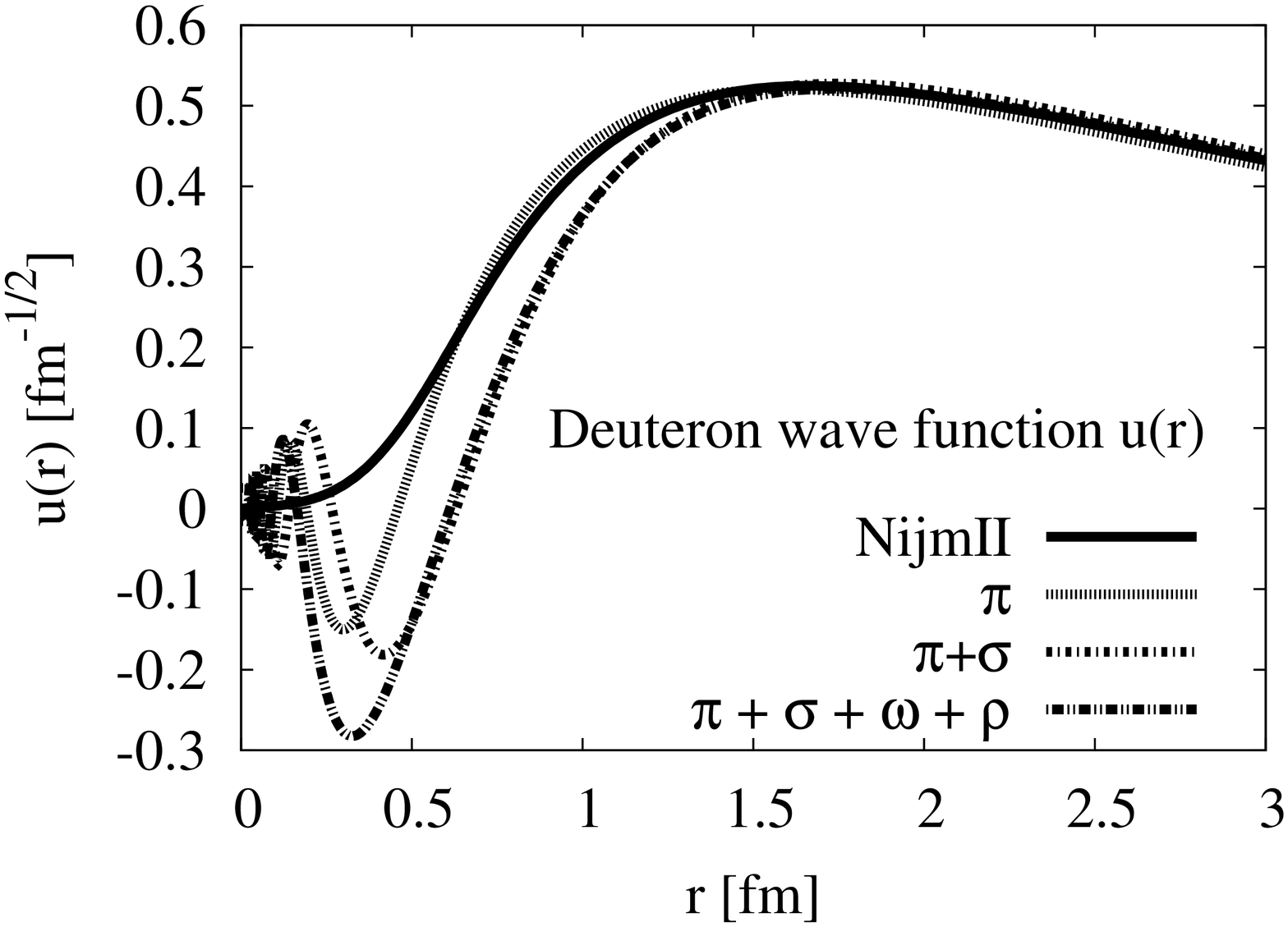}
\includegraphics[height=.3\textheight,angle=270]{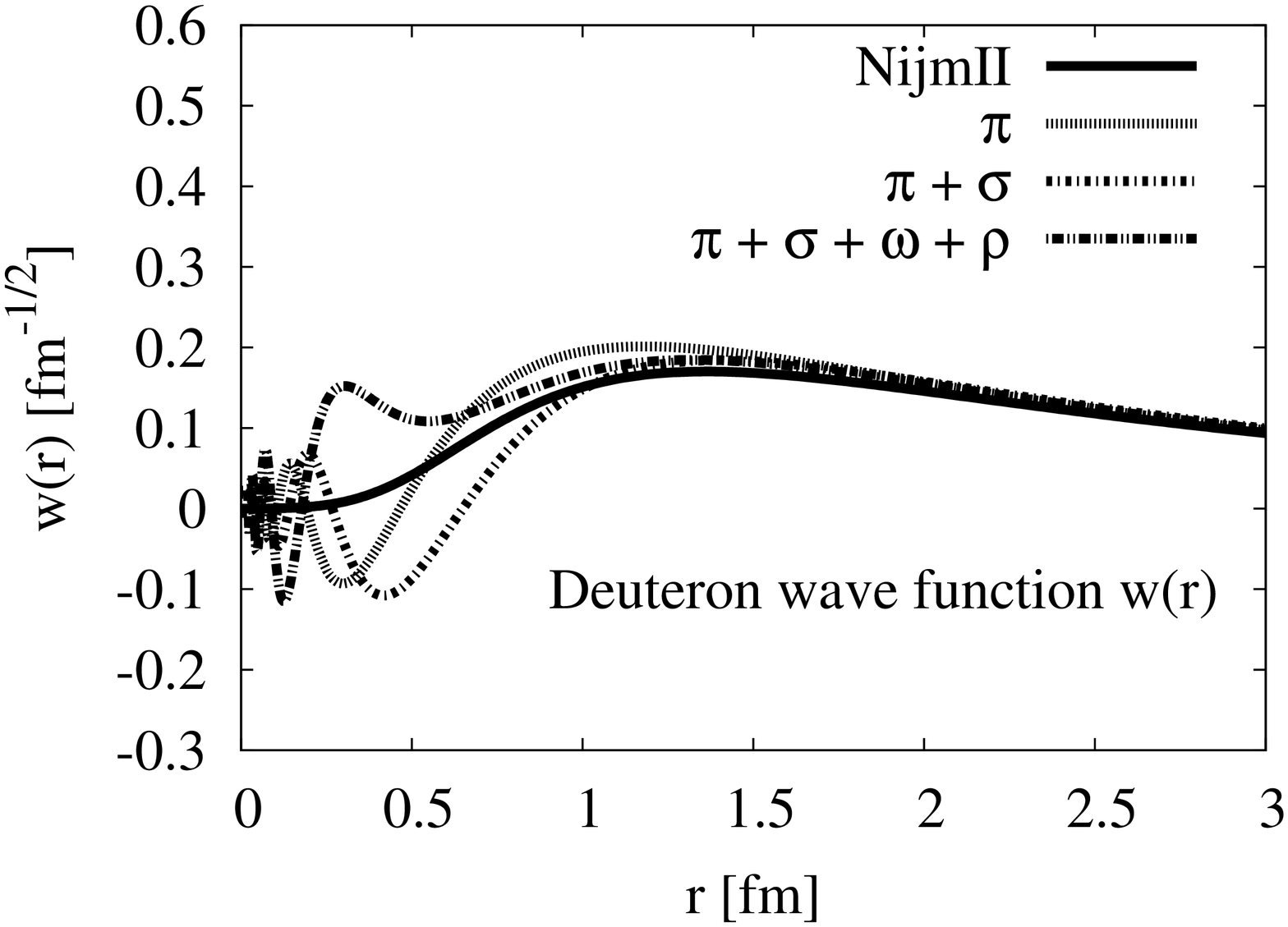}
\end{center}
\caption{Deuteron wave functions, u (left) and w (right), as a
  function of the distance (in {\rm fm}) in the OBE. We show $1\pi$,
  $1\pi+1\sigma$ and $\pi+\sigma+\rho+\omega$ compared to the Nijmegen
  II wave functions~\cite{Stoks:1994wp}. The asymptotic normalization
  $u \to e^{-\gamma r}$ has been adopted and the asymptotic D/S ratio
  is obtained to be $\eta_{\pi} = 0.2633$ and $\eta_{\pi\sigma \omega
    \rho} = 0.2597$ (see table~\ref{tab:table_triplet}).}
\label{fig:u+w_obe}
\end{figure*}

The resulting coordinate space deuteron wave functions, $u$ and $w$,
are depicted in Fig.~\ref{fig:u+w_obe} for the case of $\pi$ only
(Ref.~\cite{PavonValderrama:2005gu}), $\pi+\sigma$ and
$\pi+\sigma+\rho+\omega$ and compared to the wave functions of the
high quality Nijmegen potential~\cite{Stoks:1994wp}. As we see, after
inclusion of the scalar and vector mesons, the agreement is quite
remarkable in the region above $1.4-1.8 {\rm fm}$, their declared
range of validity.  Similarly to the singlet case, we observe
oscillations in the region below 1 fm. The first node is allowed since
we are dealing with a bound state, the second node occurs already
below $0.5 {\rm fm}$ indicating, similarly to the $^1S_0$ channel, the
appearance of infinitely many spurious bound states, as we see from
the short distance oscillatory behaviour of the wave function,
Eq.~(\ref{eq:short_bc_reg}). To compute such states we proceed
similarly to the singlet channel.  We solve Eq.~(\ref{eq:sch_coupled})
with negative energy $E_B=- \gamma_B^2 /M$, the asymptotic behaviour
in Eq.~(\ref{eq:deut-asymptotic}) and impose the regularity
conditions, Eq~(\ref{eq:bc6}), as well as orthogonality to the
deuteron state, namely
\begin{eqnarray}
u_\gamma u_B' - u_\gamma' u_B + w_\gamma
w_B' - w_\gamma ' w_B \Big|_{r=r_c} = 0 \, , 
\end{eqnarray} 
from which $\gamma_B$ can be determined. For instance, for the scalar
parameters in Eq.~(\ref{eq:sigma-fit}) and $f_{\rho NN}=15.5$ we
identify the first spurious bound state $(u_{B1}, w_{B1})$ having one
node less than the deuteron wave functions $(u_d, w_d)$ taking place
at $\gamma_{B1} = 3.438 {\rm fm}^{-1}$. The corresponding energy is
$E_{B1} = - \gamma_{B1}^2 /M = -490 {\rm MeV}$, S-wave normalization
$A_{B1}=13.58 {\rm fm}^{-1/2}$ matter radius $r_{B1} = 0.49 {\rm fm}$
and asymptotic D/S ratio $\eta_{B1}=0.1656$. This state is clearly
beyond the range of applicability of the present
framework. Subtracting this pole to the $^3S_1$ amplitude would
result, according to Eq.~(\ref{eq:delta-r0}), in $\Delta r_0 = 0.01
{\rm fm}$. The next spurious state has $E_{B2}<-18{\rm GeV}$!!.  Note
that if the scale where the second unphysical node takes place was to
be interpreted as a (``hard core'') breakdown distance scale of our
approach for the deuteron, it is certainly beyond the accessible
region at the maximal energy in elastic NN scattering. This issue is
relevant for the calculation of phase shifts where such oscillations
also occur. The variation of the observables from this breakdown scale
to the origin, could be interpreted as a source of systematic error
coming from the fact that there is only one bound state and not
infinitely many. As we see from Fig.~\ref{fig:finite_cutoff} the
effect is indeed small.

Numerical results for renormalized quantities can be looked up in
Table~\ref{tab:table_triplet}. As we see, the inclusion of $\sigma$
provides some overall improvement while $\rho$ and $\omega$ yield a
fairly accurate description of the deuteron for the choice $f_{\rho
NN}=15.5$ and $g_{\omega NN}=9$ (this latter value complies to the
SU(3) relation $g_{\omega NN} = 3 g_{\rho NN} $ when $g_{\rho NN} \sim
2.9$).

\begin{figure*}[ttt]
\includegraphics[height=5cm,width=5cm,angle=270]{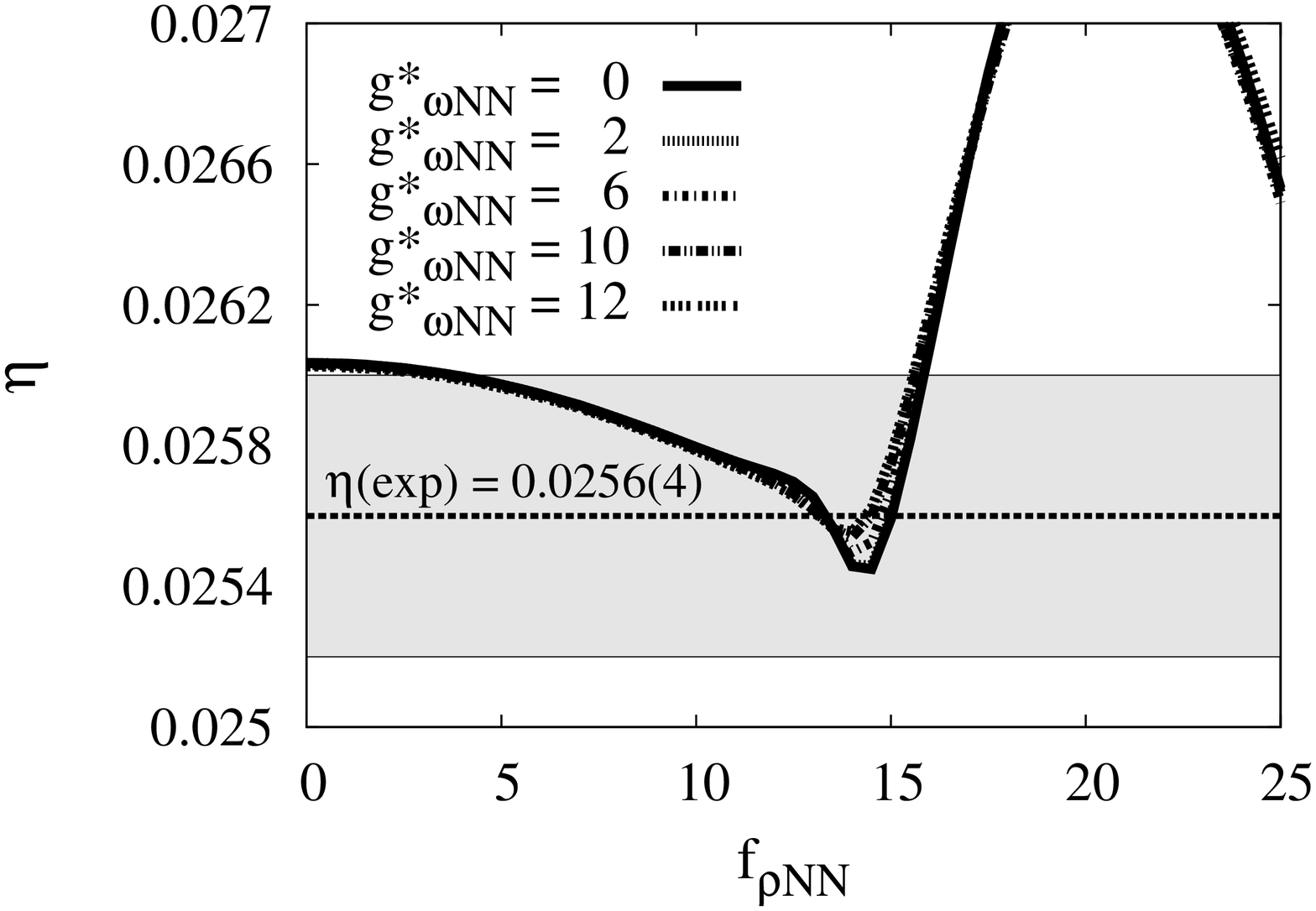}
\includegraphics[height=5cm,width=5cm,angle=270]{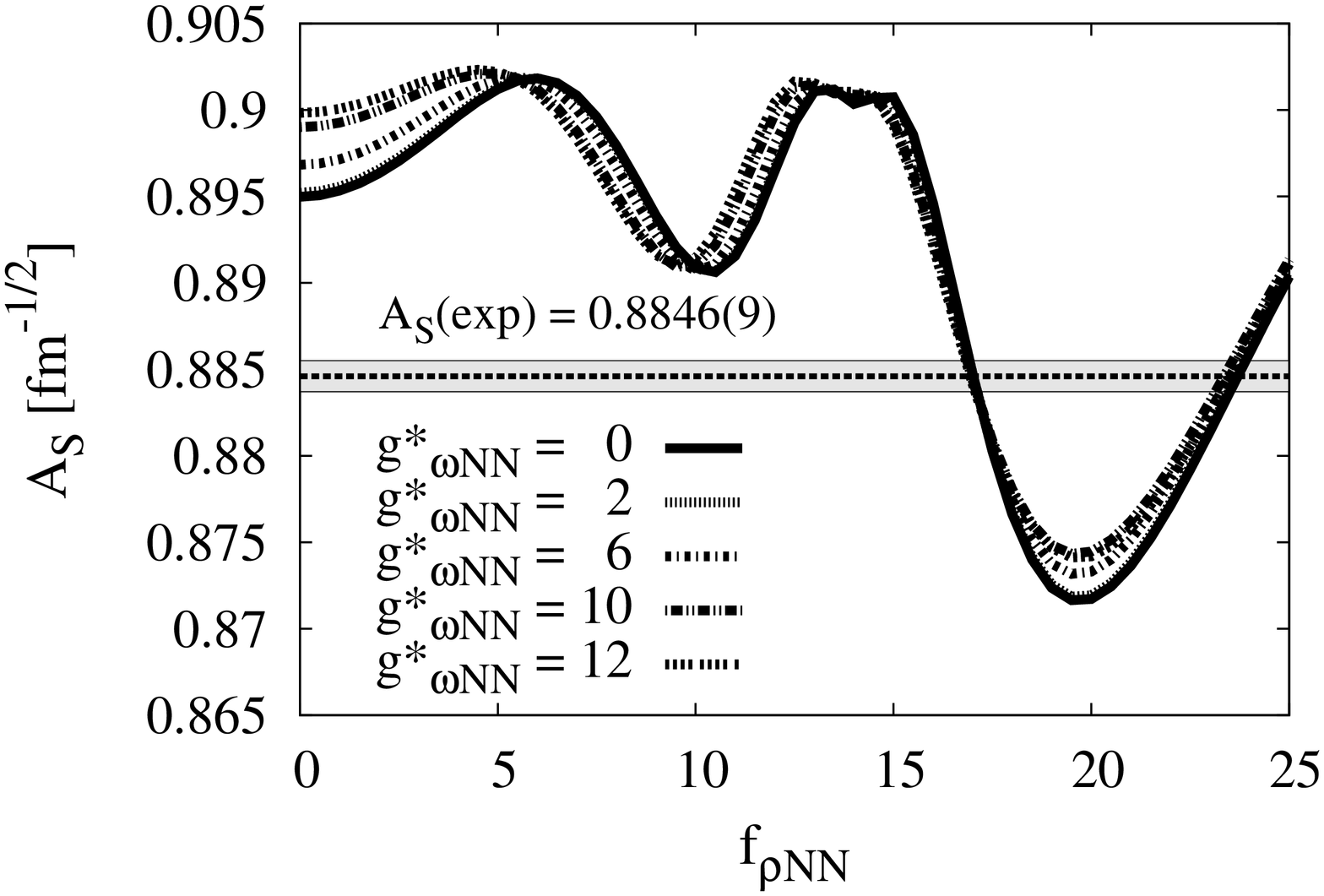}
\includegraphics[height=5cm,width=5cm,angle=270]{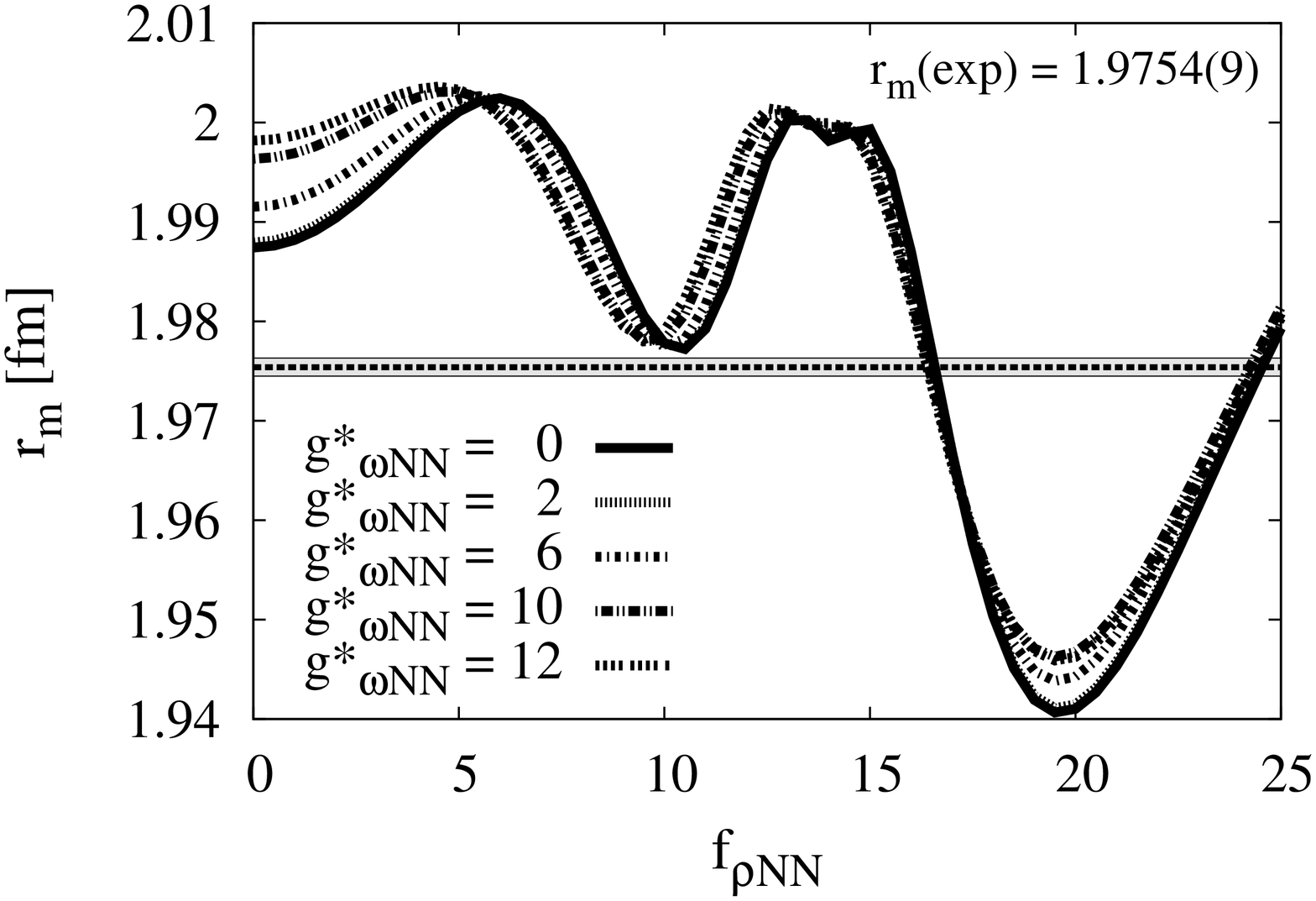}\\ 
\includegraphics[height=5cm,width=5cm,angle=270]{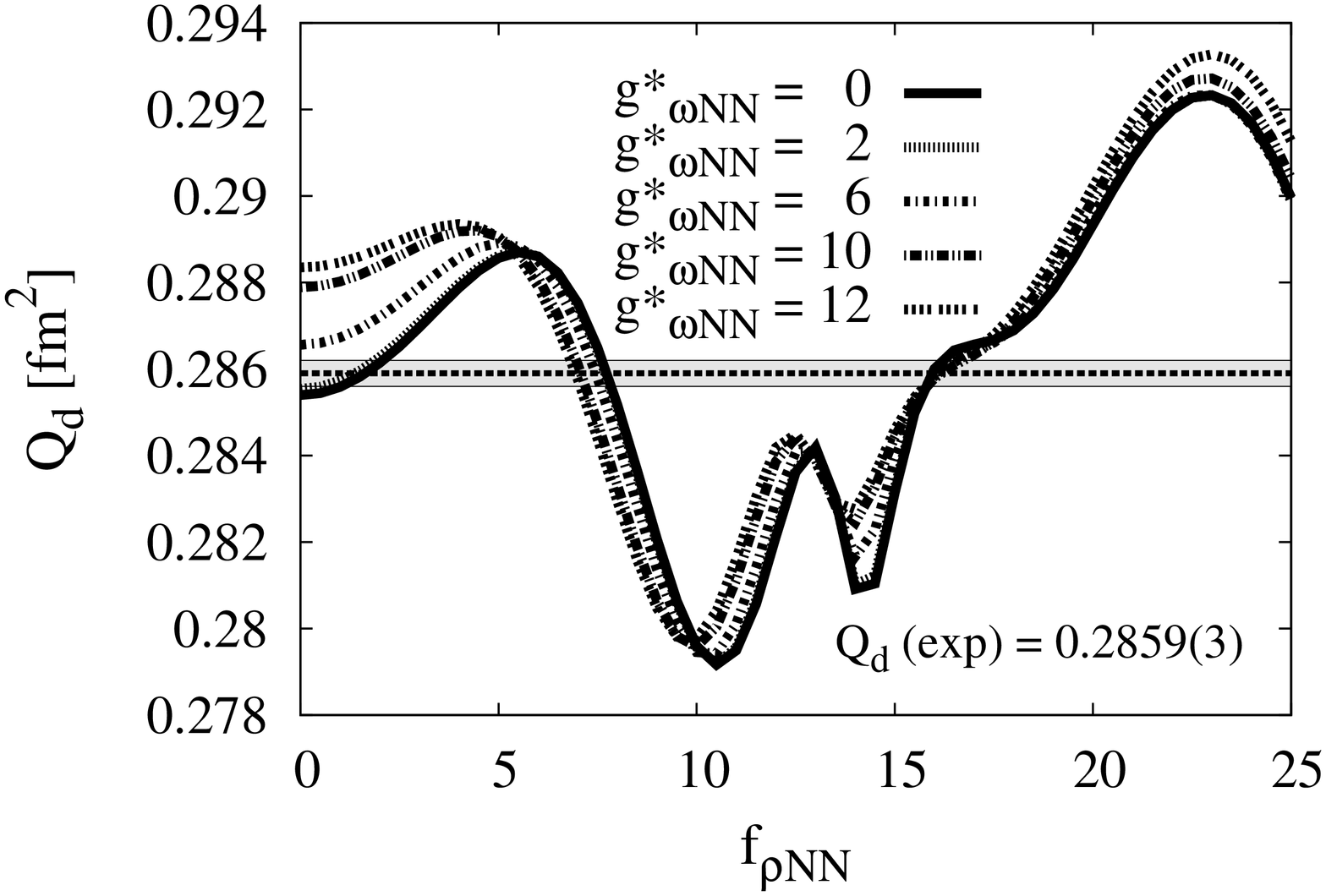}
\includegraphics[height=5cm,width=5cm,angle=270]{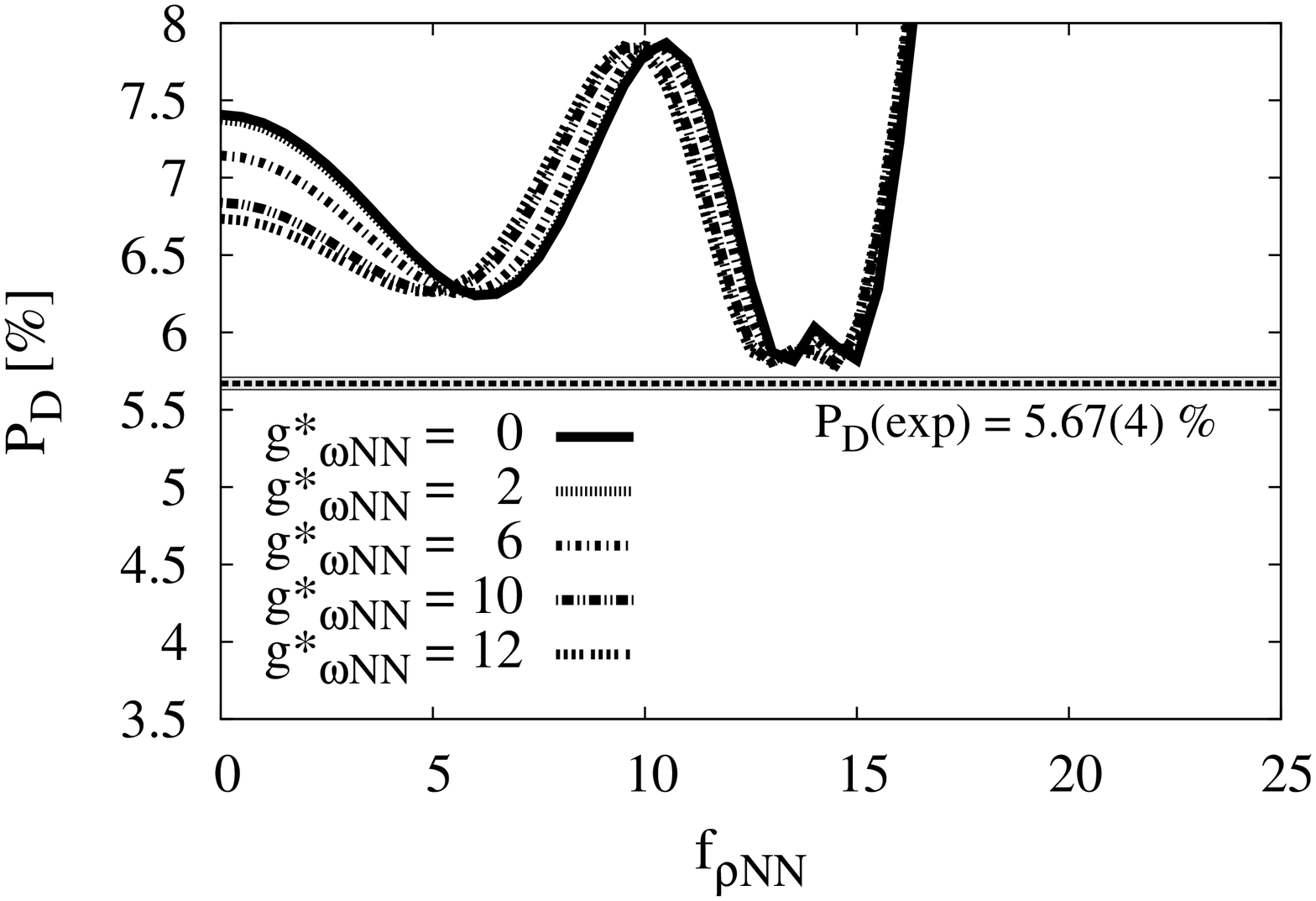}
\includegraphics[height=5cm,width=5cm,angle=270]{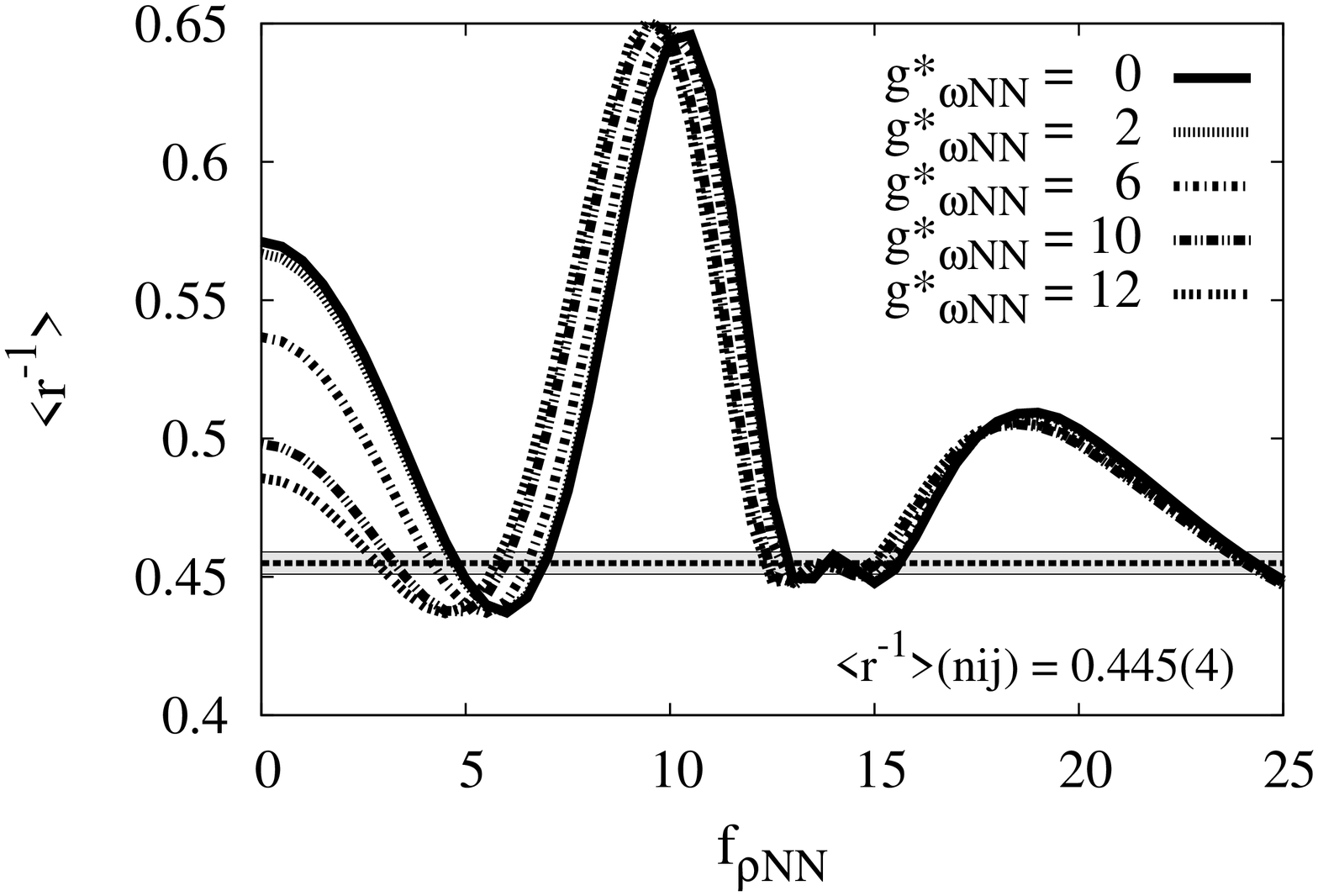}
\caption{Dependence of the deuteron observables 
as a function of $f_{\rho NN}$ for several values of the the effective
coupling constant $g_{\omega NN}^* = \sqrt{g_{\omega NN}^2- f_{\rho
NN}^2 m_\rho^2 / 2 M_N^2}$. $g_{\sigma NN}$ and $m_\sigma$ are always
readjusted to fit the $^1S_0$ phase shift. We
show the dependence of the asymptotic D/S normalization $\eta$ (upper
left panel), the S-wave normalization $A_S$ (in ${\rm fm}^{-1/2} $,
upper middle panel), the matter radius $r_m$ (in ${\rm fm} $, upper
right panel), the quadrupole moment $ Q_d $ (in ${\rm fm}^2 $, lower
left panel), the $D$-state probability (lower middle panel) and the
inverse radius $\langle r^{-1} \rangle $ (in ${\rm fm}^{-1}$ lower
right panel). The leading $N_c$
contributions to the OBE ($\sigma+\pi+\rho+\omega$) potential are considered.
Experimental or recommended
  values can be traced from Ref.~\cite{deSwart:1995ui}.}
\label{fig:obs-frNN}
\end{figure*}

\begin{table*}
\caption{\label{tab:table_triplet} Deuteron properties and low energy
parameters in the $^3S_1-^3D_1$ channel for OBE potentials including
$\pi$, $\pi+\sigma$, $\pi+\sigma+\rho+\omega$. We use the
non-relativistic relation $ \gamma= \sqrt{ 2 \mu_{np} B} $ with
$B=2.224575(9)$ and take $m=138.03 {\rm MeV} $, and $g_{\pi NN}
=13.1083 $ \cite{deSwart:1997ep}.  From a fit to the $^1S_0$ channel
we have $m_\sigma=501 {\rm MeV}$ and $g_{\sigma NN}=9.1$. The
simplifying relation $g_{\omega NN}= f_{\rho NN} m_\rho / \sqrt{2}
M_N$ is used throughout. $\pi\sigma\rho\omega$ corresponds to take
$f_{\rho NN}=15.5$ and $g_{\omega NN}=9.857$ while
$\pi\sigma\rho\omega^*$ corresponds to take $f_{\rho NN}=17.0$ and
$g_{\omega NN}=10.147$}
\begin{ruledtabular}
\begin{tabular}{|c|c|c|c|c|c|c|c|c|c|c|c|}
\hline & $\gamma ({\rm fm}^{-1})$ & $\eta$ & $A_S ( {\rm fm}^{-1/2}) $
& $r_m ({\rm fm})$ & $Q_d ( {\rm fm}^2) $ & $P_D $ & $\langle r^{-1}
\rangle $ & $ \alpha_0 ({\rm fm}) $ & $\alpha_{02} ({\rm fm}^3) $ & $
\alpha_2 ({\rm fm}^5) $ & $r_0 ({\rm fm} ) $ \\ \hline
%
%
%
$\pi$ & Input & 0.02633 & 0.8681 & 1.9351 & 0.2762
& 7.88\% & 0.476 & 5.335 & 1.673 & 6.169 & 1.638 \\ 
\hline 
$\pi\sigma$ & Input & 0.02599 & 0.9054 & 2.0098 & 0.2910
& 6.23\% & 0.432 & 5.335 & 1.673 & 6.169 & 1.638 \\ 
\hline 
$\pi\sigma\rho\omega$ & Input & 0.02597 & 0.8902 & 1.9773 & 0.2819
& 7.22\% & 0.491 & 5.444 & 1.745 & 6.679 & 1.788 \\ 
$\pi\sigma\rho\omega$$^*$ 
& Input & 0.02625 & 0.8846 & 1.9659 & 0.2821
& 9.09\% & 0.497 & 5.415 & 1.746 & 6.709 & 1.748 \\ 
\hline 
NijmII & Input & 0.02521 & 0.8845(8) & 1.9675 & 0.2707 & 
5.635\% &  0.4502 & 5.418 & 1.647 & 6.505 & 1.753 \\
Reid93 & Input & 0.02514 & 0.8845(8) & 1.9686 & 0.2703 & 
5.699\% & 0.4515 & 5.422 & 1.645 & 6.453 & 1.755 \\ \hline 
Exp. \footnotemark[1] &  0.231605 &  0.0256(4)  & 0.8846(9) & 1.9754(9)  &
0.2859(3) & 5.67(4)  & & 5.419(7) & & & 1.753(8) \\ \hline 
\end{tabular}
\end{ruledtabular}
\footnotetext[1]{(Non relativistic). See
e.g. Ref.~\cite{deSwart:1995ui} and references therein.}
\end{table*}

We show in Fig.~(\ref{fig:obs-frNN}) the dependence of several
properties when both the vector mesons $\rho$ and $\omega$ are
simultaneosuly considered. In Fig.~\ref{fig:obs-frNN} we plot the
dependendence of (renormalized) deuteron properties as a function of
$f_{\rho NN}$ for several values of the effective coupling constant
$g_{\omega NN}^* = \sqrt{g_{\omega NN}^2- f_{\rho NN}^2 m_\rho^2 / 2
M_N^2}$ featuring the strong correlation in the $^1S_0$ channel
pointed out in Section~\ref{sec:singlet}. The scalar coupling
$g_{\sigma NN}$ and scalar mass $m_\sigma$ are always readjusted to
fit the $^1S_0$ phase shift. As we see, for the asymptotic D/S ratio,
$\eta$, there is a wide range of possible values within the
experimental uncertainties but we obtain the bounds $f_{\rho NN} \le
15$ and $g_{\omega NN} \le 15$. It is amazing that the value of the
tensor-$\rho$ coupling is so well determined to be $f_{\rho NN} \sim
16-17$ and corresponds to the strong $\kappa_\rho$ situation described
by Machleidt and Brown~\cite{Brown:1994pq}. Note that results depend
in a moderate fashion on $f_ {\rho NN}$ for not too large values, as
one would expect from the short range of the $\rho-$meson.

\begin{figure}[ttt]
\includegraphics[height=4cm,width=4cm,angle=270]{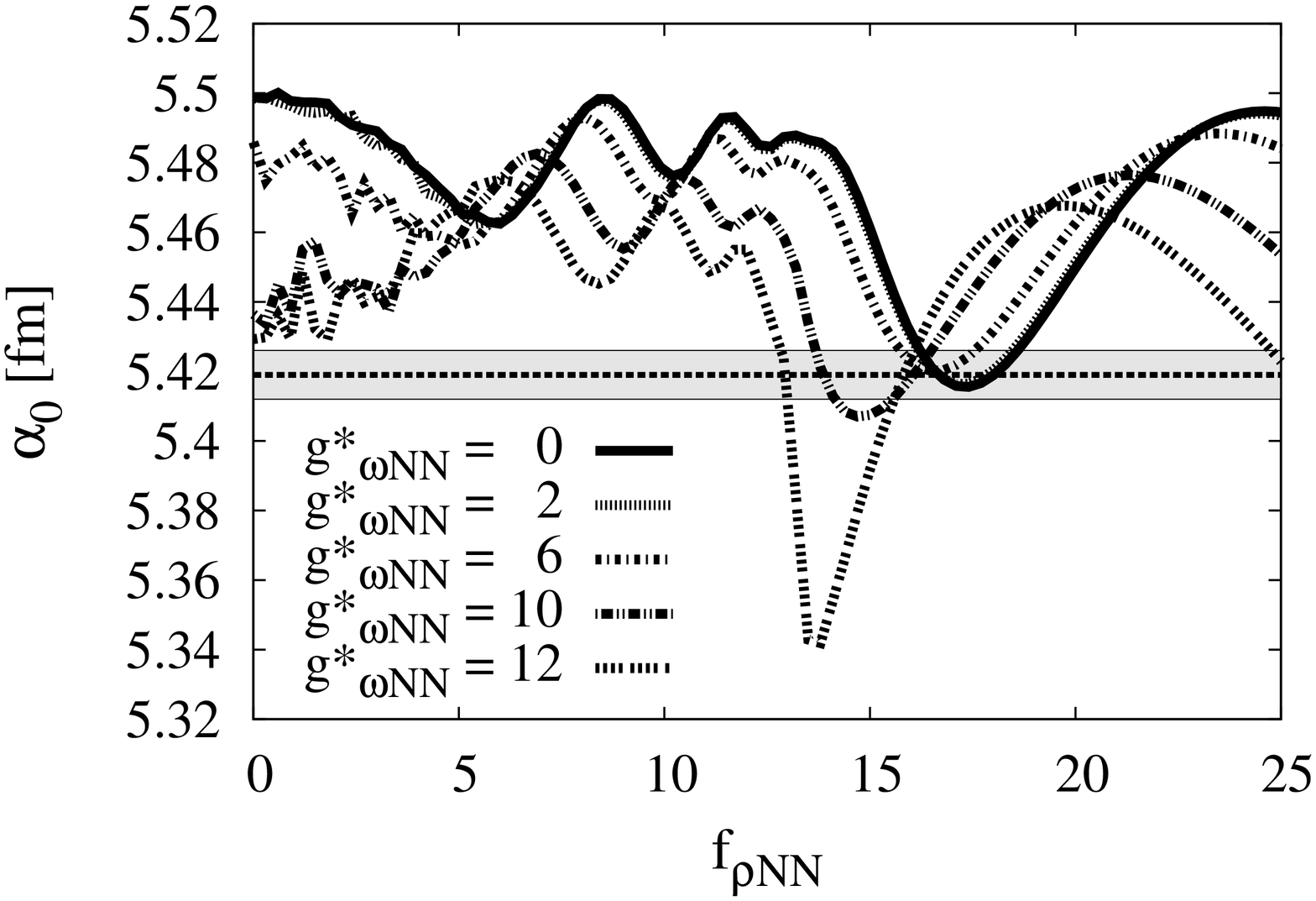}
\includegraphics[height=4cm,width=4cm,angle=270]{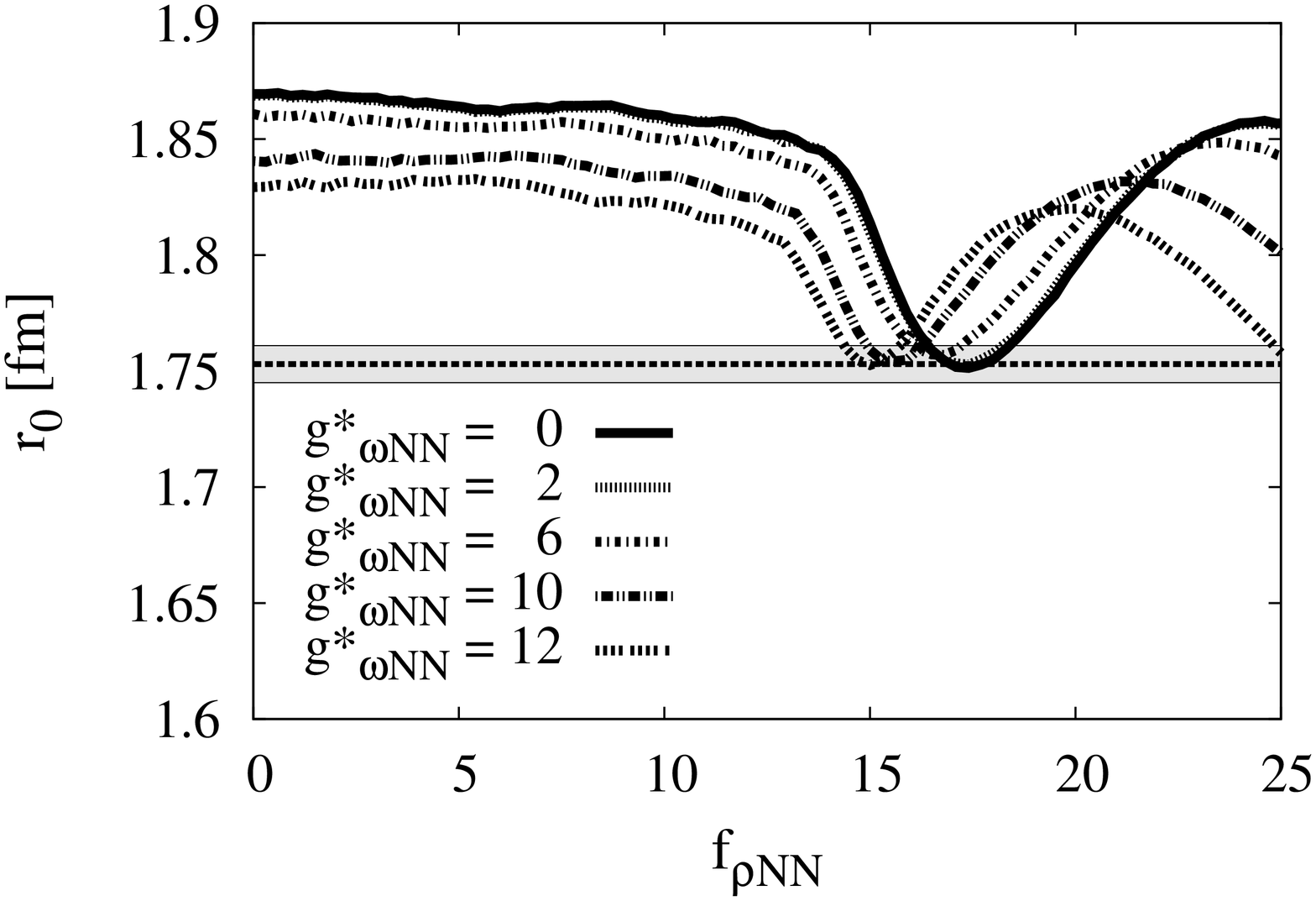}
\includegraphics[height=4cm,width=4cm,angle=270]{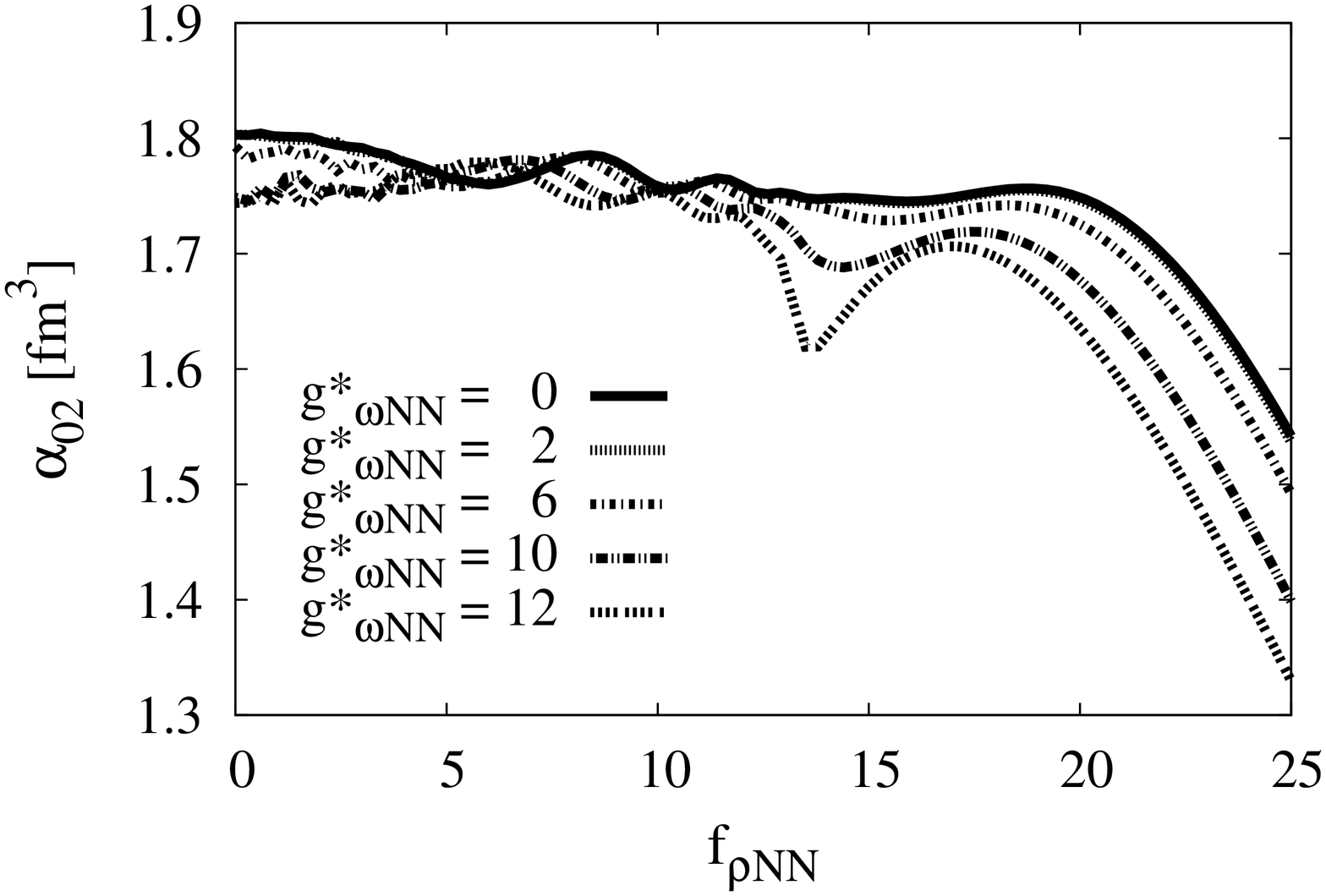}
\includegraphics[height=4cm,width=4cm,angle=270]{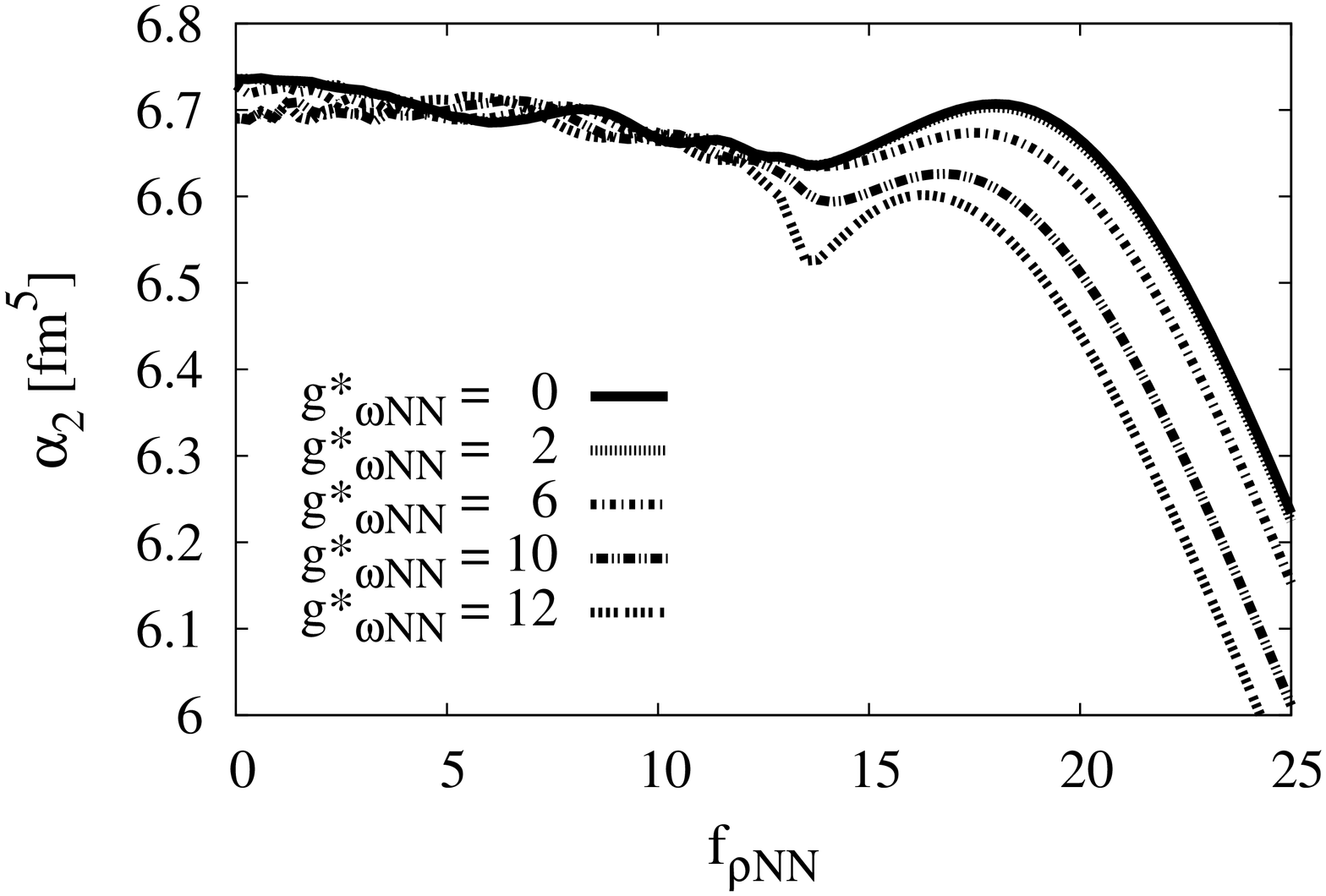}
\caption{Dependendence of the low energy parameters as a function of
$f_{\rho NN}$ for several values of the the effective coupling
constant $g_{\omega NN}^* = \sqrt{g_{\omega NN}^2- f_{\rho NN}^2
m_\rho^2 / 2 M_N^2}$. $g_{\sigma NN}$ and $m_\sigma$ are always
readjusted to fit the $^1S_0$ phase shift. We show the dependence of
the $^3S_1$ scattering length $\alpha_0$ (in fm) and effective range
$r_0$ (in fm) as well as $\alpha_{02}$ (in ${\rm fm}^3$) and
$\alpha_2$ (in ${\rm fm}^5$) compared to the experimental values or
the Nijm2 and Reid93 potentials (horizontal straight lines) The
leading $N_c$ contributions to the OBE ($\sigma+\pi+\rho+\omega$)
potential are considered.}
\label{fig:lep-frNN}
\end{figure}

\subsection{Zero energy}
\label{sec:zero-energy}

At zero energy, the asymptotic solutions to the coupled
equations~(\ref{eq:sch_coupled}) are given by 
\begin{eqnarray}
u_{0,\alpha} (r) & \to & 1- \frac{r}{\alpha_0} \, , \nonumber \\
w_{0,\alpha} (r) & \to & \frac{3 \alpha_{02}}{\alpha_0 r^2 } \, ,
\nonumber \\ 
u_{0,\beta} (r) &\to & \frac{r}{\alpha_0} \, , \nonumber
\end{eqnarray} 
\begin{eqnarray} 
w_{0,\beta} (r) &\to& \left( \frac{\alpha_2}{\alpha_{02}} -
\frac{\alpha_{02}}{\alpha_0} \right) \frac{3}{r^2}- \frac{r^3}{15
\alpha_{02}} \, , 
\label{eq:zero_energy}
\end{eqnarray} 
where $\alpha_0$, $\alpha_2$ and $\alpha_{02}$ are low energy
parameters obtained from the phase shifts (see
Sect.~\ref{sec:phase-shifts}). Using these zero energy solutions one
can determine the effective range. The $^3S_1$ effective range
parameter is given by
\begin{eqnarray} 
r_0 &=& 2 \int_0^\infty \left[ \left(1-\frac{r}{\alpha_0} \right)^2 -
u_{0,\alpha} (r)^2 - w_{0,\alpha} (r)^2 \right] dr \, . \nonumber \\ 
\label{eq:r0_triplet} 
\end{eqnarray} 
Using the superposition principle of boundary conditions
we may write the solutions as  
\begin{eqnarray} 
u_{0,\alpha} (r) &=& u_1 (r) - \frac{1}{\alpha_0} u_2 (r) + \frac{3
\alpha_{02}}{\alpha_0} u_3 (r) \, , \nonumber \\  
w_{0,\alpha} (r) &=& w_1
(r) - \frac{1}{\alpha_0} w_2 (r) + \frac{3 \alpha_{02}}{\alpha_0} w_3
(r) \, , \nonumber 
\end{eqnarray} 
\begin{eqnarray} 
 u_{0,\beta} (r) &=& \frac{1}{\alpha_0} u_2 (r) +
\left( \frac{3 \alpha_2}{\alpha_{02}} - \frac{3\alpha_{02}}{\alpha_0}
\right) u_3 (r) -\frac1{15 \alpha_{02}} u_4 (r) \, , \nonumber \\
w_{0,\beta} (r) &=& \frac{1}{\alpha_0} w_2 (r) + \left( \frac{3
\alpha_2}{\alpha_{02}} - \frac{3\alpha_{02}}{\alpha_0} \right) w_3 (r)
-\frac1{15\alpha_{02}} w_4 (r) \, , \nonumber \\
\label{eq:sup_zero}
\end{eqnarray}
where the functions $u_{1,2,3,4}$ and $w_{1,2,3,4}$ are independent on
$\alpha_0$, $\alpha_{02}$ and $\alpha_2$ and fulfill suitable boundary
conditions. The orthogonality constraints for the $\alpha$ and $\beta$
states read in this case
\begin{eqnarray}
u_\gamma u_{0,\alpha}' - u_\gamma' u_{0,\alpha} + w_\gamma
w_{0,\alpha}' - w_\gamma ' u_{0,\alpha} \Big|_{r=r_c} &=& 0 \nonumber \\
u_\gamma u_{0,\beta}' - u_\gamma' u_{0,\beta} + w_\gamma w_{0,\beta}'
- w_\gamma ' u_{0,\beta} \Big|_{r=r_c} &=& 0 \nonumber \\
\label{eq:orth_triplet} 
\end{eqnarray} 
A further condition which should be satisfied is the $\alpha-\beta$
orthogonality
\begin{eqnarray}
u_{0,\alpha} u_{0,\beta}' - u_{0,\alpha}' u_{0,\beta} + w_{0,\alpha}
w_{0,\beta}' - w_{0,\alpha} ' u_{0,\beta} \Big|_{r=r_c} &=& 0
\nonumber \\
\label{eq:orth_alpha-beta_0} 
\end{eqnarray} 
as well as the short distance regularity conditions,
Eq.~(\ref{eq:bc6}). In all we have an over-determined system with 5
equations and three unknowns $\alpha_{02}$, $\alpha_2$ and $\alpha_0$.
Solving the equations in triplets we have checked the numerical
compatibility at the $0.01\%$ level for the shortest cut-offs, $r_c
\sim 0.02 {\rm fm}$ typically used.  Using the superposition principle
decomposition of the bound state, Eq.~(\ref{eq:sup_bound}), and for
the zero energy states, Eq.~(\ref{eq:sup_zero}), one can make the
orthogonality relations explicit in $\alpha_0$, $\alpha_{02}$,
$\alpha_2$. The values of $\alpha_{02}$ and $\alpha_2$ are not so well
known although they have been determined from potential models in
Ref.~\cite{PavonValderrama:2005ku}.

In Fig.~\ref{fig:lep-frNN} we show the dependence of the low energy
parameters of the leading $N_c$ contributions to the OBE
($\sigma+\pi+\rho+\omega$) potential as a function of $f_{\rho NN}$
for several values of the the effective coupling constant $g_{\omega
NN}^* = \sqrt{g_{\omega NN}^2- f_{\rho NN}^2 m_\rho^2 / 2 M_N^2}$
being $g_{\sigma NN}$ and $m_\sigma$ always readjusted to fit the
$^1S_0$ phase shift. Similarly to the deuteron case we observe
stronger dependence on $f_{\rho NN}$ and a relative insensitivity on
the effective coupling $g_{\omega NN}^* $. We remind that along any of
these curves the $^1S_0$ phase shift is well reproduced with an
acceptable $\chi^2/{\rm DOF} < 1$. As we see, the values $f_{\rho
NN}=17.0$ and $g_{\omega NN}^* =0$ reproduce quite well the low energy
parameters, corresponding to the reasonable $g_{\omega NN} =10.4$.

Numerical results for the low energy parameters are shown in
Table~\ref{tab:table_triplet}. Again, the inclusion of $\sigma$
provides some overall improvement while $\rho$ and $\omega$ yield a
better description of the deuteron for the choice $f_{\rho NN}=15.5$
and $g_{\omega NN}=9.0$. There is nonetheless a small mismatch to the
experimental or recommended potential values when the zero energy wave
functions are obtained from the orthogonality relations to the
deuteron, Eq.~(\ref{eq:orth_triplet}). As one can see further
improvement is obtained when $f_{\rho NN}=17.0$ and $g_{\omega
NN}=10.3$. In this case we get a SU(3) violation; $g_{\omega NN} = 3.5
g_{\rho NN} $, which actually agrees with the expectations from
radiative decays $\omega \to e^+ e^-$ and $\rho \to e^+ e^-$ (see
e.g.~\cite{Dumbrajs:1983jd}). 

\begin{figure*}[ttt]
\begin{center}
\includegraphics[height=5cm,width=5cm,angle=270]{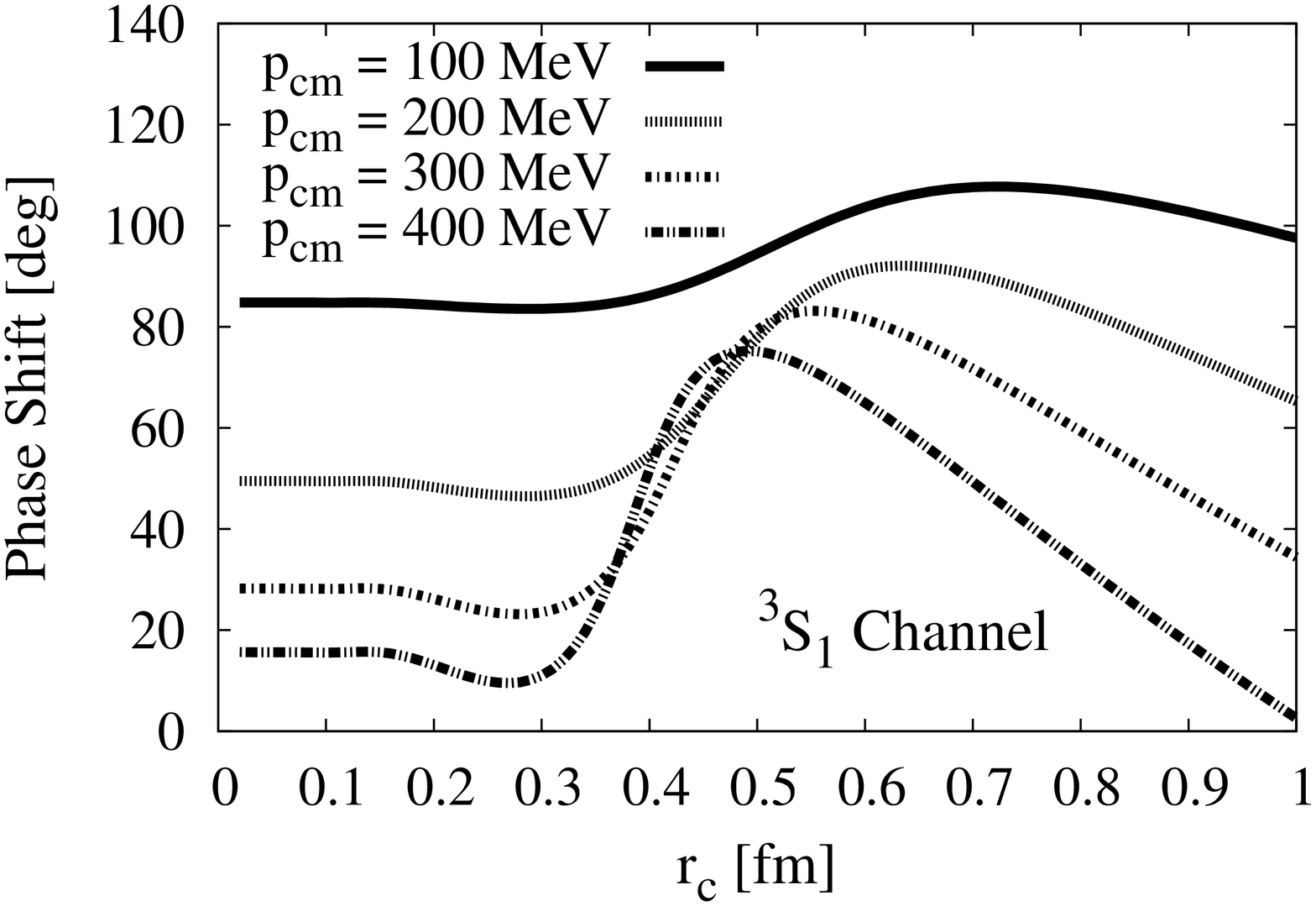}
\includegraphics[height=5cm,width=5cm,angle=270]{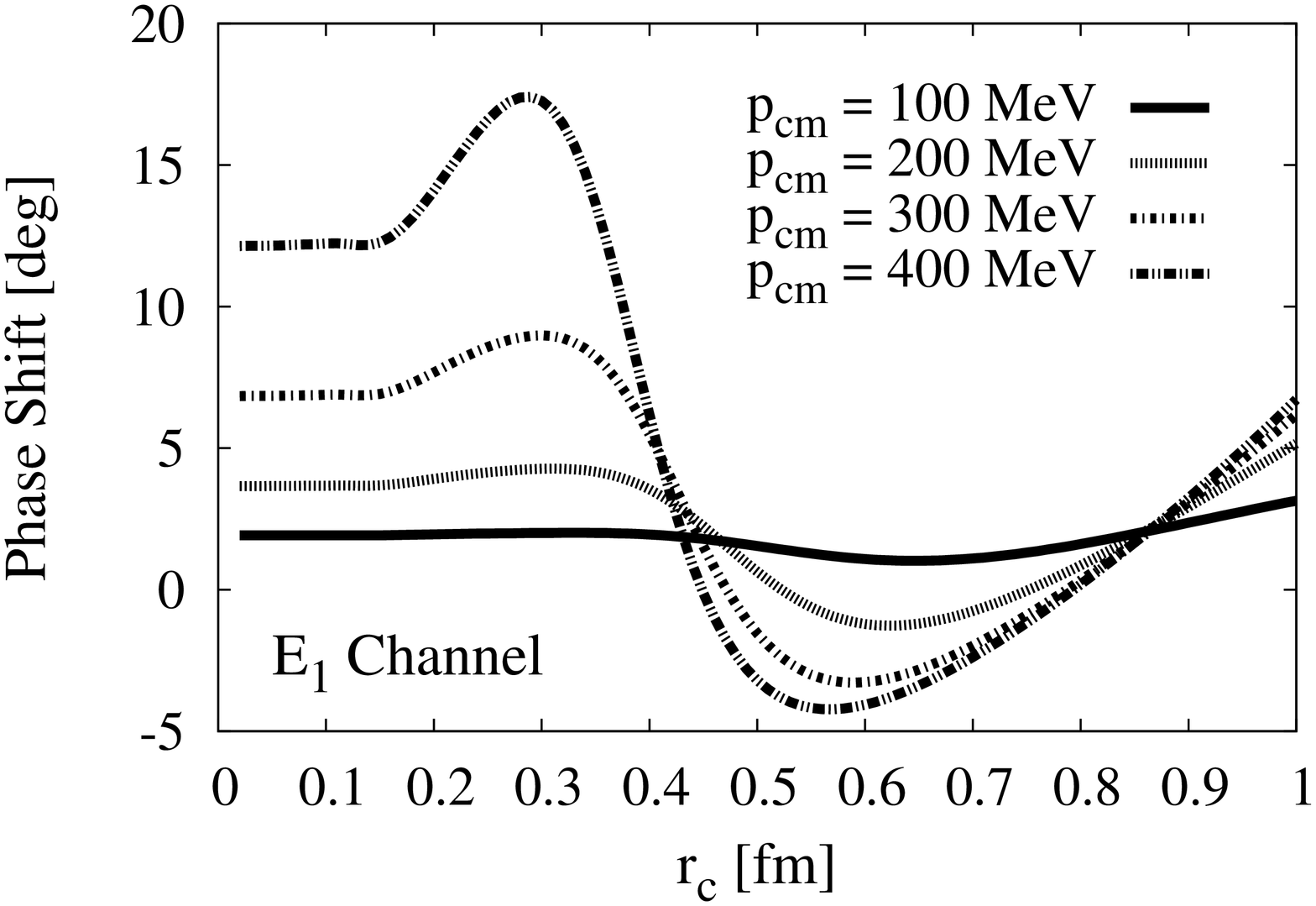}
\includegraphics[height=5cm,width=5cm,angle=270]{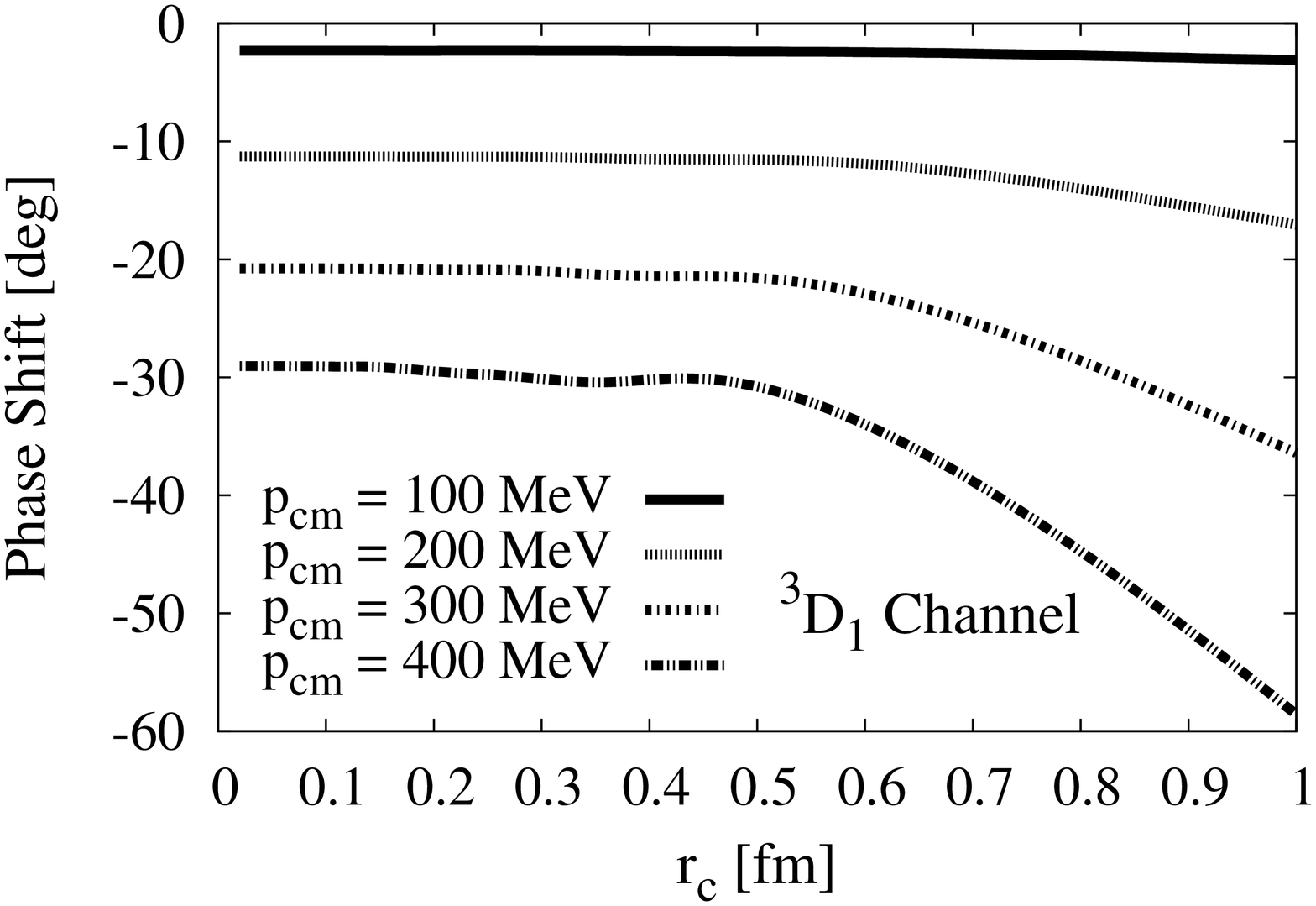}
\end{center}
\caption{Convergence of the np Spin Triplet Eigen phase shifts for the
total angular momentum $j=1$ as a function of the
short distance cut-off radius $r_c$ (in fm) for several fixed values
of the CM momentum $p=100,200,300$ and $400 {\rm MeV}$.}
\label{fig:phase-conv}
\end{figure*}

\begin{figure*}[ttt]
\begin{center}
\includegraphics[height=5cm,width=5cm,angle=270]{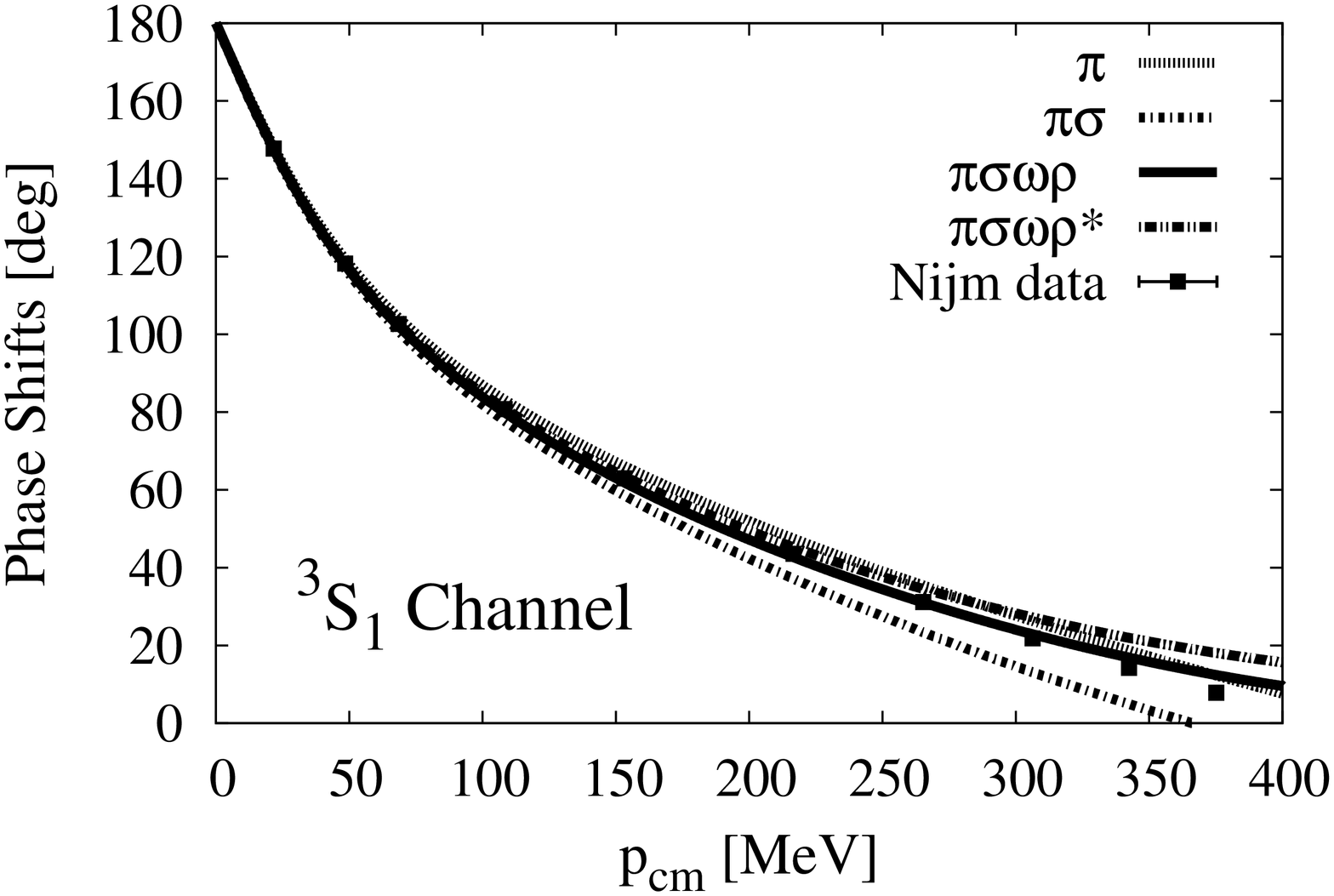}
\includegraphics[height=5cm,width=5cm,angle=270]{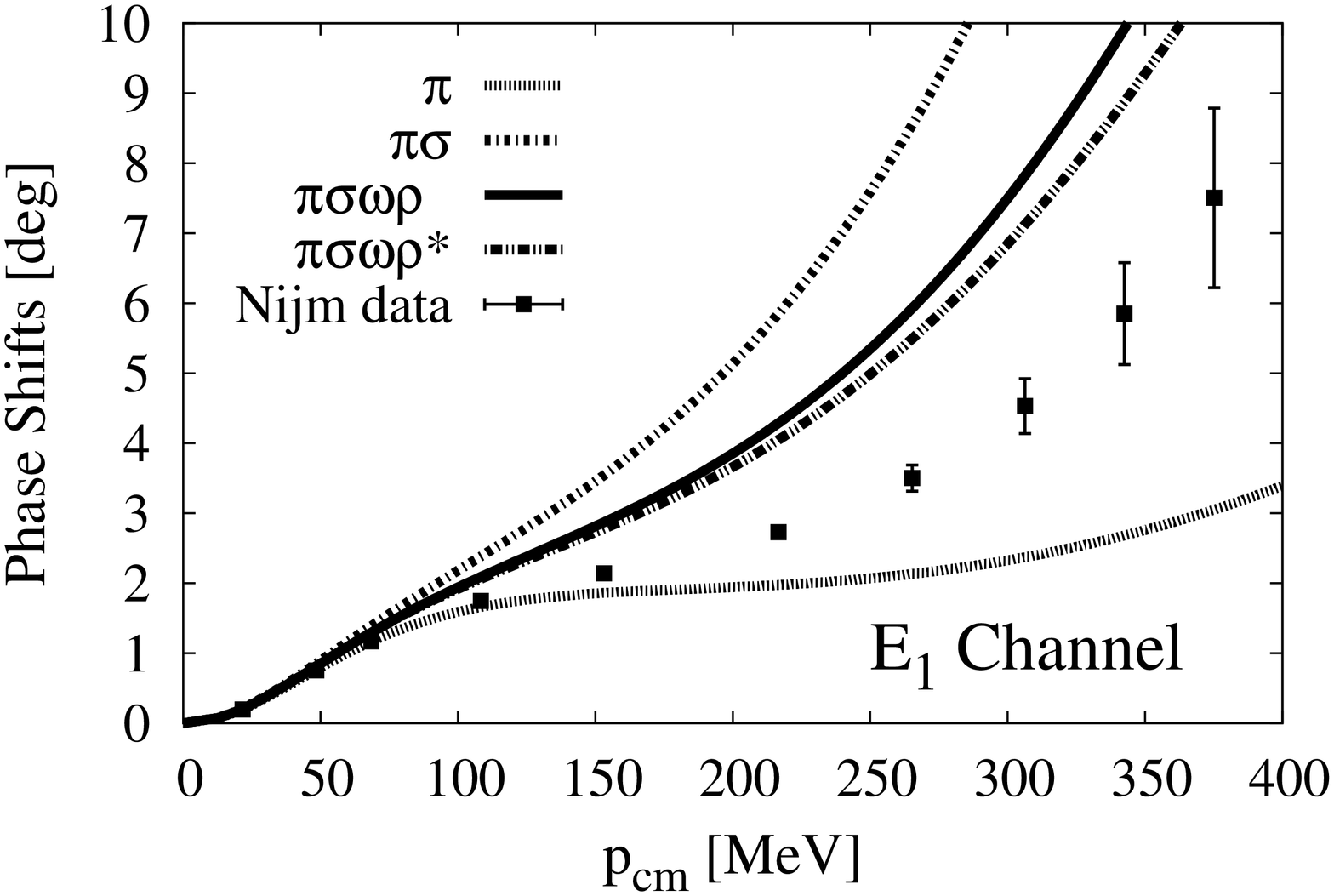}
\includegraphics[height=5cm,width=5cm,angle=270]{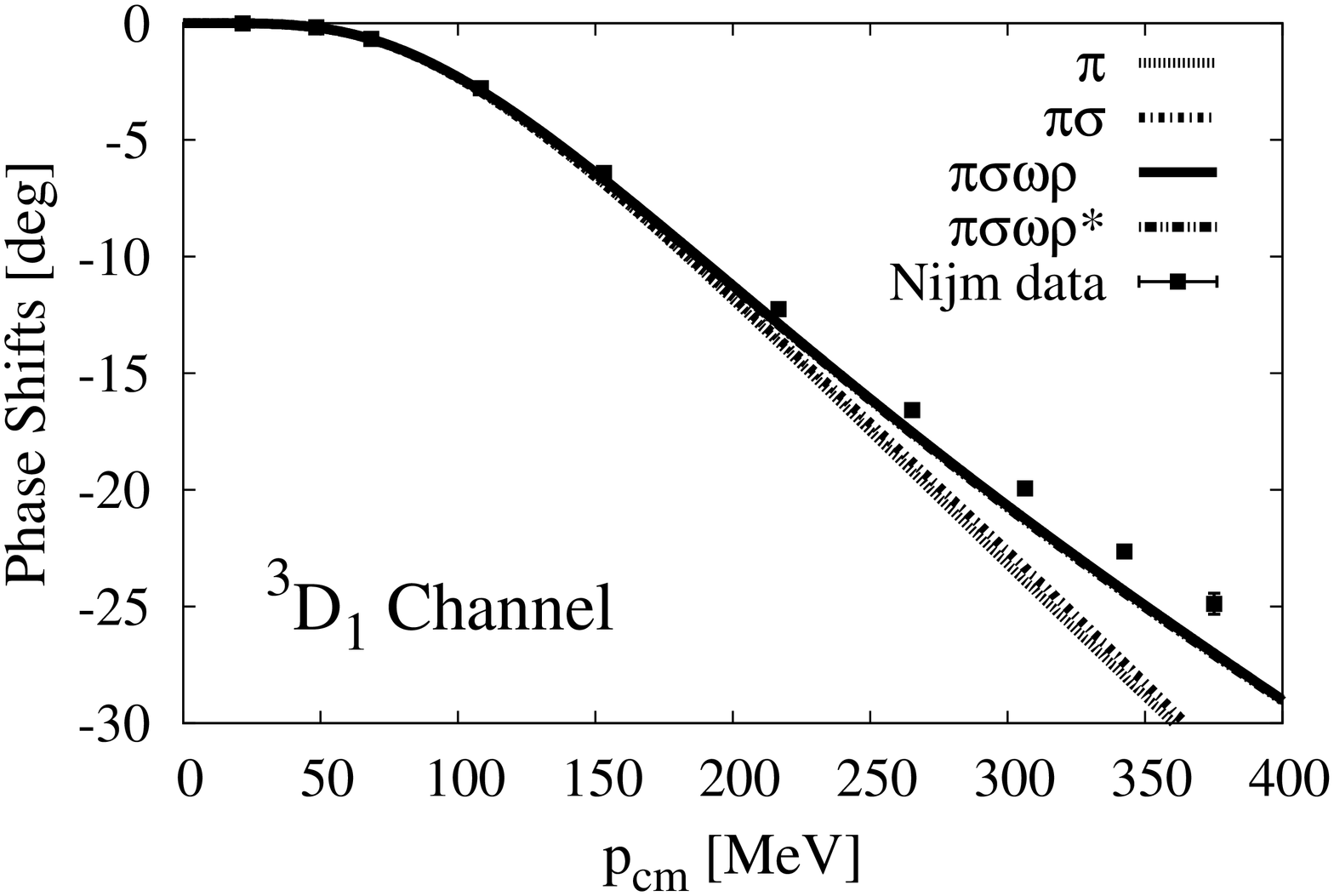}
\end{center}
\caption{np Spin Triplet Eigen phase shifts for the total angular
momentum $j=1$ as a function of as a function of the CM momentum.  We
show $\pi$, $\pi+\sigma$ and $\pi+\sigma+\rho+\omega$ compared to an
average of the Nijmegen partial wave analysis and high quality
potential models~\cite{Stoks:1993tb,Stoks:1994wp}.  We take $(f_{\rho
NN}, g_{\omega NN}) =(15.5,9.857)$, $(f_{\rho NN}, g_{\omega NN}) =(17.0,10.147)$,
.}
\label{fig:phase-triplet}
\end{figure*}

\subsection{Phase shifts} 
\label{sec:phase-shifts}

Finally, in the case of positive energy we consider
Eq.~(\ref{eq:sch_coupled}) with
\begin{eqnarray}
E= \frac{p^2}{M} \, , 
\end{eqnarray} 
with $p$ the corresponding CM momentum.  We solve
Eq.~(\ref{eq:sch_coupled}) for the $\alpha$ and $\beta$ positive
energy scattering states and choose the asymptotic normalization
\begin{eqnarray}
u_{k,\alpha} (r) &\to & \frac{\cos \epsilon}{\sin \delta_1}\Big( \hat
j_0 (kr) \cos \delta_1 - \hat y_0 (kr) \sin \delta_1 \Big) \, ,
\nonumber \\ w_{k,\alpha} (r) &\to & \frac{\sin \epsilon}{\sin
\delta_1}\Big( \hat j_2 (kr) \cos \delta_1 - \hat y_2(kr) \sin
\delta_1 \Big) \, , \nonumber \\ 
u_{k,\beta} (r) & \to & -\frac1{\sin \delta_1}\Big( \hat j_0 (kr) \cos \delta_2 - \hat y_0 (kr)
\sin \delta_2 \Big) \, ,  \nonumber \\ 
w_{k,\beta} (r) &\to & \frac{\tan \epsilon}{\sin \delta_1}\Big( \hat
j_2 (kr) \cos \delta_2 - \hat y_2(kr) \sin \delta_2 \Big) \, , \nonumber \\ 
\label{eq:phase_triplet}
\end{eqnarray} 
where $ \hat j_l (x) = x j_l (x) $ and $ \hat y_l (x) = x y_l (x) $
are the reduced spherical Bessel functions and $\delta_1$ and
$\delta_2$ are the eigen-phases in the $^3S_1$ and $^3D_1$ channels,
and $\epsilon$ is the mixing angle $E_1$. To carry out the
renormalization program, we use the superposition principle of
boundary conditions which makes the discussion more transparent. Let
us define the four auxiliary problems
\begin{widetext}
\begin{eqnarray}
\begin{pmatrix} u_{k,1} \\ w_{k,1} \end{pmatrix} \to 
\begin{pmatrix} \hat j_0 (kr) \\ 0 \end{pmatrix} \, , 
\qquad 
\begin{pmatrix} u_{k,2} \\ w_{k,2} \end{pmatrix} \to 
\begin{pmatrix} \hat y_0 (kr) \\ 0 \end{pmatrix} \, , 
\qquad 
\begin{pmatrix} u_{k,3} \\ w_{k,3} \end{pmatrix} \to 
\begin{pmatrix} 0 \\ \hat j_2 (kr)  \end{pmatrix} 
\, , 
\qquad 
\begin{pmatrix} u_{k,4}  \\ w_{k,4} \end{pmatrix} \to 
\begin{pmatrix} 0 \\  \hat y_2 (kr)  \end{pmatrix} \, ,  
\end{eqnarray} 
which depend solely on the potential and can be obtained by
integrating in. Thus, the general solution satisfying the $\alpha$
and $\beta$ asymptotic conditions can be written as
\begin{eqnarray}
u_{k,\alpha} (r) = \sum_{i=1}^4 c_{i,\alpha} u_{k,i}(r) \, ,\quad 
w_{k,\alpha} (r) = \sum_{i=1}^4 c_{i,\alpha} w_{k,i}(r) \, ,\quad 
u_{k,\beta} (r) = \sum_{i=1}^4 c_{i,\beta} u_{k,i}(r) \, ,\quad 
w_{k,\beta} (r) = \sum_{i=1}^4 c_{i,\beta} w_{k,i}(r)  \, .
\end{eqnarray} 
\end{widetext}
Fixing the constants to the asymptotic conditions
Eq.~(\ref{eq:phase_triplet}) we get 
\begin{eqnarray}
u_{k,\alpha} (r) & = & \frac{\cos \epsilon}{\sin \delta_\alpha}\Big( 
u_1(r) \cos \delta_\alpha - u_2(r) \sin \delta_\alpha \Big) \,  \nonumber
\\  
&+&\frac{\sin \epsilon}{\sin \delta_\alpha}\Big( \cos \delta_\alpha u_3 (r) -
u_4 (r) \sin \delta_\alpha \Big) \, , \nonumber \\ 
w_{k,\alpha} (r) & = & \frac{\cos \epsilon}{\sin \delta_\alpha}\Big( 
w_1(r) \cos \delta_\alpha - w_2(r) \sin \delta_\alpha \Big) \,  \nonumber
\\  
&+&\frac{\sin \epsilon}{\sin \delta_\alpha}\Big( \cos \delta_\alpha w_3 (r) -
w_4 (r) \sin \delta_\alpha \Big) \, , \nonumber 
\end{eqnarray} 
\begin{eqnarray} 
u_{k,\beta} (r) & = & \frac1{\sin \delta_\alpha}\Big( 
u_1(r) \cos \delta_\beta - u_2(r) \sin \delta_\beta \Big) \,  \nonumber
\\  
&-&\frac{\tan \epsilon}{\sin \delta_\alpha}\Big( \cos \delta_\beta u_3 (r) -
u_4 (r) \sin \delta_\beta \Big) \, , \nonumber  \\ 
w_{k,\beta} (r) & = & \frac1{\sin \delta_\alpha}\Big( 
w_1(r) \cos \delta_\beta - w_2(r) \sin \delta_\beta \Big) \,  \nonumber
\\  
&-&\frac{\tan \epsilon}{\sin \delta_\alpha}\Big( \cos \delta_\beta w_3 (r) -
w_4 (r) \sin \delta_\beta \Big) \, . \nonumber \\ 
\label{eq:phase_triplet_bis}
\end{eqnarray} 

\begin{figure*}[]
\begin{center}
\includegraphics[height=7.5cm,width=7.1cm,angle=270]{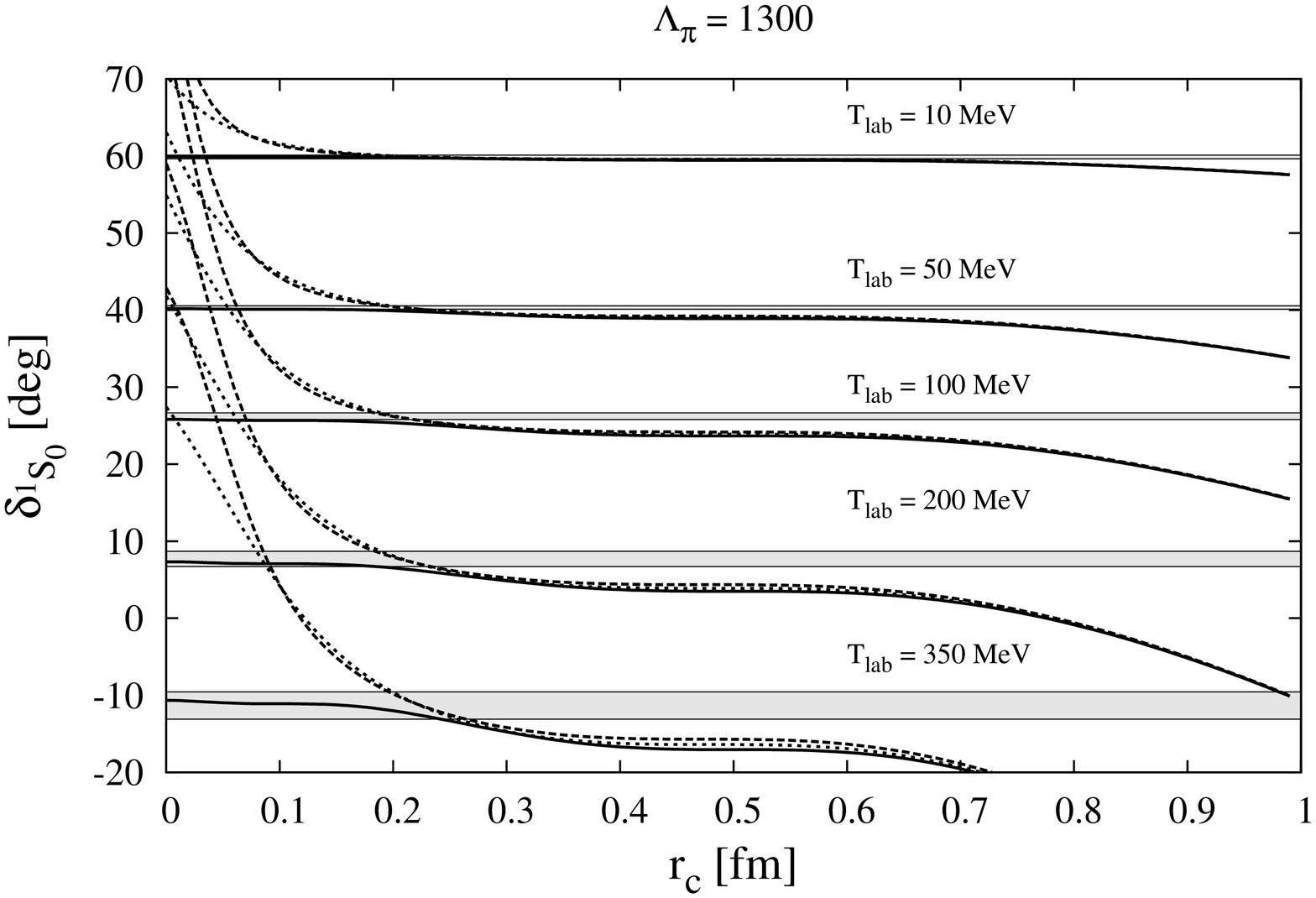} 
\includegraphics[height=7.5cm,width=7.1cm,angle=270]{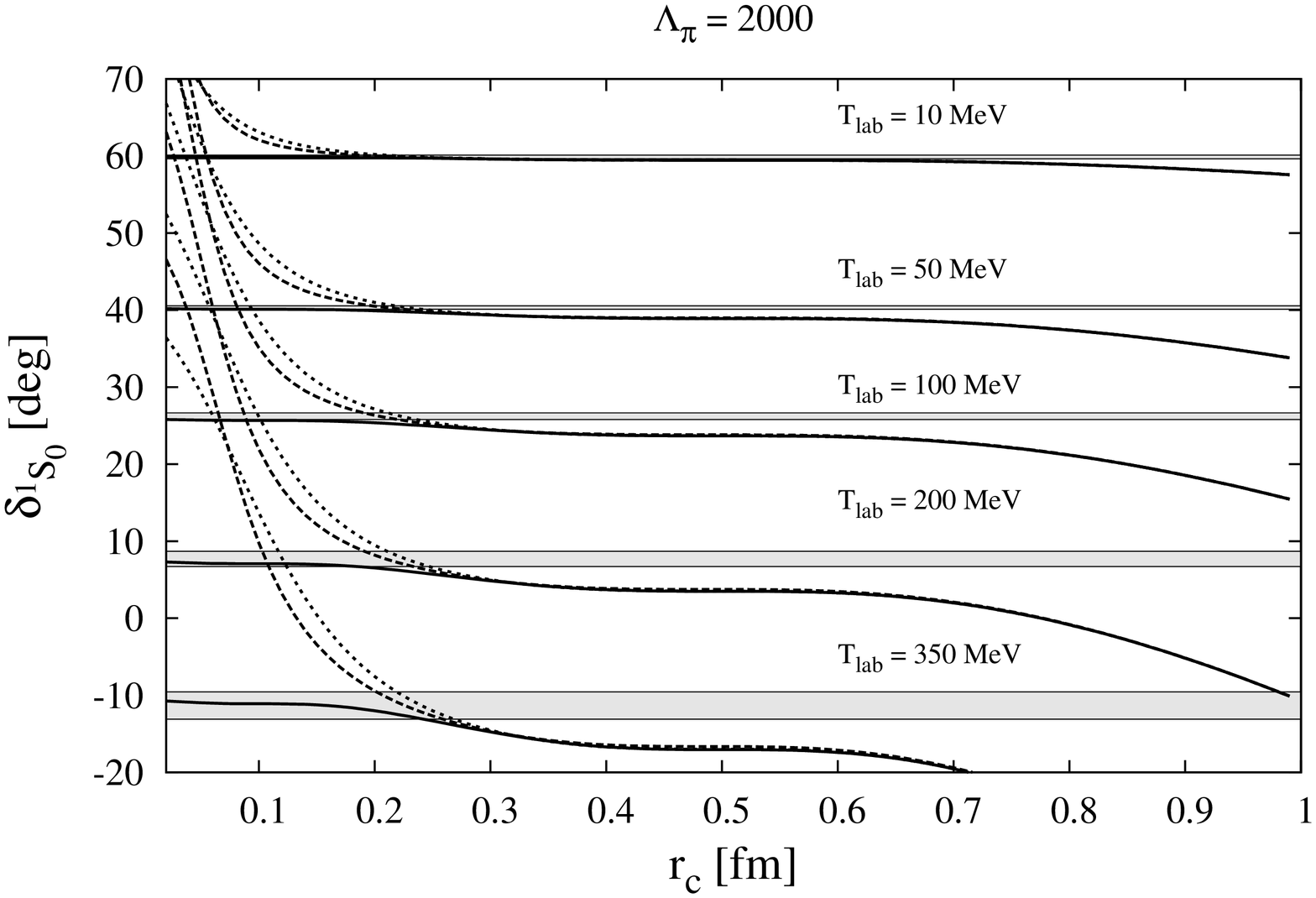}
\end{center}
\caption{Convergence of the np Spin Singlet $^1S_0$ phase shift (in
  degrees) as a function of the short distance cut-off radius $r_c$
  (in fm) for several fixed values of the LAB energy when the
  scattering length $\alpha_0=-23.74 {\rm fm}$ is also fixed. We
  compare the purely renormalized case with no form factors (solid
  line) with the renormalization {\it also} including the exponential
  (dotted line) and monopole (dashed line) form factors and for the
  cut-off values $\Lambda_{\pi NN}=1300 {\rm MeV}$ (left panel) and
  $\Lambda_{\pi NN}=2000 {\rm MeV}$ (right panel), all others fixed to
  $\Lambda_{\sigma NN}= \Lambda_{\rho NN}= \Lambda_{\omega NN}= 2 {\rm
    GeV}$. We also add the error bands related to the Nijmegen PWA and
  high quality potentials~\cite{Stoks:1993tb,Stoks:1994wp}. All other
  meson parameters in the OBE potential are kept the same.}
\label{fig:phase-conv-1S0}
\end{figure*}

In the low energy limit $\epsilon \to -\alpha_{02} k^3$ $\delta_\alpha
\to -\alpha_{0} k$ and $\delta_\beta \to -(\alpha_{2}-
\alpha_{02}^2/\alpha_0) k^5$ and the zero energy solutions discussed
in Sec.~\ref{sec:zero-energy} are reproduced.  The use of the
superposition principle for boundary conditions as well as the
orthogonality constraints to the deuteron wave analogous to
Eq.~(\ref{eq:orth_singlet}) yields
\begin{eqnarray}
u_\gamma u_{k,\alpha}' - u_\gamma' u_{k,\alpha} + w_\gamma
w_{k,\alpha}' - w_\gamma ' u_{k,\alpha} \Big|_{r=r_c} &=& 0 \nonumber \\
u_\gamma u_{k,\beta}' - u_\gamma' u_{k,\beta} + w_\gamma w_{k,\beta}'
- w_\gamma ' u_{k,\beta} \Big|_{r=r_c} &=& 0 \nonumber \\
\label{eq:orth_triplet_k} 
\end{eqnarray} 
which together with the short distance regularity conditions,
Eq.~(\ref{eq:bc6}) allow us to deduce the corresponding $^3S_1-^3D_1$
phase-shifts. A further condition is the $\alpha-\beta$ orthogonality 
\begin{eqnarray}
u_{k,\alpha} u_{k,\beta}' - u_{k,\alpha}' u_{k,\beta} + w_{k,\alpha}
w_{k,\beta}' - w_{k,\alpha} ' u_{k,\beta} \Big|_{r=r_c} &=& 0
\nonumber \\
\label{eq:orth_alpha-beta_k} 
\end{eqnarray} 
In all we have an over-determined system with 5 equations and three
unknowns. We have checked that almost any choice yields equivalent
results with an accuracy of $0.001^o$ for the highest CM momenta and
the shortest cut-off, $r_c \sim 0.02 {\rm fm}$.

\begin{table*}[ttt]
\caption{\label{tab:table_fits_bc} Fits to the Renormalized $^1S_0$
  phase shift of the Nijmegen group~\cite{Stoks:1994wp} using the OBE
  potential without or with strong exponential and monopole form
  factor.  We fix $\alpha_0 = -23.74 {\rm fm}$ and take $m=138.03 {\rm
    MeV} $, and $g_{\pi NN} =13.1083 $ \cite{deSwart:1997ep} and
  $m_\rho=m_\omega=770 {\rm MeV}$ and fit $m_\sigma$, $g_{\sigma NN}$
  and fix $g_{\omega NN}^*=0$. We use $\Lambda_{\pi NN}=1300 {\rm
    MeV}$ and $\Lambda_{\sigma NN}= \Lambda_{\rho NN}= \Lambda_{\omega
    NN}= 2000 {\rm MeV}$. $E_B$ represents the energy of the
  (spurious) bound state when it does exist.}
\begin{ruledtabular}
\begin{tabular}{|c|c|c|c|c|c|c|c|c|}
 & $ r_c ({\rm fm})$ &  $m_\sigma ({\rm MeV})$ & $g_{\sigma NN}$ &  $g_{\omega NN}^*$ & $\chi^2 /DOF $ & $\alpha_0 ({\rm fm })$ &  $r_0 ({\rm fm})$ & 
$E_B ({\rm MeV}) $\\ 
\hline 
$\Gamma (q^2)=1$       & 0  & 501(25) & 9(1) & 0(3) & 0.12 & Input & 2.695& -777  \\ 
\hline 
$\Gamma (q^2)= \Gamma^{\rm exp} (q^2) $ & 0 & 526(20)    & 10.4(8) & 0(3) & 0.19 & Input & 2.692& -790 \\ 
\hline 
$\Gamma (q^2)= \Gamma^{\rm exp} (q^2) $ & 0.1 & 523(27)    & 10.2(1.1) & 0(3) & 0.18 & Input & 2.491& -834 \\ 
\hline 
$\Gamma (q^2)= \Gamma^{\rm mon} (q^2) $ & 0  & 532(20)    & 10.7(7) & 0(3) & 0.20 & Input & 2.691& -796  \\  
\hline 
$\Gamma (q^2)= \Gamma^{\rm mon} (q^2) $ & 0.1  & 528(28)    & 10.5(1.1) & 0(3) & 0.19 & Input  &2.490 & -853\\  
\end{tabular}
\end{ruledtabular}
\end{table*}

As we have mentioned already, the numerical solution of the problem
requires taking care of spurious amplification of the undesired
growing exponential at any step of the calculation. The situation is
aggravated by the fact that for the phase shifts the maximum momentum
$p=400 {\rm MeV}$ explores the region around $0.1-0.5 {\rm fm}$, so it
is important to make sure that we do not see cut-off effects in this
region. To provide a handle on the numerical uncertainties we show in
Fig.~(\ref{fig:phase-conv}) the results for the phase shifts
$\delta_1$, $\delta_2$ and $\epsilon$ as a function of the cut-off
radius, $r_c$ and for several fixed CM pn momenta, $p=100,200,300,400
{\rm MeV}$. As we see, there appear
clear plateaus between $0.1-0.2 {\rm fm}$ which somewhat steadily
shrink when the momentum is increased. Note that these values of the
short distance cut-off translates into a CM momentum space cut-off
range $\Lambda = \pi / (2 r_c) = 1.5-3 {\rm GeV}$.

The results for the $^3S_1-^3D_1$ phase shifts as a function of the CM
momentum are depicted in Fig.~\ref{fig:phase-triplet} for $\pi$,
$\pi+\sigma$ and $\pi+\sigma+\rho+\omega$ and compared to the Nijmegen
analysis~\cite{Stoks:1993tb,Stoks:1994wp}. We use $g_{\sigma NN}=9.1
$, $m_\sigma=501 {\rm MeV}$ and when vector mesons are included we
take $f_{\rho NN}=15.5$ and $g_{\omega NN}=9$ or $f_{\rho NN}=17.0$
and $g_{\omega NN}=10.147$ corresponding to Sets $\pi\sigma\rho\omega$
and $\pi\sigma\rho\omega^*$ in Table~\ref{tab:table_triplet}
respectively. On a first sight we see an obvious improvement in both
the $^3S_1$ and $^3D_1$ phases and not so much in the mixing angle
$E_1$ as compared to the simple OPE case. One should note, however,
that besides describing by construction the single phase shift $^1S_0$
(see Fig.~(\ref{fig:elipsefits}) ) we also improve on the deuteron
(see Table~\ref{tab:table_triplet}). Obviously, it would be possible
to provide a better description of triplet phase shifts, however, at
the expense of worsening the deuteron properties and the singlet
channel.  Clearly, there is room for improvement, and our results call
for consideration of sub-leading large $N_c$ corrections in the OBE
potential. This would incorporate, the relative to leading $1/N_c^2$
relativistic corrections, spin-orbit effects, finite meson widths,
non-localities, etc.

\section{Influence of Strong form factors in the renormalization process} 
\label{sec:formfac} 

Given the reasonable phenomenological success of the renormalization
approach one may naturally wonder what would be the effect of the form
factors in our calculation. In this section we discuss the influence
of strong form factors in the calculated properties {\it on top of the
renormalization process}. Our main quest is to find out whether they
lead to observable physical effects {\it after} renormalization. An
equivalent way of posing the question is to determine whether finite
nucleon size effects can be disentangled explicitly in NN scattering
in the elastic region.

To analyze this important issue in detail, in
Fig.~\ref{fig:phase-conv-1S0} we show the phase shift in the $^1S_0$
channel for fixed LAB energy values as a function of the short
distance cut-off radius $r_c$ when the scattering length is fixed to
its experimental value, $\alpha_0=-23.74 {\rm fm}$ as we explained in
Section~ \ref{sec:singlet}. We use the same parameters as for the
renormalized solution without vertex function, for several fixed
values of the LAB energy and for the cut-off values $\Lambda_{\pi
NN}=1300 {\rm MeV}$ and $\Lambda_{\pi NN}=2000 {\rm MeV}$, all others
fixed to $\Lambda_{\sigma NN}= \Lambda_{\rho NN}= \Lambda_{\omega NN}=
2 {\rm GeV}$. As one clearly sees strong form factors are invisible
for $r_c > 0.3 {\rm fm}$. For lower values of the short distance
cut-off $r_c$ both monopole and exponential form factors agree with
each other but deviate strongly from the Nijmegen database. Note that
the lines should be supplemented with estimates of theoretical errors,
not shown to avoid cluttering of the plot.  When those errors are
included the Nijmegen data are basically compatible with the
theoretical curves in the flat preasymptotic region around $0.3-0.5
{\rm fm}$ (see also the discussion around Fig.~\ref{fig:chi2-rc}).

\begin{figure*}[ttt]
\begin{center}
\includegraphics[height=5cm,width=5cm,angle=270]{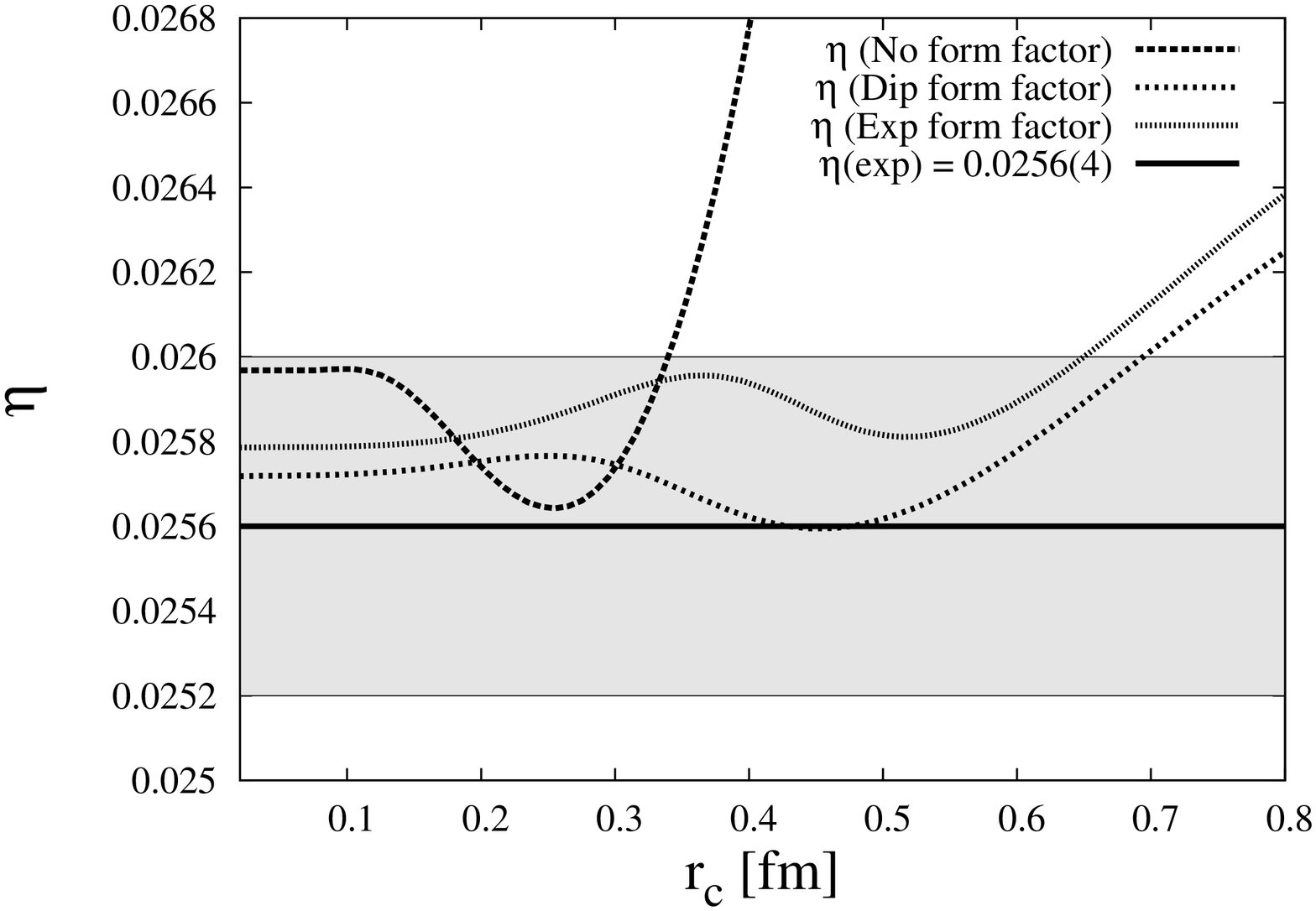}
\includegraphics[height=5cm,width=5cm,angle=270]{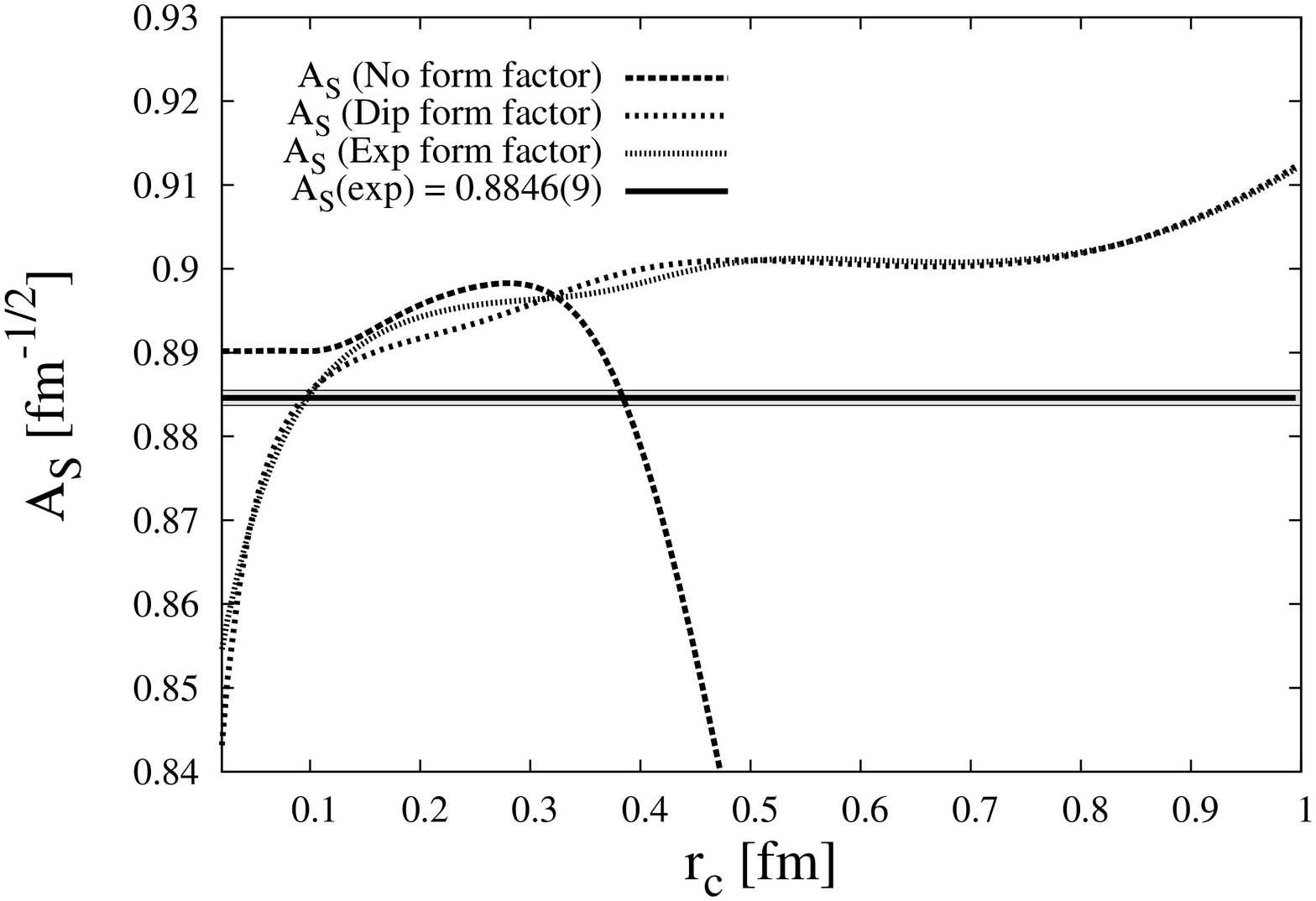}
\includegraphics[height=5cm,width=5cm,angle=270]{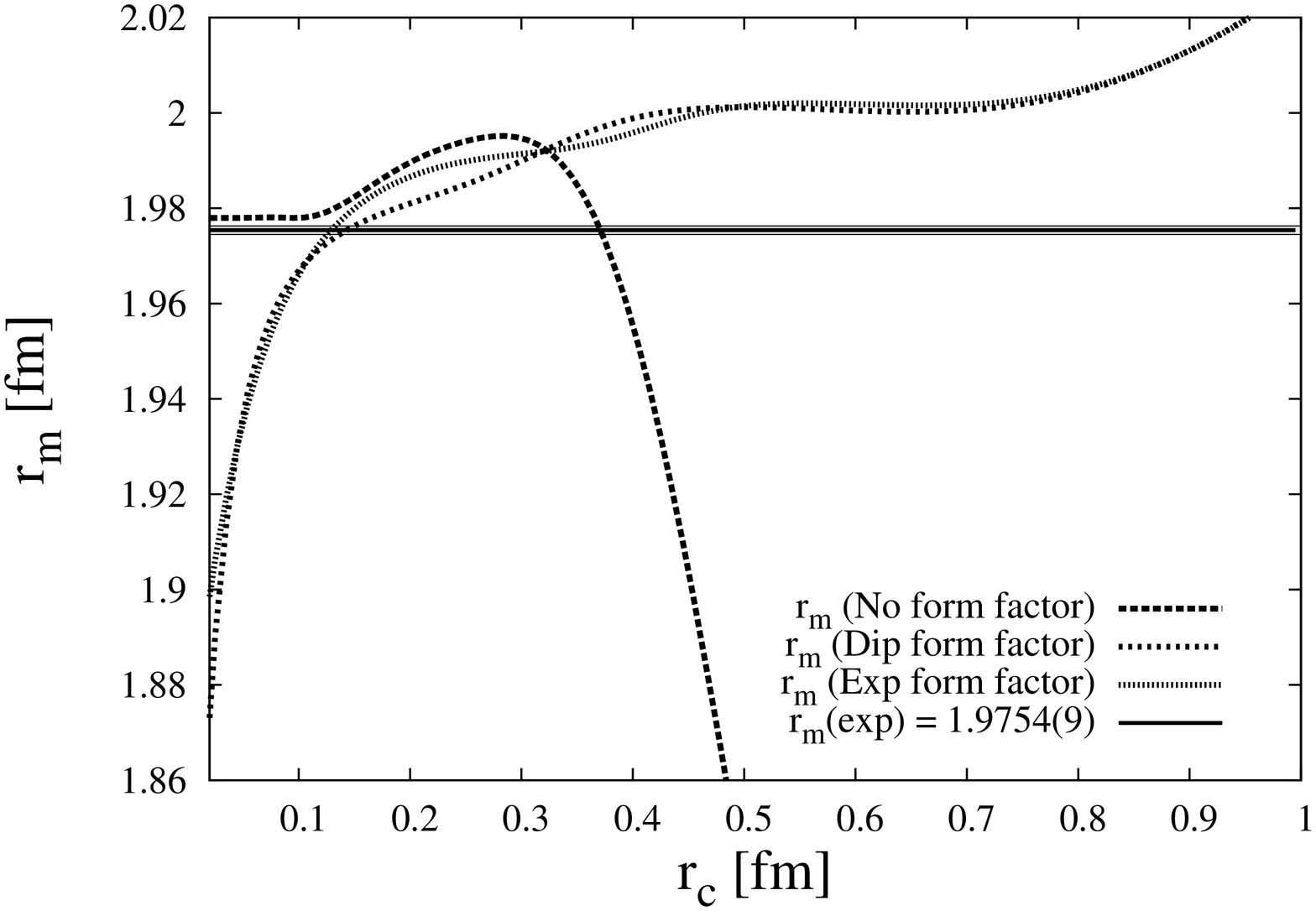}\\ 
\includegraphics[height=5cm,width=5cm,angle=270]{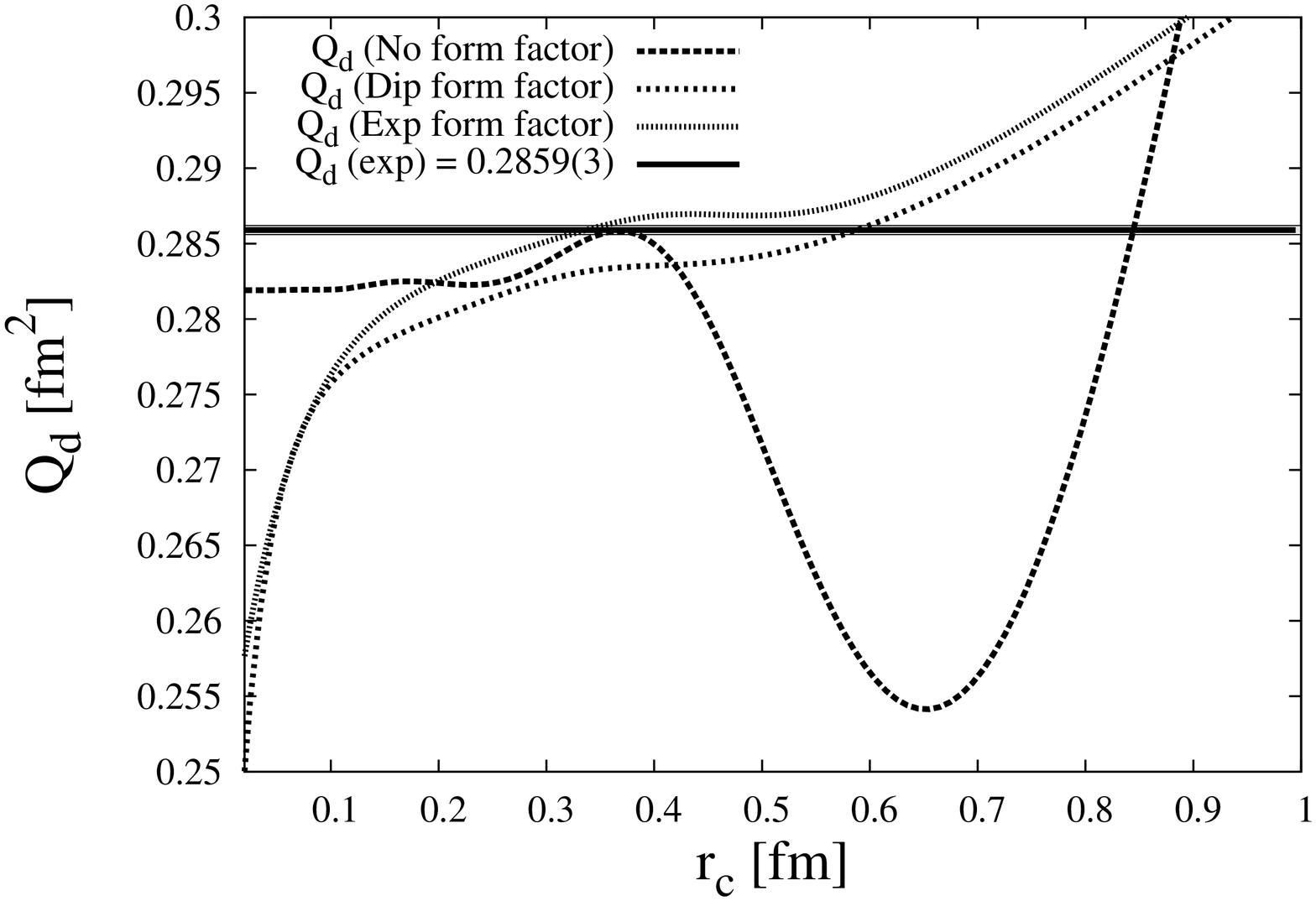}
\includegraphics[height=5cm,width=5cm,angle=270]{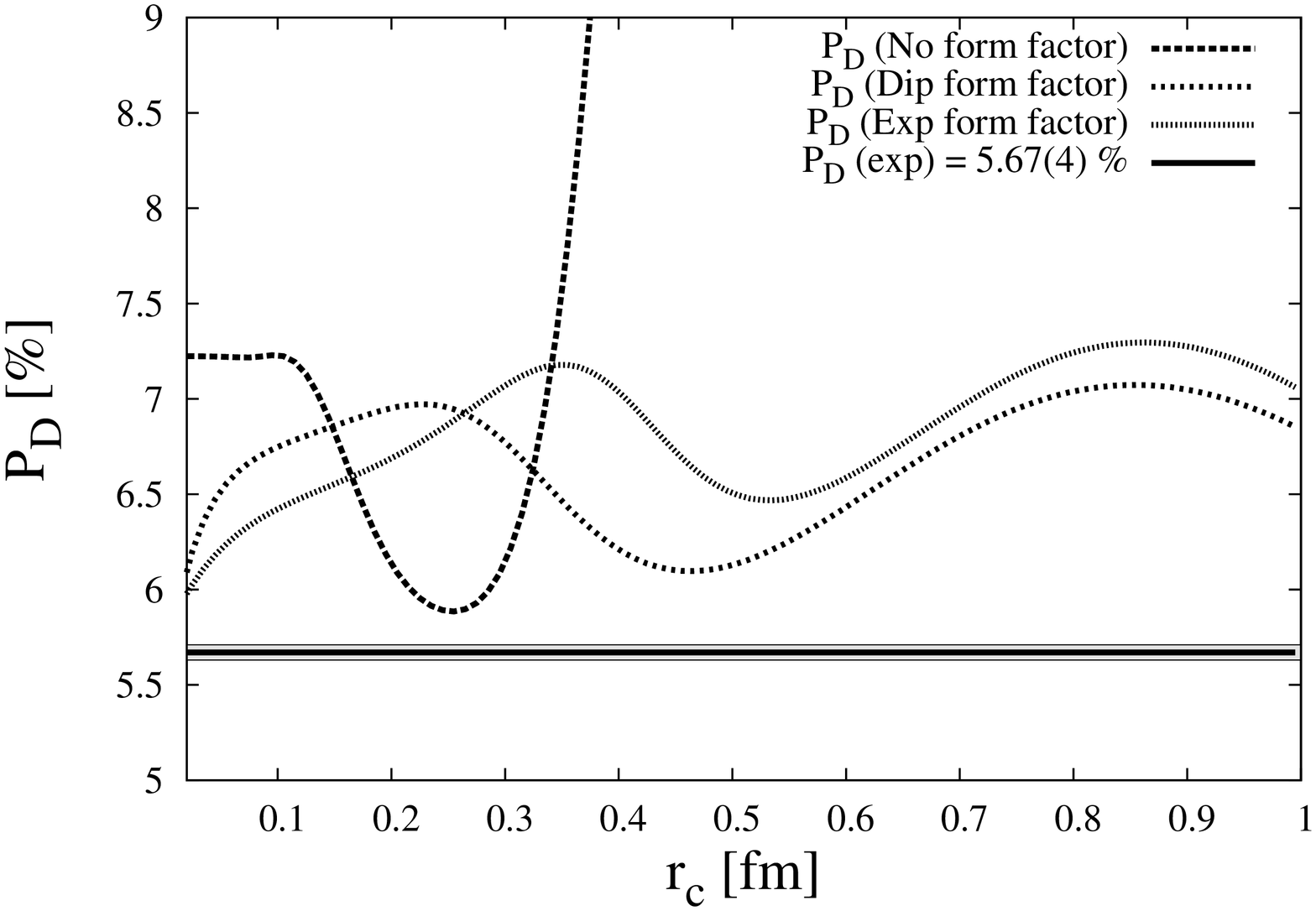}
\includegraphics[height=5cm,width=5cm,angle=270]{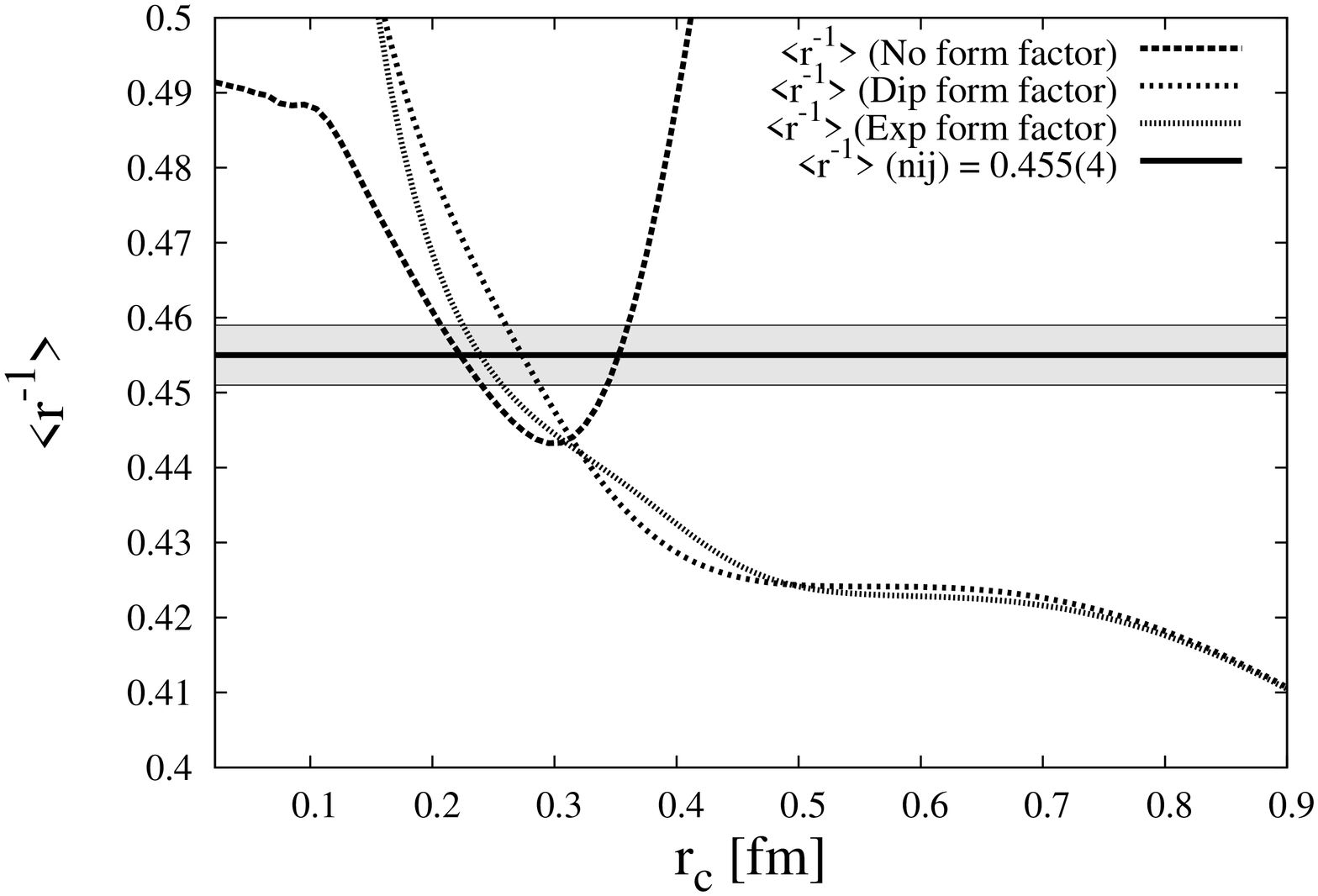}
\end{center}
\caption{Short distance cut-off $r_c$ dependence of deuteron
  properties for the $\pi\sigma\rho\omega$ case (see
  table~\ref{tab:table_triplet}). We compare the purely renormalized
  calculation with the cases for both exponential,
  Eq.~(\ref{eq:exponential}), and monopole Eq.~(\ref{eq:monopole})
  form factors taking $\Lambda_{\pi NN}=1300 {\rm MeV}$, all other
  cut-offs being kept to $\Lambda_{\sigma NN}= \Lambda_{\rho NN}=
  \Lambda_{\omega NN}= 2000 {\rm MeV}$.  We show the dependence of the
  asymptotic D/S normalization $\eta$ (upper left panel), the S-wave
  normalization $A_S$ (in ${\rm fm}^{-1/2} $, upper middle panel), the
  matter radius $r_m$ (in ${\rm fm} $, upper right panel), the
  quadrupole moment $ Q_d $ (in ${\rm fm}^2 $, lower left panel), the
  $D$-state probability (lower middle panel) and the inverse radius
  $\langle r^{-1} \rangle $ (in ${\rm fm}^{-1}$ lower right panel).
Experimental or recommended
  values can be traced from Ref.~\cite{deSwart:1995ui}.}
\label{fig:finite_cutoff-formfac}
\end{figure*}

Of course, one may attribute the discrepancy to the choice of
parameters, which have been chosen to fit the renormalized solution
without form factors. A somewhat complementary way of seeing this is
by refitting the parameters using both exponential and monopole vertex
functions but fixing by construction the scattering length
$\alpha_0$. The results for $\Lambda_{\pi NN}=1.3 {\rm GeV}$ and
$\Lambda_{\sigma NN}= \Lambda_{\rho NN}= \Lambda_{\omega NN}= 2 {\rm
  GeV}$ are displayed in Table~\ref{tab:table_fits_bc}. As we see, the
parameters change almost within the uncertainties, showing the
marginal effect of the vertex functions {\it after
  renormalization}. Due to the presence of non-linear correlations,
difficult to handle by standard means, we have fixed $g_{\omega NN}^*$
to its minimum value (compatible with zero) and estimated its error by
varying it independently from its mean value to values still giving an
acceptable fit, yielding $g_{\omega NN}^*=0(3)$. We also show the
effect of the short distance cut-off $r_c$ which, as we see, is rather
small. Overall, these results provide a further confirmation of our
naive expectations; nucleon finite size effects and vector mesons do
not provide the bulk in NN scattering in central waves, and actually
cannot be clearly resolved. Of course, this should be checked in
higher partial waves, but those are expected in fact to be less
sensitive to short distances.

Finally, in Fig.~\ref{fig:finite_cutoff-formfac} the influence of the
vertex functions is analyzed for some of the computed deuteron
properties. As we see there is a fair coincidence of the purely
renormalized solution with no form factors with the equally
renormalized solution including the form factors in the potential in
the region around $r_c \sim 0.3-0.6 {\rm fm}$. The deviation below 0.3
fm signals the onset of the irregular $D$-wave solution, which behaves
as $w(r) \sim r^{-2}$ at small distances and hence yields eventually a
divergent result. Note that in order to have a smooth behaviour at
short distances when renormalization is over-imposed to the potential
with form factors we should choose the regular D-wave solution $w(r)
\sim r^{3}$ but then the potential parameters, either couplings or
form factor cut-off parameters should also be {\it fine-tuned}.

While it is fairly clear that vertex functions do exist and are of
fundamental importance, it is also true that they start playing a role
as soon as the probing wavelength resolves the finite nucleon size.
Our calculations suggest on a quantitative level that provided the NN
scattering data are properly described with form factors, they will be
effectively irrelevant under the renormalization process, and for CM
momenta below $400 {\rm MeV}$, vertex functions are expected to play a
marginal role.

\section{Conclusions and Outlook} 
\label{sec:concl}

In the present paper we have analyzed the OBE potential from a
renormalization point of view. As we have shown, the meson-nucleon
Lagrangean does not predict the S-matrix beyond perturbation theory.
The non-perturbative nature of low partial waves and the deuteron in
the NN problem suggests resuming OBE diagrams by extracting the
corresponding potential. The OBE potential, however, presents short
distance divergences which make the solution of the corresponding
Schr\"odinger equation ambiguous. The traditional remedy for this
problem has been the inclusion of phenomenological form factors which
parameterize the vertex functions in the meson exchange picture. We
have shown that the meson exchange potential with form factors
generates spurious deeply bound states for natural values of the
coupling constants. The price to remove those is to fine tune the
potential at all distances, and in particular at short distances.
Thus, while it is claimed that vertex functions implement the finite
nucleon size, it is very difficult to disentangle this from meson
dressing and many other effects where the meson theory does not
hold.

The renormalization approach suggests that extracting this detailed
short distance information may in fact be unnecessary for the purposes
of Nuclear Physics and the verification of the meson exchange picture.
Contrarily to what one might naively think, renormalization is a
practical and feasible way of minimizing short distance ambiguities,
by imposing conditions which are fixed by low energy
data independently on the potential.  We have argued that within this
approach we face from the start our inability to pin down the short
distance physics {\it below} the smallest de Broglie wave length
probed in NN scattering. Indeed, the central scattering waves and the
deuteron can be described reasonably well and with natural values of
the meson-nucleon couplings.  Within the standard approach this could
only be achieved by fine tuning meson parameters or postulating the
meson exchange picture to even shorter ranges than $0.5 {\rm fm}$. In
our case the inclusion of shorter range mesons induces moderate
changes, due to the expected short distance insensitivity embodied by
renormalization, {\it despite} the short distance singularity and {\it
  without} introducing strong meson-nucleon-nucleon vertex functions.
If phenomenological vertex functions are added on top of the
renormalized calculations minor effects are observed confirming the
naive expectation that finite nucleon size $\sim 0.5 {\rm fm}$ need
not be explicitly introduced within the OBE calculations for CM
momenta corresponding to the minimal wavelength $1/p \sim 0.5 {\rm
  fm}$.

The renormalization process introduces spurious deeply bound states
regardless on whether or not the potential is regular or
singular. This can be appreciated in the excessive number of nodes of
the wave function close to the origin, in the region below $0.5 {\rm
fm}$. We have checked that the corresponding CM energies are in
absolute values much higher than the maximum scattering CM energies,
and hence the role played by these spurious states is completely
irrelevant. We remind that within the standard approach with form
factors those spurious bound states also take place when natural
values of the coupling constants are taken.

One of the problems with potential model calculations is the ambiguity
in form of the potential, since it is determined from the on-shell
S-matrix in the Born approximation and an off-shell extrapolation
becomes absolutely necessary. In the large $N_c$ limit the
spin-isospin and kinematic structure of the NN potential simplifies
tremendously yielding a non-relativistic and uniquely defined local
and energy independent function. Relativistic effects, spin-orbit,
non-localities as well as meson widths or other mesons enter as
sub-leading corrections to the potential with a relative order
$1/N_c^2$. However, it consists of an infinite tower of multi-meson
exchanged states, which range is given by the Compton wavelength of
the total multi-meson mass.  One of the advantages of the large $N_c$
expansion is that it is not particularly restricted for low energies.
This is exemplified by several recent calculations of NN potentials
using the holographic principle based on the AdS/CFT
correspondence~\cite{Kim:2008iy,Kim:2009sr,Hashimoto:2009ys}~\footnote{In
this calculations only $\pi$,$\rho$,$\omega$ and $A_1$ mesons and
their radial excitations contribute. Note, however, that the {\it
only} contribution to the central force $V_C$ stems from the tower
$\omega,\omega',\omega'', \dots$ which is generally repulsive.}.  A 
truncation of the infinite number and range of exchanged mesons is
based on the assumption that the hardly accessible high mass states
are irrelevant for NN energies below the inelastic pion production
threshold. This need not be the case, unless a proper renormalization
scheme makes this short distance insensitivity manifest. Actually,
within such a scheme the counterterms include all unknown short
distance effects, but enter as free parameters which do not follow
from the potential and which must be fixed directly from NN scattering
data or deuteron properties. In the present work we have implemented a
boundary condition regularization and carried out the necessary
renormalization. This allows, within the OBE potential to keep only
$\pi$, $\sigma$, $\rho$ and $\omega$ mesons and neglect effectively
higher mass effects for the lowest central s-waves as well as the
deuteron wave function. In many regards we see improvements which come
with very natural choices of the couplings, and are compatible with
determinations from other sources. From this viewpoint, the leading
$N_c$ contribution to the OBE potential where $\pi$, $\sigma$, $\rho$
and $\omega$ mesons appear on equal footing, seems superior than the
leading chiral contribution which consists just on $\pi$.

The value of the $\sigma$ mass was fixed by a fit to the $^1S_0$ phase
shift yielding $m_\sigma = 501 (25) {\rm MeV}$. The values
obtained from the coupling constants reproducing the $^1S_0$ and
$^3S_1-^3D_1$ channels are very reasonable taking into account the
approximate nature of our calculation, $g_{\sigma NN}=9(1)$,
$g_{\omega NN}=9.5(5)$ and $f_{\rho NN}=16.3(7)$; the range is
compatible with the putative $10\%$ accuracy of the $1/N_c^2$
corrections. For the accepted value $g_{\rho NN} = 2.9(1)$ this yields
$g_{\omega NN} / g_{\rho NN}=3.27(17)$ a value in between the $SU(3)$
prediction $g_{\omega NN} / g_{\rho NN}|_{SU(3)}=3$ and the one from
the $e^+ e^- \to \rho$ and $e^+ e^- \to \omega$ decay ratios,
$g_{\omega NN} / g_{\rho NN}|_{e^+ e^-}=3.5$. We also get $f_{\rho NN}
/ g_{\rho NN}= \kappa_\rho = 5.6(3)$; a value in agreement from tensor
coupling studies. It is noteworthy that the repulsion triggered by the
$\omega$ meson is not as strong and important as required in the
conventional OBE approach where usually a strong violation of the
$SU(3)$ relation is observed as well. The reason is that, unlike the
traditional approach, the renormalization viewpoint stresses the
irrelevance of small distances. This is done by the introduction of
counterterms which are fixed by threshold scattering parameters at any
given short distance cut-off scale $r_c$. For the minimal de Broglie
wavelength probed in NN scattering below pion production threshold,
$1/p \sim 0.5 {\rm fm}$, a stable result is obtained generally when
$r_c =0.1-0.2 {\rm fm}$. Any mismatch to the observables can then be
attributed to missing physical effects. While the present calculations
are encouraging there is of course room for improvement.

One serious source of complications and limitations for
renormalization in general lies in its difficult marriage with the
variational principle~\cite{Feynman:1987hv}. The existence of two-body
spurious deeply bound states drives naturally the energy of the system
to its lowest energy state, if allowed to. On the other hand, one
should recognize that the existence of a minimum is tightly linked to
a subtle balance between kinetic and potential energy, which
undoubtely exists but may well take place beyond the applicability
range of the meson exchange picture requiring an artificial fine
tuning. This clearly influences the three, four, etc. body problems if
they would be treated in the standard and variational fashion but not
necessarily so if the few body problem is consistently
renormalized. Our results show that one has to choose between fine
tuning and renormalization. The standard approach has traditionally
been sensitive to short distance details and has required fine tuning
meson coupling constants, in particular those corresponding to vector
mesons, to unnatural values. In contrast, the renormalization approach
is free of fine tuning, and allows to fix meson constant from other
sources to their natural values.

While we have been using the leading large $N_c$ contributions to the
full OBE as a simplifying book-keeping reduction, we do not expect
that such an approximation becomes crucial regarding the main
conclusions on form factors. However, the most speculative prospective
of the present calculation lies in the possibility of promoting it to
a model independent large $N_c$ result. One should bear in mind,
however, that we have only kept leading $N_c$ OBE contributions. There
is, of course, the delicate question on {\it what} $2\pi$, $3\pi$ and
$\Delta$ contributions should be considered, firstly to avoid double
counting with the collective $\sigma$, $\rho$ and $\omega$ states, and
second to comply with the large $N_c$ requirements. To our knowledge,
the expectations of Ref.~\cite{Banerjee:2001js} of a large $N_c$
consistent multimeson exchange picture have not been explicitly
realized for the chiral potentials without~\cite{Kaiser:1997mw} and
with~\cite{Kaiser:1998wa} $\Delta$-isobar contributions as they do not
scale properly with $N_c$; one has $g_A \sim N_c$, $f_\pi \sim
\sqrt{N_c}$ and there are terms scaling as $ V_{\rm 2\pi}^{\rm ChPT}
\sim g_A^4 / f_\pi^4 \sim N_c^2 $ and not as $\sim N_c$ as found in
Refs.~\cite{Witten:1979kh,Kaplan:1996rk}.  Our results suggest a
scenario where the multimeson contributions invoked in
Ref.~\cite{Banerjee:2001js} would indeed be small, but this should be
checked explicitly. One further complication comes from the fact that
in the large $N_c$ limit the nucleon-delta splitting becomes small,
and in fact lighter than the pion mass. According to the Regge theory
formula $M_\Delta^2 - M_N^2 = m_\rho^2 -
m_\pi^2$~\cite{Ademollo:1969nx} and assuming the scaling $M_N = N_c
m_\rho /2$, the crossover between both mass parameters happens at
about $N_c \sim 6$. Actually, in the strict limit one should
consider not only $NN$ but at least also $N\Delta$ and $\Delta\Delta$
channels as well, as they become degenerate. The calculation of
\cite{Kaplan:1995yg,Kaplan:1996rk} only includes the restriction of
the baryon-baryon interaction to the $NN$ sector. In a more elaborate
treatment one should include the $\Delta$ as intermediate dynamical
states which in the elastic $NN$ region contribute as sub-threshold
effects~\cite{Savage:1996tb} which decouple for large $N\Delta$
splitting but which become degenerated when the $N\Delta$ splitting is
driven to zero.  In addition, it would also be interesting, still
within the OBE framework, to see what is the effect of the relative
$1/N_c^2$ corrections, which include in particular relativistic,
non-local, finite-size, spin-orbit, finite meson width corrections as
well as other mesons.

Finally, let us also note that besides the many improvements mentioned
above to the present calculation, the possibility of making a good
phenomenology while avoiding strong form factors in the NN potential
has further and important benefits.  In particular it makes the
discussion of gauge invariance much simpler, as we are effectively
dealing with local theories with no cut-off. Under this circumstance
the cumbersome gauging procedures involving path-dependent link
operators and which becomes necessary in order to minimally implement
gauge invariance would not be needed. In a recent
communication~\cite{Arriola:2009bg} we have evaluated electromagnetic
deuteron form factors in the impulse approximation and using the
renormalization scheme presented in this paper, with a reasonable
momentum transfer dependent behaviour up to about $q \sim 800 {\rm
MeV}$ and definitely improving over OPE.  Actually, these form factors
as well as some of the presently computed deuteron properties are
expected to have significant corrections from MEC. Let us remind that
MEC are a genuine consequence of the Meson Exchange picture in the NN
interaction, but in fairness also require constructing exact NN wave
functions from the corresponding Hamiltonian, as we have done
here. The present paper shows that renormalization for the OBE
potential is not only feasible as a previous and theoretically
appealing step to evaluate matrix elements of electroweak currents but
also and perhaps surprisingly yields a sound phenomenologically. It
also helps in reducing the impact of the hardly accesible short
distance region of the nucleon-nucleon interaction, thereby reducing
standard and much debated ambiguities. It remains to be seen if this
holds true also for low energy electroweak reactions where the meson
exchange picture is traditionally expected to work.

\begin{acknowledgments}

We thank M. Pav\'on Valderrama and D. R. Entem for many
discussions and a critical reading of the ms.

This work has been partially supported by the Spanish DGI and FEDER
funds with grant FIS2008-01143/FIS, Junta de Andaluc{\'\i}a grant
FQM225-05, and EU Integrated Infrastructure Initiative Hadron Physics
Project contract RII3-CT-2004-506078.

\end{acknowledgments}

\appendix

\section{Overview of coupling constants}
\label{sec:couplings}

A crucial point in the present framework corresponds to the choice of
coupling constants, $g_{\pi NN}$, $g_{\sigma NN}$, $f_{\rho NN}$ and
$g_{\omega NN}$ (for an older review see
e.g. Ref.~\cite{Dumbrajs:1983jd}) and masses, $m_\pi$, $m_\sigma$,
$m_\rho$ and $m_\omega$, entering the calculation. We review here
reasonable ranges on the basis of several sources, but bearing in mind
that we are keeping only the leading $N_c$ contributions to the OBE
potential. 

\begin{itemize} 
\item $g_{\pi NN}$. 
According to the Goldberger-Treiman relation (subjected to pion mass
corrections and/or higher meson states) the pion nucleon coupling
constant should be $g_{\pi NN} = g_A M_N /f_\pi = 12.8$ for the axial
coupling constant $g_A=1.26$.  A phase shift analysis of NN
scattering~\cite{deSwart:1997ep} yields $ g_{\pi
NN}=13.1083$. Nevertheless, the latest determinations from the
Goldberger-Miyazawa-Oehme (GMO) sum rule~\cite{Ericson:2000md} yields
the value $ g_{\pi NN} =13.3158$ this variation at the $5 \% $ level
dominates the uncertainties in the $1\pi$ exchange calculations.

\item $g_{\sigma NN}$. 
For the scalar coupling constant, the Goldberger-Treiman relation for
scalar mesons yields, $g_{\sigma NN} =M_N / f_\pi= 10.1$. However, if
we consider contributions from excited scalar mesons we may expect a a
somewhat different number. Actually, QCD sum rules
yield~\cite{Erkol:2006eq} $g_{\sigma NN}=14.4 \pm 3.7$ for the Ioffe
current nucleon interpolator and a smaller value $g_{\sigma NN}= 7 \pm
3 $ for more general interpolators~\cite{Aliev:2006xm}. A recent quark
model calculation yields $g_{\sigma NN} = 14.5 \pm
2$~\cite{FernandezCarames:2008en}.

\item $g_{\rho NN}$. 
The vector $g_{\rho NN}$ coupling constant is after Sakurai's
universality $g_{\rho NN} = g_{\rho \pi\pi} /2 $ while the
current-algebra KSFR relation provides $ g_{\rho \pi \pi} = m_\rho /(
\sqrt{2} f_\pi )$, yielding $g_{\rho NN}=2.9$. The $\rho NN$ Vertex in
Vector Dominance Models was also determined in the old
analysis~\cite{Hohler:1974ht} yielding yields $g_\rho = 2.9 (1)$ a
value confirmed in Ref.~\cite{Grein:1977id}. 

\item $f_{\rho NN}$. 
The tensor $f_{\rho NN}$ coupling is usually given by the ratio to the
vector coupling$ f_{\rho NN} = \kappa_\rho g_{\rho NN} $. In single
vector meson dominance models $ \kappa_\rho= \mu_p - \mu_n -1 $ with
$\mu_p=2.79$ and $\mu_n=-1.91$ the magnetic moments (in nuclear
magneton units $e/(2M_p)$) of proton and neutron respectively,
yielding $\kappa_\rho=3.7 $ and hence $f_{\rho NN} = 10.7(4)$ for
$g_{\rho NN}=2.9(1)$.

\item $g_{\omega NN}$. 
The relation $g_{\omega NN} = 3 g_{\rho NN}$ ($=8.7(3)$ for $g_{\rho
NN}=2.9(1)$) is the SU(3) prediction for the ideal $\omega-\phi$
mixing case corresponding to the OZI rule where $g_{\phi NN}=0$ as
well. Vector meson e.m. decays $\omega \to e^+ e^-$ and $\rho \to e^+ e^-$.
account for SU(3) breaking as $g_{\omega NN} = 3.5
g_{\rho NN}$ ($=10.2(4)$ for $g_{\rho NN}=2.9(1)$). 

\item $f_{\omega NN}$. 
The tensor $f_{\omega NN}$ coupling is also given by its the ratio to
the vector coupling$ f_{\omega NN} = \kappa_\omega g_{\omega NN} $. In
single vector meson dominance models $ \kappa_\omega= \mu_p + \mu_n -1
$ yielding $\kappa_\omega=-0.12$ and hence $f_{\omega NN} = -0.3(1)$ for
$g_{\omega NN}=3-3.5$.

\end{itemize}

Nucleon Electromagnetic Form factors with high energy QCD constraints
also provide information on vector meson
couplings. Ref.~\cite{Mergell:1995bf} yields $g_{\omega
NN}=20.86(25)$ and $f_{\omega NN}=-3.41(24)$ and $\kappa_\rho = 6.1
(2)$, and more recently~\cite{Belushkin:2006qa} it was found
$g_{\omega NN}=20(3)$ and $f_{\omega NN}=3(7)$. On the other hand, QCD
sum rules yield for the $\rho NN$ coupling a spread of values is
$g_{\rho NN}=2.4 \pm 0.6 $ and $f_{\rho NN}= 7.7 \pm
1.9$~\cite{Erkol:2006sa} and $g_{\rho NN}=3.2 \pm 0.9 $ and $f_{\rho
NN}+ g_{\rho NN} = 36.8 \pm 13.0$~\cite{Wang:2007yt}

Phase-shift analyzes of NN scattering below 160 MeV based on the
$\epsilon_1$ mixing angle were argued to be an indication for a strong
tensor force~\cite{Henneck:1993zz}, an issue further qualified in
Ref.~\cite{Brown:1994pq}. The strong tensor coupling is $\kappa_\rho =
f_{\rho NN} / g_{\rho NN}=6.1(6)$ and the weak is $\kappa_\rho = \mu_p
- 1 - \mu_n = 3.7$ corresponding to vector meson dominance saturated
with a single state. Note that the value $f_{\rho NN}=g_{\pi NN}=13.1$
for which the tensor force $1/r^3$ singularity disappears corresponds
to $\kappa_\rho=4.5(2)$ a value in between weak and strong.

\begin{table*}[ttt]
\caption{Deuteron properties for the exceptional case $f_{\rho NN}=
  g_{\pi NN}$ of non-singular large $N_c$ OBE potentials. In all cases
  we take $r_c=0.001{\rm fm}$. We compare renormalized vs. regular
  solutions for similar choices of parameters.  We use $ \gamma=
  \sqrt{ 2 \mu_{np} B_d} $ with $B_d=2.224575(9)$ and take $g_{\pi NN}
  =13.1083 $, $m_\pi=138.03 {\rm MeV} $, $m_\rho=m_\omega=782 {\rm
    MeV} $.  The fit to the $^1S_0$ phase shift gives $m_\sigma=501
  {\rm MeV}$ and $g_{\sigma NN}=9.1$. Experimental or recommended
  values can be traced from Ref.~\cite{deSwart:1995ui}.}
\begin{tabular}{|c|c|c|c|c|c|c|c|c|c|c|}
\hline & $g_{\omega NN}^*$ &  $ r_c \frac{u'(r_c)}{u(r_c)} $ & $ r_c \frac{w'(r_c)}{w(r_c)} $ 
& $\gamma ({\rm fm}^{-1})$ & $\eta$ & $A_S ( {\rm fm}^{-1/2}) $
& $r_m ({\rm fm})$ & $Q_d ( {\rm fm}^2) $ & $P_D $ & $\langle r^{-1}
\rangle $  \\ \hline
Renormalized  
& 0 & -0.1274 & 3& Input & 0.02567  & 0.8986 & 1.9949 &0.2830 & 5.87\% & 0.470  \\ 
Regular  
& 0 & 1 &  3& 0.6615 &  1.1502 & 0.0925 & 2.2523 &0.1215 & 10.77\% & 0.851  \\ 
Renorm.=Reg.  
&  3.74 & 1 & 3 & Input & 0.02567 & 0.8979 & 1.9935 & 0.2827 & 5.88\% & 0.491  \\ 
Renorm.  
&  2x3.74 & 0.0297 & 3 & Input   & 0.02569 & 0.8957 & 1.9890 & 0.2817 & 5.92\% &  0.517 \\ 
\hline 
\hline 
NijmII(\cite{Stoks:1994wp}) &- &-  & - & Input & 0.02521 & 0.8845(8) & 1.9675 & 0.2707 & 
5.635\% &  0.4502  \\
Reid93(\cite{Stoks:1994wp}) &- &- & - & Input & 0.02514 & 0.8845(8) & 1.9686 & 0.2703 & 
5.699\% & 0.4515  \\ \hline 
%
Exp. (\cite{deSwart:1995ui}) &-  &- & - &  0.231605 &  0.0256(4)  & 0.8846(9) & 1.9754(9)  & 0.2859(3) & 5.67(4)  & \\ \hline 
\end{tabular}
\label{tab:table_triplet_exceptional}
\end{table*}

\section{The exceptional non-singular case}
\label{sec:exceptional}

As already mentioned in Section~\ref{sec:OBE-largeNc} there is an
exceptional situation $f_{\rho NN}=g_{\pi NN}$ where the OBE potential
is not singular, Eq.~(\ref{eq:1/r^3}), and the use of form factors
would not be necessary. If we keep $g_{\pi NN}=13.1$ that means
$f_{\rho NN}=13.1$, a not completely unrealistic value lying in
between the single vector meson dominance estimate and the usual OBE
value (see Appendix~\ref{sec:couplings}), so it is worth analyzing
this case separately. Since the singularity affects mainly the coupled
spin triplet channel, one may wonder what would be the consequences
for the deuteron. We will show that our conclusions are not ruled out
by this exceptional case~\footnote{A compelling scenario where the
  singularity cancels might happen for an infinite tower of exchanged
  mesons fulfilling the sum rule $g_{\pi NN}^2 + g_{\pi' NN}^2 + \dots
  = f_{\rho NN}^2 + f_{\rho' NN}^2 + \dots = $. Even if this was the
  case the implications {\it after renormalization} are meager.}.

Note that within the renormalization approach this particular
situation has been scanned through in Fig.~\ref{fig:obs-frNN} where
nothing particularly noticeable happens. Actually, at short distances
we have a coupled channel Coulomb problem where the short distance
behaviour can generally be written as a linear admixture or regular
and irregular solutions,
\begin{eqnarray}
u(r) &\sim& a_1 r + a_2 \nonumber \\  
w(r) &\sim& b_1 r^3 + b_2 r^{-2}   
\end{eqnarray}
In order to get a normalizable wave function we {\it must} impose the
regular solution for the $D-$wave, meaning $b_2=0$.  The {\it
  renormalized} solution corresponds then to fix the deuteron binding
energy as explained in detail in Section~\ref{sec:triplet} and
integrate in with the result that the $S-$wave may have an admixture
of the irregular solution. The {\it regular } solution takes the value
$a_2=0$. The bound state properties are now {\it predicted} completely
from the potential. 

In practice we deal with arbitrarily small but finite cut-offs, $r_c
\to 0$.  In this situation it is simplest to use the superposition
principle of boundary conditions given by Eq.~(\ref{eq:sup_bound}) for
a {\it given} energy or $\gamma$. From the regularity condition of the
D-wave we get
\begin{eqnarray}
r_c \frac{w'(r_c)}{w(r_c)}=3 , \qquad ({\rm regular \quad D-wave})
\end{eqnarray}
which yields  the asymptotic $D/S$-ratio 
\begin{eqnarray}
\eta(r_c) = \frac{-3 w_S(r_c) + r_c w_S'(r_c)}{3 w_D(r_c) + r_c w_D'(r_c)}
\end{eqnarray}
This provides a relation between $\gamma$ and $\eta$. The renormalized
condition yields an arbitrary value of $u$ at the origin, so the
energy may be fixed arbitrarily, and thus 
\begin{eqnarray}
r_c \frac{u'(r_c)}{u(r_c)} \neq 1,  \qquad ({\rm irregular \quad S-wave})
\end{eqnarray}
The regular solution corresponds to
\begin{eqnarray}
r_c \frac{u'(r_c)}{u(r_c)}=1,  \qquad ({\rm regular \quad S-wave})
\end{eqnarray}
which in general will not be satisfied by the physical deuteron
binding energy. Thus, for the regular solution we will have either a
wrong value of the energy or the potential parameters must be
readjusted.  A value of $r_c=0.001 {\rm fm}$ proves more than enough.

\begin{figure*}[ttt]
\begin{center}
\includegraphics[height=.3\textheight,angle=270]{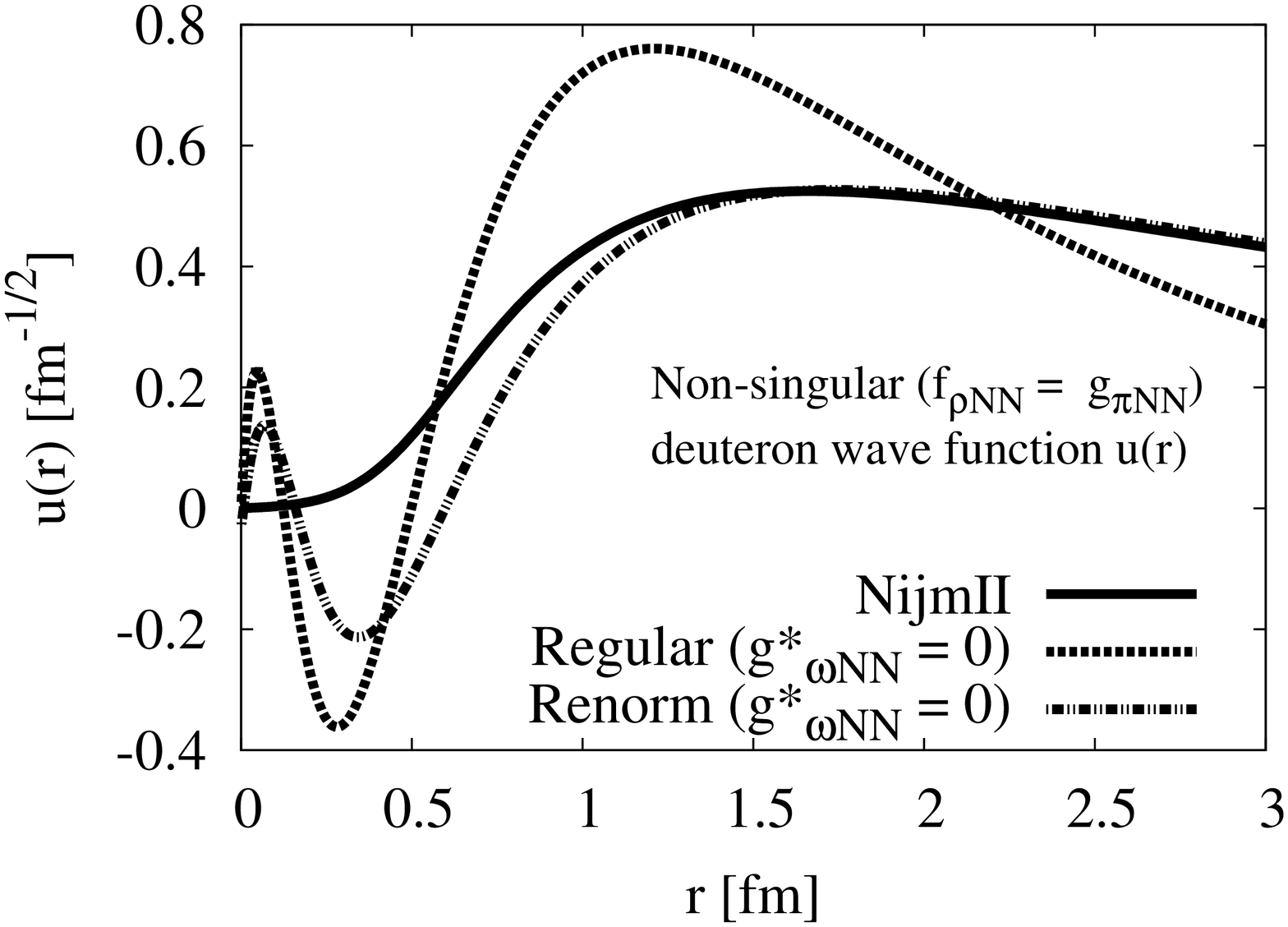}
\includegraphics[height=.3\textheight,angle=270]{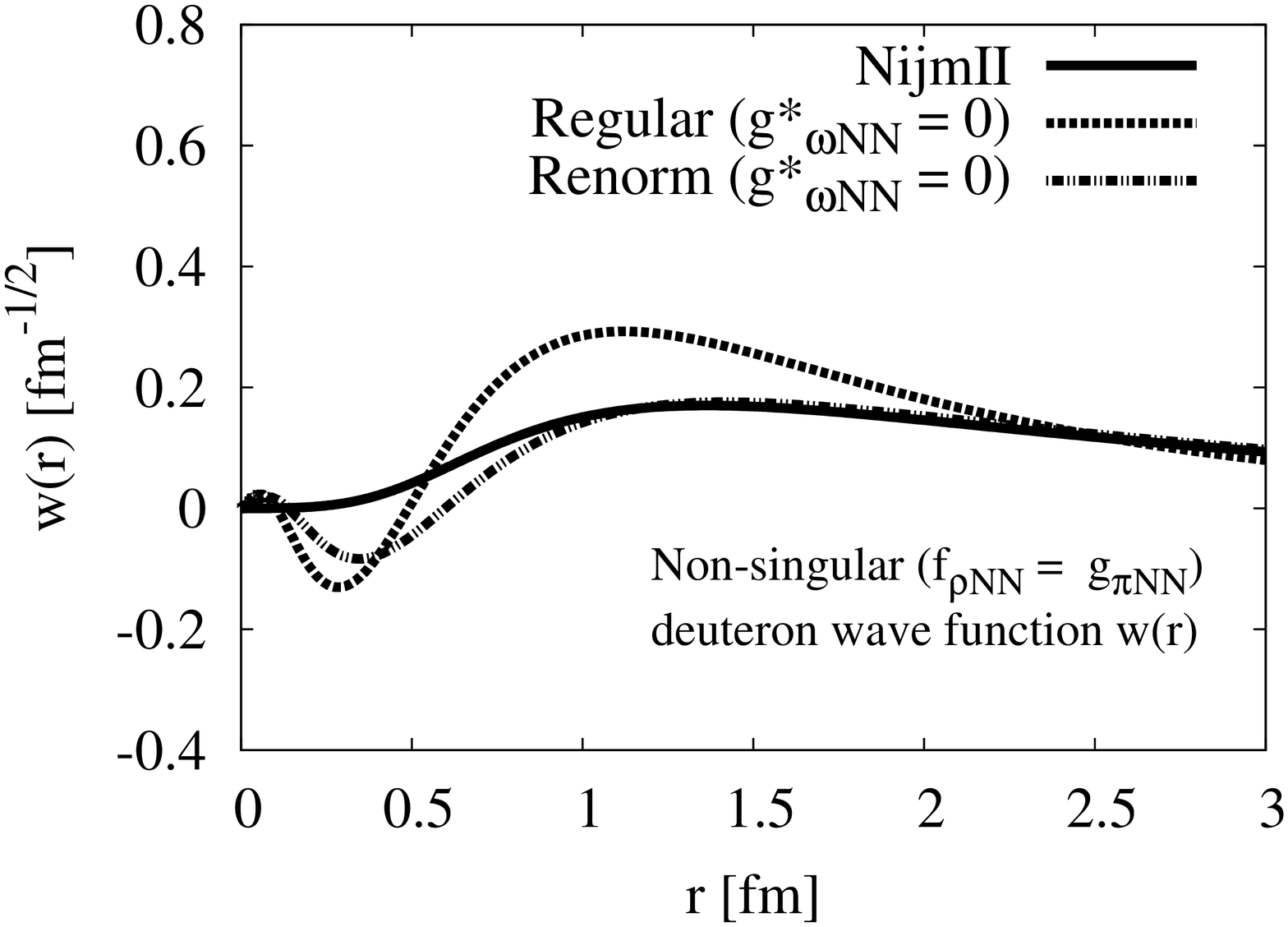}
\end{center}
\caption{Normalized Deuteron wave functions, u (left) and w (right),
  as a function of the distance (in {\rm fm}) in the OBE for the
  exceptional non-singular case $f_{\rho NN}= g_{\pi NN}$.  We show
  $\pi+\sigma+\rho+\omega$ both renormalized and the regular solution
  with the {\it same} parameters $g_{\omega NN}^*=0$. We compare to
  the Nijmegen II wave functions~\cite{Stoks:1994wp}(see
  table~\ref{tab:table_triplet_exceptional}).}
\label{fig:u+w_obe_exceptional}
\end{figure*}

Numerical results for a fixed parameter choice with $g_{\omega
  NN}^*=0$ are presented in
table~\ref{tab:table_triplet_exceptional}. As we see the regular
solution generates a bound state with $E_B \sim -16 {\rm MeV}$ which
is clearly off the deuteron with equally bad properties.  In order to
achieve the correct deuteron binding energy we just increase the
coupling to $g_{\omega NN}^*=3.75$ in the regular solution case.  In
this case both renormalized and regular solution would coincide {\it
  accidentally}. However, if we increase to twice this value
$g_{\omega NN}^*=2 \times 3.75$ we observe {\it tiny} changes in the
deuteron properties as compared to the $g_{\omega NN}^*=0$ case when
the renormalized solution is considered whereas the regular solution
becomes unbound. These results illustrate further the sharp
distinction between regular and renormalized solutions where one
chooses between fine tuning and short distance insensitivity
respectively. The corresponding wave functions to both the
renormalized and regular solutions with the {\it same} meson
parameters are depicted in Fig.~\ref{fig:u+w_obe_exceptional}. In both
cases inner nodes of the wave functions exihibit the existence of
deeply bound states, as dictated by the oscillation theorem.

Finally, we might try to analyze the consequences of taking
$V_{^3S_1}(r)= V_{^1S_0}(r)$ in the exceptional case $f_{\rho
  NN}=g_{\pi NN}=13.1$ and other parameters from the case with no form
factor, $\Gamma=1$, of table~\ref{tab:table_fits_reg} for
the $^1S_0$ channel. Let us remind that two possible scenarious arise
in such a case, one with no boud state and another one with a spurious
deeply bound state. For the $^3S_1-^3D_1$ channel, this complies to
the standard picture that the deuteron becomes bound due to the
additional binding introduced by the small tensor force mixing with
the $D-$wave, basically shifting the $S-$wave potential to an
effective one $V_{^3S_1}(r) \sim V_{^1S_0}(r)+ W_T(r)^2/V_{^3D_1}(r) $
. While in the case with no spurious bound state for the $^1S_0$ we
{\it do not} get any deuteron bound state, in the case with the
spurious bound state the binding energy is $E_{\rm B} \sim -50 {\rm
  MeV}$. This is another manifestation of the fine-tuning discussed at
length in Section~\ref{sec:fine-tining}.

In summary, although the $1/r^3$ singularity makes renormalization
process mandatory to implement the physical requirement of short
distance insensitivity, the important aspect here is that this
requirement remains equally valid even if there are no singularities
at all.


\begin{thebibliography}{90}
\expandafter\ifx\csname natexlab\endcsname\relax\def\natexlab#1{#1}\fi
\expandafter\ifx\csname bibnamefont\endcsname\relax
  \def\bibnamefont#1{#1}\fi
\expandafter\ifx\csname bibfnamefont\endcsname\relax
  \def\bibfnamefont#1{#1}\fi
\expandafter\ifx\csname citenamefont\endcsname\relax
  \def\citenamefont#1{#1}\fi
\expandafter\ifx\csname url\endcsname\relax
  \def\url#1{\texttt{#1}}\fi
\expandafter\ifx\csname urlprefix\endcsname\relax\def\urlprefix{URL }\fi
\providecommand{\bibinfo}[2]{#2}
\providecommand{\eprint}[2][]{\url{#2}}

\bibitem[{\citenamefont{Yukawa}(1935)}]{Yukawa:1935xg}
\bibinfo{author}{\bibfnamefont{H.}~\bibnamefont{Yukawa}},
  \bibinfo{journal}{Proc. Phys. Math. Soc. Jap.} \textbf{\bibinfo{volume}{17}},
  \bibinfo{pages}{48} (\bibinfo{year}{1935}).

\bibitem[{\citenamefont{Johnson and Teller}(1955)}]{PhysRev.98.783}
\bibinfo{author}{\bibfnamefont{M.~H.} \bibnamefont{Johnson}} \bibnamefont{and}
  \bibinfo{author}{\bibfnamefont{E.}~\bibnamefont{Teller}},
  \bibinfo{journal}{Phys. Rev.} \textbf{\bibinfo{volume}{98}},
  \bibinfo{pages}{783} (\bibinfo{year}{1955}).

\bibitem[{\citenamefont{Bryan and Scott}(1964)}]{Bryan:1964zz}
\bibinfo{author}{\bibfnamefont{R.~A.} \bibnamefont{Bryan}} \bibnamefont{and}
  \bibinfo{author}{\bibfnamefont{B.~L.} \bibnamefont{Scott}},
  \bibinfo{journal}{Phys. Rev.} \textbf{\bibinfo{volume}{135}},
  \bibinfo{pages}{B434} (\bibinfo{year}{1964}).

\bibitem[{\citenamefont{Bryan and Scott}(1967)}]{Bryan:1967zz}
\bibinfo{author}{\bibfnamefont{R.~A.} \bibnamefont{Bryan}} \bibnamefont{and}
  \bibinfo{author}{\bibfnamefont{B.~L.} \bibnamefont{Scott}},
  \bibinfo{journal}{Phys. Rev.} \textbf{\bibinfo{volume}{164}},
  \bibinfo{pages}{1215} (\bibinfo{year}{1967}).

\bibitem[{\citenamefont{Bryan and Scott}(1969)}]{Bryan:1969mp}
\bibinfo{author}{\bibfnamefont{R.}~\bibnamefont{Bryan}} \bibnamefont{and}
  \bibinfo{author}{\bibfnamefont{B.~L.} \bibnamefont{Scott}},
  \bibinfo{journal}{Phys. Rev.} \textbf{\bibinfo{volume}{177}},
  \bibinfo{pages}{1435} (\bibinfo{year}{1969}).

\bibitem[{\citenamefont{Partovi and Lomon}(1970)}]{Partovi:1969wd}
\bibinfo{author}{\bibfnamefont{M.~H.} \bibnamefont{Partovi}} \bibnamefont{and}
  \bibinfo{author}{\bibfnamefont{E.~L.} \bibnamefont{Lomon}},
  \bibinfo{journal}{Phys. Rev.} \textbf{\bibinfo{volume}{D2}},
  \bibinfo{pages}{1999} (\bibinfo{year}{1970}).

\bibitem[{\citenamefont{Nagels et~al.}(1978)\citenamefont{Nagels, Rijken, and
  de~Swart}}]{Nagels:1977ze}
\bibinfo{author}{\bibfnamefont{M.~M.} \bibnamefont{Nagels}},
  \bibinfo{author}{\bibfnamefont{T.~A.} \bibnamefont{Rijken}},
  \bibnamefont{and} \bibinfo{author}{\bibfnamefont{J.~J.}
  \bibnamefont{de~Swart}}, \bibinfo{journal}{Phys. Rev.}
  \textbf{\bibinfo{volume}{D17}}, \bibinfo{pages}{768} (\bibinfo{year}{1978}).

\bibitem[{\citenamefont{Ueda and Green}(1968)}]{Ueda:1969er}
\bibinfo{author}{\bibfnamefont{T.}~\bibnamefont{Ueda}} \bibnamefont{and}
  \bibinfo{author}{\bibfnamefont{A.~E.~S.} \bibnamefont{Green}},
  \bibinfo{journal}{Phys. Rev.} \textbf{\bibinfo{volume}{174}},
  \bibinfo{pages}{1304} (\bibinfo{year}{1968}).

\bibitem[{\citenamefont{Erkelenz}(1974)}]{Erkelenz:1974uj}
\bibinfo{author}{\bibfnamefont{K.}~\bibnamefont{Erkelenz}},
  \bibinfo{journal}{Phys. Rept.} \textbf{\bibinfo{volume}{13}},
  \bibinfo{pages}{191} (\bibinfo{year}{1974}).

\bibitem[{\citenamefont{Machleidt et~al.}(1987)\citenamefont{Machleidt,
  Holinde, and Elster}}]{Machleidt:1987hj}
\bibinfo{author}{\bibfnamefont{R.}~\bibnamefont{Machleidt}},
  \bibinfo{author}{\bibfnamefont{K.}~\bibnamefont{Holinde}}, \bibnamefont{and}
  \bibinfo{author}{\bibfnamefont{C.}~\bibnamefont{Elster}},
  \bibinfo{journal}{Phys. Rept.} \textbf{\bibinfo{volume}{149}},
  \bibinfo{pages}{1} (\bibinfo{year}{1987}).

\bibitem[{\citenamefont{Machleidt}(1989)}]{Machleidt:1989tm}
\bibinfo{author}{\bibfnamefont{R.}~\bibnamefont{Machleidt}},
  \bibinfo{journal}{Adv. Nucl. Phys.} \textbf{\bibinfo{volume}{19}},
  \bibinfo{pages}{189} (\bibinfo{year}{1989}).

\bibitem[{\citenamefont{Machleidt}(2007)}]{Machleidt:2007ms}
\bibinfo{author}{\bibfnamefont{R.}~\bibnamefont{Machleidt}}
  (\bibinfo{year}{2007}), \eprint{0704.0807}.

\bibitem[{\citenamefont{Machleidt}(2001)}]{Machleidt:2000ge}
\bibinfo{author}{\bibfnamefont{R.}~\bibnamefont{Machleidt}},
  \bibinfo{journal}{Phys. Rev.} \textbf{\bibinfo{volume}{C63}},
  \bibinfo{pages}{024001} (\bibinfo{year}{2001}), \eprint{nucl-th/0006014}.

\bibitem[{\citenamefont{Stoks et~al.}(1993)\citenamefont{Stoks, Kompl,
  Rentmeester, and de~Swart}}]{Stoks:1993tb}
\bibinfo{author}{\bibfnamefont{V.~G.~J.} \bibnamefont{Stoks}},
  \bibinfo{author}{\bibfnamefont{R.~A.~M.} \bibnamefont{Kompl}},
  \bibinfo{author}{\bibfnamefont{M.~C.~M.} \bibnamefont{Rentmeester}},
  \bibnamefont{and} \bibinfo{author}{\bibfnamefont{J.~J.}
  \bibnamefont{de~Swart}}, \bibinfo{journal}{Phys. Rev.}
  \textbf{\bibinfo{volume}{C48}}, \bibinfo{pages}{792} (\bibinfo{year}{1993}).

\bibitem[{\citenamefont{Stoks and Rijken}(1997)}]{Stoks:1996yj}
\bibinfo{author}{\bibfnamefont{V.~G.~J.} \bibnamefont{Stoks}} \bibnamefont{and}
  \bibinfo{author}{\bibfnamefont{T.~A.} \bibnamefont{Rijken}},
  \bibinfo{journal}{Nucl. Phys.} \textbf{\bibinfo{volume}{A613}},
  \bibinfo{pages}{311} (\bibinfo{year}{1997}), \eprint{nucl-th/9611002}.

\bibitem[{\citenamefont{Furnstahl et~al.}(1997)\citenamefont{Furnstahl, Serot,
  and Tang}}]{Furnstahl:1996wv}
\bibinfo{author}{\bibfnamefont{R.~J.} \bibnamefont{Furnstahl}},
  \bibinfo{author}{\bibfnamefont{B.~D.} \bibnamefont{Serot}}, \bibnamefont{and}
  \bibinfo{author}{\bibfnamefont{H.-B.} \bibnamefont{Tang}},
  \bibinfo{journal}{Nucl. Phys.} \textbf{\bibinfo{volume}{A615}},
  \bibinfo{pages}{441} (\bibinfo{year}{1997}), \eprint{nucl-th/9608035}.

\bibitem[{\citenamefont{Papazoglou et~al.}(1999)}]{Papazoglou:1998vr}
\bibinfo{author}{\bibfnamefont{P.}~\bibnamefont{Papazoglou}}
  \bibnamefont{et~al.}, \bibinfo{journal}{Phys. Rev.}
  \textbf{\bibinfo{volume}{C59}}, \bibinfo{pages}{411} (\bibinfo{year}{1999}),
  \eprint{nucl-th/9806087}.

\bibitem[{\citenamefont{Janssen et~al.}(1994)\citenamefont{Janssen, Holinde,
  and Speth}}]{Janssen:1994kh}
\bibinfo{author}{\bibfnamefont{G.}~\bibnamefont{Janssen}},
  \bibinfo{author}{\bibfnamefont{K.}~\bibnamefont{Holinde}}, \bibnamefont{and}
  \bibinfo{author}{\bibfnamefont{J.}~\bibnamefont{Speth}},
  \bibinfo{journal}{Phys. Rev. Lett.} \textbf{\bibinfo{volume}{73}},
  \bibinfo{pages}{1332} (\bibinfo{year}{1994}).

\bibitem[{\citenamefont{Janssen et~al.}(1996)\citenamefont{Janssen, Holinde,
  and Speth}}]{Janssen:1996kx}
\bibinfo{author}{\bibfnamefont{G.}~\bibnamefont{Janssen}},
  \bibinfo{author}{\bibfnamefont{K.}~\bibnamefont{Holinde}}, \bibnamefont{and}
  \bibinfo{author}{\bibfnamefont{J.}~\bibnamefont{Speth}},
  \bibinfo{journal}{Phys. Rev.} \textbf{\bibinfo{volume}{C54}},
  \bibinfo{pages}{2218} (\bibinfo{year}{1996}).

\bibitem[{\citenamefont{Case}(1950)}]{Case:1950an}
\bibinfo{author}{\bibfnamefont{K.~M.} \bibnamefont{Case}},
  \bibinfo{journal}{Phys. Rev.} \textbf{\bibinfo{volume}{80}},
  \bibinfo{pages}{797} (\bibinfo{year}{1950}).

\bibitem[{\citenamefont{Frank et~al.}(1971)\citenamefont{Frank, Land, and
  Spector}}]{Frank:1971xx}
\bibinfo{author}{\bibfnamefont{W.}~\bibnamefont{Frank}},
  \bibinfo{author}{\bibfnamefont{D.~J.} \bibnamefont{Land}}, \bibnamefont{and}
  \bibinfo{author}{\bibfnamefont{R.~M.} \bibnamefont{Spector}},
  \bibinfo{journal}{Rev. Mod. Phys.} \textbf{\bibinfo{volume}{43}},
  \bibinfo{pages}{36} (\bibinfo{year}{1971}).

\bibitem[{\citenamefont{{Woloshyn} and {Jackson}}(1972)}]{1972NuPhA.185..131W}
\bibinfo{author}{\bibfnamefont{R.~M.} \bibnamefont{{Woloshyn}}}
  \bibnamefont{and} \bibinfo{author}{\bibfnamefont{A.~D.}
  \bibnamefont{{Jackson}}}, \bibinfo{journal}{Nuclear Physics A}
  \textbf{\bibinfo{volume}{185}}, \bibinfo{pages}{131} (\bibinfo{year}{1972}).

\bibitem[{\citenamefont{Gari and Kaulfuss}(1984)}]{Gari:1984pq}
\bibinfo{author}{\bibfnamefont{M.}~\bibnamefont{Gari}} \bibnamefont{and}
  \bibinfo{author}{\bibfnamefont{U.}~\bibnamefont{Kaulfuss}},
  \bibinfo{journal}{Phys. Lett.} \textbf{\bibinfo{volume}{B136}},
  \bibinfo{pages}{139} (\bibinfo{year}{1984}).

\bibitem[{\citenamefont{Kaulfuss and Gari}(1983)}]{Kaulfuss:1984tw}
\bibinfo{author}{\bibfnamefont{U.}~\bibnamefont{Kaulfuss}} \bibnamefont{and}
  \bibinfo{author}{\bibfnamefont{M.}~\bibnamefont{Gari}},
  \bibinfo{journal}{Nucl. Phys.} \textbf{\bibinfo{volume}{A408}},
  \bibinfo{pages}{507} (\bibinfo{year}{1983}).

\bibitem[{\citenamefont{Flender and Gari}(1995)}]{Flender:1994uh}
\bibinfo{author}{\bibfnamefont{J.}~\bibnamefont{Flender}} \bibnamefont{and}
  \bibinfo{author}{\bibfnamefont{M.~F.} \bibnamefont{Gari}},
  \bibinfo{journal}{Phys. Rev.} \textbf{\bibinfo{volume}{C51}},
  \bibinfo{pages}{1619} (\bibinfo{year}{1995}).

\bibitem[{\citenamefont{Schutz et~al.}(1996)\citenamefont{Schutz, Haidenbauer,
  and Holinde}}]{Schutz:1995dj}
\bibinfo{author}{\bibfnamefont{C.}~\bibnamefont{Schutz}},
  \bibinfo{author}{\bibfnamefont{J.}~\bibnamefont{Haidenbauer}},
  \bibnamefont{and} \bibinfo{author}{\bibfnamefont{K.}~\bibnamefont{Holinde}},
  \bibinfo{journal}{Phys. Rev.} \textbf{\bibinfo{volume}{C54}},
  \bibinfo{pages}{1561} (\bibinfo{year}{1996}), \eprint{nucl-th/9508021}.

\bibitem[{\citenamefont{Bockmann et~al.}(1999)\citenamefont{Bockmann, Hanhart,
  Krehl, Krewald, and Speth}}]{Bockmann:1999nu}
\bibinfo{author}{\bibfnamefont{R.}~\bibnamefont{Bockmann}},
  \bibinfo{author}{\bibfnamefont{C.}~\bibnamefont{Hanhart}},
  \bibinfo{author}{\bibfnamefont{O.}~\bibnamefont{Krehl}},
  \bibinfo{author}{\bibfnamefont{S.}~\bibnamefont{Krewald}}, \bibnamefont{and}
  \bibinfo{author}{\bibfnamefont{J.}~\bibnamefont{Speth}},
  \bibinfo{journal}{Phys. Rev.} \textbf{\bibinfo{volume}{C60}},
  \bibinfo{pages}{055212} (\bibinfo{year}{1999}), \eprint{nucl-th/9905043}.

\bibitem[{\citenamefont{Bryan et~al.}(1980)\citenamefont{Bryan, Dominguez, and
  VerWest}}]{Bryan:1979ve}
\bibinfo{author}{\bibfnamefont{R.~A.} \bibnamefont{Bryan}},
  \bibinfo{author}{\bibfnamefont{C.~A.} \bibnamefont{Dominguez}},
  \bibnamefont{and} \bibinfo{author}{\bibfnamefont{B.~J.}
  \bibnamefont{VerWest}}, \bibinfo{journal}{Phys. Rev.}
  \textbf{\bibinfo{volume}{C22}}, \bibinfo{pages}{160} (\bibinfo{year}{1980}).

\bibitem[{\citenamefont{Cohen}(1986)}]{Cohen:1986ux}
\bibinfo{author}{\bibfnamefont{T.~D.} \bibnamefont{Cohen}},
  \bibinfo{journal}{Phys. Rev.} \textbf{\bibinfo{volume}{D34}},
  \bibinfo{pages}{2187} (\bibinfo{year}{1986}).

\bibitem[{\citenamefont{Holzwarth and Machleidt}(1997)}]{Holzwarth:1996bc}
\bibinfo{author}{\bibfnamefont{G.}~\bibnamefont{Holzwarth}} \bibnamefont{and}
  \bibinfo{author}{\bibfnamefont{R.}~\bibnamefont{Machleidt}},
  \bibinfo{journal}{Phys. Rev.} \textbf{\bibinfo{volume}{C55}},
  \bibinfo{pages}{1088} (\bibinfo{year}{1997}), \eprint{nucl-th/9610041}.

\bibitem[{\citenamefont{Alberto et~al.}(1990)}]{Alberto:1990ru}
\bibinfo{author}{\bibfnamefont{P.}~\bibnamefont{Alberto}} \bibnamefont{et~al.},
  \bibinfo{journal}{Z. Phys.} \textbf{\bibinfo{volume}{A336}},
  \bibinfo{pages}{449} (\bibinfo{year}{1990}).

\bibitem[{\citenamefont{Christov et~al.}(1996)}]{Christov:1995vm}
\bibinfo{author}{\bibfnamefont{C.~V.} \bibnamefont{Christov}}
  \bibnamefont{et~al.}, \bibinfo{journal}{Prog. Part. Nucl. Phys.}
  \textbf{\bibinfo{volume}{37}}, \bibinfo{pages}{91} (\bibinfo{year}{1996}),
  \eprint{hep-ph/9604441}.

\bibitem[{\citenamefont{Meissner}(1995)}]{Meissner:1995ra}
\bibinfo{author}{\bibfnamefont{T.}~\bibnamefont{Meissner}},
  \bibinfo{journal}{Phys. Rev.} \textbf{\bibinfo{volume}{C52}},
  \bibinfo{pages}{3386} (\bibinfo{year}{1995}), \eprint{nucl-th/9506030}.

\bibitem[{\citenamefont{Coon and Scadron}(1990)}]{Coon:1990fh}
\bibinfo{author}{\bibfnamefont{S.~A.} \bibnamefont{Coon}} \bibnamefont{and}
  \bibinfo{author}{\bibfnamefont{M.~D.} \bibnamefont{Scadron}},
  \bibinfo{journal}{Phys. Rev.} \textbf{\bibinfo{volume}{C42}},
  \bibinfo{pages}{2256} (\bibinfo{year}{1990}).

\bibitem[{\citenamefont{Liu et~al.}(1995)\citenamefont{Liu, Dong, Draper, and
  Wilcox}}]{Liu:1994dr}
\bibinfo{author}{\bibfnamefont{K.~F.} \bibnamefont{Liu}},
  \bibinfo{author}{\bibfnamefont{S.~J.} \bibnamefont{Dong}},
  \bibinfo{author}{\bibfnamefont{T.}~\bibnamefont{Draper}}, \bibnamefont{and}
  \bibinfo{author}{\bibfnamefont{W.}~\bibnamefont{Wilcox}},
  \bibinfo{journal}{Phys. Rev. Lett.} \textbf{\bibinfo{volume}{74}},
  \bibinfo{pages}{2172} (\bibinfo{year}{1995}), \eprint{hep-lat/9406007}.

\bibitem[{\citenamefont{Alexandrou et~al.}(2007)\citenamefont{Alexandrou,
  Koutsou, Leontiou, Negele, and Tsapalis}}]{Alexandrou:2007zz}
\bibinfo{author}{\bibfnamefont{C.}~\bibnamefont{Alexandrou}},
  \bibinfo{author}{\bibfnamefont{G.}~\bibnamefont{Koutsou}},
  \bibinfo{author}{\bibfnamefont{T.}~\bibnamefont{Leontiou}},
  \bibinfo{author}{\bibfnamefont{J.~W.} \bibnamefont{Negele}},
  \bibnamefont{and} \bibinfo{author}{\bibfnamefont{A.}~\bibnamefont{Tsapalis}},
  \bibinfo{journal}{Phys. Rev.} \textbf{\bibinfo{volume}{D76}},
  \bibinfo{pages}{094511} (\bibinfo{year}{2007}).

\bibitem[{\citenamefont{Melde et~al.}(2009)\citenamefont{Melde, Canton, and
  Plessas}}]{Melde:2008dg}
\bibinfo{author}{\bibfnamefont{T.}~\bibnamefont{Melde}},
  \bibinfo{author}{\bibfnamefont{L.}~\bibnamefont{Canton}}, \bibnamefont{and}
  \bibinfo{author}{\bibfnamefont{W.}~\bibnamefont{Plessas}},
  \bibinfo{journal}{Phys. Rev. Lett.} \textbf{\bibinfo{volume}{102}},
  \bibinfo{pages}{132002} (\bibinfo{year}{2009}), \eprint{0811.0277}.

\bibitem[{\citenamefont{Janssen et~al.}(1993)\citenamefont{Janssen, Durso,
  Holinde, Pearce, and Speth}}]{Janssen:1993nj}
\bibinfo{author}{\bibfnamefont{G.}~\bibnamefont{Janssen}},
  \bibinfo{author}{\bibfnamefont{J.~W.} \bibnamefont{Durso}},
  \bibinfo{author}{\bibfnamefont{K.}~\bibnamefont{Holinde}},
  \bibinfo{author}{\bibfnamefont{B.~C.} \bibnamefont{Pearce}},
  \bibnamefont{and} \bibinfo{author}{\bibfnamefont{J.}~\bibnamefont{Speth}},
  \bibinfo{journal}{Phys. Rev. Lett.} \textbf{\bibinfo{volume}{71}},
  \bibinfo{pages}{1978} (\bibinfo{year}{1993}).

\bibitem[{\citenamefont{Holinde and Thomas}(1990)}]{Holinde:1990fe}
\bibinfo{author}{\bibfnamefont{K.}~\bibnamefont{Holinde}} \bibnamefont{and}
  \bibinfo{author}{\bibfnamefont{A.~W.} \bibnamefont{Thomas}},
  \bibinfo{journal}{Phys. Rev.} \textbf{\bibinfo{volume}{C42}},
  \bibinfo{pages}{1195} (\bibinfo{year}{1990}).

\bibitem[{\citenamefont{Haidenbauer et~al.}(1994)\citenamefont{Haidenbauer,
  Holinde, and Thomas}}]{Haidenbauer:1994zz}
\bibinfo{author}{\bibfnamefont{J.}~\bibnamefont{Haidenbauer}},
  \bibinfo{author}{\bibfnamefont{K.}~\bibnamefont{Holinde}}, \bibnamefont{and}
  \bibinfo{author}{\bibfnamefont{A.~W.} \bibnamefont{Thomas}},
  \bibinfo{journal}{Phys. Rev.} \textbf{\bibinfo{volume}{C49}},
  \bibinfo{pages}{2331} (\bibinfo{year}{1994}).

\bibitem[{\citenamefont{Ueda}(1992)}]{Ueda:1991ca}
\bibinfo{author}{\bibfnamefont{T.}~\bibnamefont{Ueda}}, \bibinfo{journal}{Phys.
  Rev. Lett.} \textbf{\bibinfo{volume}{68}}, \bibinfo{pages}{142}
  (\bibinfo{year}{1992}).

\bibitem[{\citenamefont{Ericson and Weise}(1988)}]{Ericson:1988gk}
\bibinfo{author}{\bibfnamefont{T.~E.~O.} \bibnamefont{Ericson}}
  \bibnamefont{and} \bibinfo{author}{\bibfnamefont{W.}~\bibnamefont{Weise}},
  \emph{\bibinfo{title}{Pions and Nuclei}} (\bibinfo{publisher}{Oxford, UK:
  Clarendon (1988)}, \bibinfo{year}{1988}).

\bibitem[{\citenamefont{Riska}(1989)}]{Riska:1989bh}
\bibinfo{author}{\bibfnamefont{D.~O.} \bibnamefont{Riska}},
  \bibinfo{journal}{Phys. Rept.} \textbf{\bibinfo{volume}{181}},
  \bibinfo{pages}{207} (\bibinfo{year}{1989}).

\bibitem[{\citenamefont{Pavon~Valderrama and
  Ruiz~Arriola}(2005)}]{PavonValderrama:2005gu}
\bibinfo{author}{\bibfnamefont{M.}~\bibnamefont{Pavon~Valderrama}}
  \bibnamefont{and}
  \bibinfo{author}{\bibfnamefont{E.}~\bibnamefont{Ruiz~Arriola}},
  \bibinfo{journal}{Phys. Rev.} \textbf{\bibinfo{volume}{C72}},
  \bibinfo{pages}{054002} (\bibinfo{year}{2005}), \eprint{nucl-th/0504067}.

\bibitem[{\citenamefont{Pavon~Valderrama and
  Ruiz~Arriola}(2006{\natexlab{a}})}]{Valderrama:2005wv}
\bibinfo{author}{\bibfnamefont{M.}~\bibnamefont{Pavon~Valderrama}}
  \bibnamefont{and}
  \bibinfo{author}{\bibfnamefont{E.}~\bibnamefont{Ruiz~Arriola}},
  \bibinfo{journal}{Phys. Rev.} \textbf{\bibinfo{volume}{C74}},
  \bibinfo{pages}{054001} (\bibinfo{year}{2006}{\natexlab{a}}),
  \eprint{nucl-th/0506047}.

\bibitem[{\citenamefont{Pavon~Valderrama and
  Ruiz~Arriola}(2006{\natexlab{b}})}]{PavonValderrama:2005uj}
\bibinfo{author}{\bibfnamefont{M.}~\bibnamefont{Pavon~Valderrama}}
  \bibnamefont{and}
  \bibinfo{author}{\bibfnamefont{E.}~\bibnamefont{Ruiz~Arriola}},
  \bibinfo{journal}{Phys. Rev.} \textbf{\bibinfo{volume}{C74}},
  \bibinfo{pages}{064004} (\bibinfo{year}{2006}{\natexlab{b}}),
  \eprint{nucl-th/0507075}.

\bibitem[{\citenamefont{Pavon~Valderrama and
  Ruiz~Arriola}(2004)}]{PavonValderrama:2004td}
\bibinfo{author}{\bibfnamefont{M.}~\bibnamefont{Pavon~Valderrama}}
  \bibnamefont{and}
  \bibinfo{author}{\bibfnamefont{E.}~\bibnamefont{Ruiz~Arriola}}
  (\bibinfo{year}{2004}), \eprint{nucl-th/0410020}.

\bibitem[{\citenamefont{Entem et~al.}(2008)\citenamefont{Entem, Ruiz~Arriola,
  Pavon~Valderrama, and Machleidt}}]{Entem:2007jg}
\bibinfo{author}{\bibfnamefont{D.~R.} \bibnamefont{Entem}},
  \bibinfo{author}{\bibfnamefont{E.}~\bibnamefont{Ruiz~Arriola}},
  \bibinfo{author}{\bibfnamefont{M.}~\bibnamefont{Pavon~Valderrama}},
  \bibnamefont{and}
  \bibinfo{author}{\bibfnamefont{R.}~\bibnamefont{Machleidt}},
  \bibinfo{journal}{Phys. Rev.} \textbf{\bibinfo{volume}{C77}},
  \bibinfo{pages}{044006} (\bibinfo{year}{2008}), \eprint{0709.2770}.

\bibitem[{\citenamefont{Witten}(1979)}]{Witten:1979kh}
\bibinfo{author}{\bibfnamefont{E.}~\bibnamefont{Witten}},
  \bibinfo{journal}{Nucl. Phys.} \textbf{\bibinfo{volume}{B160}},
  \bibinfo{pages}{57} (\bibinfo{year}{1979}).

\bibitem[{\citenamefont{Manohar}(1998)}]{Manohar:1998xv}
\bibinfo{author}{\bibfnamefont{A.~V.} \bibnamefont{Manohar}}
  (\bibinfo{year}{1998}), \eprint{hep-ph/9802419}.

\bibitem[{\citenamefont{Jenkins}(1998)}]{Jenkins:1998wy}
\bibinfo{author}{\bibfnamefont{E.~E.} \bibnamefont{Jenkins}},
  \bibinfo{journal}{Ann. Rev. Nucl. Part. Sci.} \textbf{\bibinfo{volume}{48}},
  \bibinfo{pages}{81} (\bibinfo{year}{1998}), \eprint{hep-ph/9803349}.

\bibitem[{\citenamefont{Kaplan and Manohar}(1997)}]{Kaplan:1996rk}
\bibinfo{author}{\bibfnamefont{D.~B.} \bibnamefont{Kaplan}} \bibnamefont{and}
  \bibinfo{author}{\bibfnamefont{A.~V.} \bibnamefont{Manohar}},
  \bibinfo{journal}{Phys. Rev.} \textbf{\bibinfo{volume}{C56}},
  \bibinfo{pages}{76} (\bibinfo{year}{1997}), \eprint{nucl-th/9612021}.

\bibitem[{\citenamefont{Banerjee et~al.}(2002)\citenamefont{Banerjee, Cohen,
  and Gelman}}]{Banerjee:2001js}
\bibinfo{author}{\bibfnamefont{M.~K.} \bibnamefont{Banerjee}},
  \bibinfo{author}{\bibfnamefont{T.~D.} \bibnamefont{Cohen}}, \bibnamefont{and}
  \bibinfo{author}{\bibfnamefont{B.~A.} \bibnamefont{Gelman}},
  \bibinfo{journal}{Phys. Rev.} \textbf{\bibinfo{volume}{C65}},
  \bibinfo{pages}{034011} (\bibinfo{year}{2002}), \eprint{hep-ph/0109274}.

\bibitem[{\citenamefont{Calle~Cordon and
  Ruiz~Arriola}(2008{\natexlab{a}})}]{CalleCordon:2008eu}
\bibinfo{author}{\bibfnamefont{A.}~\bibnamefont{Calle~Cordon}}
  \bibnamefont{and}
  \bibinfo{author}{\bibfnamefont{E.}~\bibnamefont{Ruiz~Arriola}},
  \bibinfo{journal}{AIP Conf. Proc.} \textbf{\bibinfo{volume}{1030}},
  \bibinfo{pages}{334} (\bibinfo{year}{2008}{\natexlab{a}}),
  \eprint{0804.2350}.

\bibitem[{\citenamefont{Calle~Cordon and
  Ruiz~Arriola}(2008{\natexlab{b}})}]{CalleCordon:2008cz}
\bibinfo{author}{\bibfnamefont{A.}~\bibnamefont{Calle~Cordon}}
  \bibnamefont{and}
  \bibinfo{author}{\bibfnamefont{E.}~\bibnamefont{Ruiz~Arriola}},
  \bibinfo{journal}{Phys. Rev.} \textbf{\bibinfo{volume}{C78}},
  \bibinfo{pages}{054002} (\bibinfo{year}{2008}{\natexlab{b}}),
  \eprint{0807.2918}.

\bibitem[{\citenamefont{Cordon and Arriola}(2009)}]{Cordon:2009ps}
\bibinfo{author}{\bibfnamefont{A.~C.} \bibnamefont{Cordon}} \bibnamefont{and}
  \bibinfo{author}{\bibfnamefont{E.~R.} \bibnamefont{Arriola}}
  (\bibinfo{year}{2009}), \eprint{0904.0421}.

\bibitem[{\citenamefont{Arriola and Cordon}(2009)}]{Arriola:2009bg}
\bibinfo{author}{\bibfnamefont{E.~R.} \bibnamefont{Arriola}} \bibnamefont{and}
  \bibinfo{author}{\bibfnamefont{A.~C.} \bibnamefont{Cordon}}
  (\bibinfo{year}{2009}), \eprint{0904.4132}.

\bibitem[{\citenamefont{Belitsky and Cohen}(2002)}]{Belitsky:2002ni}
\bibinfo{author}{\bibfnamefont{A.~V.} \bibnamefont{Belitsky}} \bibnamefont{and}
  \bibinfo{author}{\bibfnamefont{T.~D.} \bibnamefont{Cohen}},
  \bibinfo{journal}{Phys. Rev.} \textbf{\bibinfo{volume}{C65}},
  \bibinfo{pages}{064008} (\bibinfo{year}{2002}), \eprint{hep-ph/0202153}.

\bibitem[{\citenamefont{Cohen}(2002)}]{Cohen:2002im}
\bibinfo{author}{\bibfnamefont{T.~D.} \bibnamefont{Cohen}},
  \bibinfo{journal}{Phys. Rev.} \textbf{\bibinfo{volume}{C66}},
  \bibinfo{pages}{064003} (\bibinfo{year}{2002}), \eprint{nucl-th/0209072}.

\bibitem[{\citenamefont{Jenkins and Manohar}(1991)}]{Jenkins:1990jv}
\bibinfo{author}{\bibfnamefont{E.~E.} \bibnamefont{Jenkins}} \bibnamefont{and}
  \bibinfo{author}{\bibfnamefont{A.~V.} \bibnamefont{Manohar}},
  \bibinfo{journal}{Phys. Lett.} \textbf{\bibinfo{volume}{B255}},
  \bibinfo{pages}{558} (\bibinfo{year}{1991}).

\bibitem[{\citenamefont{Bernard et~al.}(1992)\citenamefont{Bernard, Kaiser,
  Kambor, and Meissner}}]{Bernard:1992qa}
\bibinfo{author}{\bibfnamefont{V.}~\bibnamefont{Bernard}},
  \bibinfo{author}{\bibfnamefont{N.}~\bibnamefont{Kaiser}},
  \bibinfo{author}{\bibfnamefont{J.}~\bibnamefont{Kambor}}, \bibnamefont{and}
  \bibinfo{author}{\bibfnamefont{U.~G.} \bibnamefont{Meissner}},
  \bibinfo{journal}{Nucl. Phys.} \textbf{\bibinfo{volume}{B388}},
  \bibinfo{pages}{315} (\bibinfo{year}{1992}).

\bibitem[{\citenamefont{Pavon~Valderrama and
  Arriola}(2006)}]{PavonValderrama:2005wv}
\bibinfo{author}{\bibfnamefont{M.}~\bibnamefont{Pavon~Valderrama}}
  \bibnamefont{and} \bibinfo{author}{\bibfnamefont{E.~R.}
  \bibnamefont{Arriola}}, \bibinfo{journal}{Phys. Rev.}
  \textbf{\bibinfo{volume}{C74}}, \bibinfo{pages}{054001}
  (\bibinfo{year}{2006}), \eprint{nucl-th/0506047}.

\bibitem[{\citenamefont{Stoks et~al.}(1994)\citenamefont{Stoks, Klomp,
  Terheggen, and de~Swart}}]{Stoks:1994wp}
\bibinfo{author}{\bibfnamefont{V.~G.~J.} \bibnamefont{Stoks}},
  \bibinfo{author}{\bibfnamefont{R.~A.~M.} \bibnamefont{Klomp}},
  \bibinfo{author}{\bibfnamefont{C.~P.~F.} \bibnamefont{Terheggen}},
  \bibnamefont{and} \bibinfo{author}{\bibfnamefont{J.~J.}
  \bibnamefont{de~Swart}}, \bibinfo{journal}{Phys. Rev.}
  \textbf{\bibinfo{volume}{C49}}, \bibinfo{pages}{2950} (\bibinfo{year}{1994}),
  \eprint{nucl-th/9406039}.

\bibitem[{\citenamefont{de~Swart et~al.}(1997)\citenamefont{de~Swart,
  Rentmeester, and Timmermans}}]{deSwart:1997ep}
\bibinfo{author}{\bibfnamefont{J.~J.} \bibnamefont{de~Swart}},
  \bibinfo{author}{\bibfnamefont{M.~C.~M.} \bibnamefont{Rentmeester}},
  \bibnamefont{and} \bibinfo{author}{\bibfnamefont{R.~G.~E.}
  \bibnamefont{Timmermans}}, \bibinfo{journal}{PiN Newslett.}
  \textbf{\bibinfo{volume}{13}}, \bibinfo{pages}{96} (\bibinfo{year}{1997}),
  \eprint{nucl-th/9802084}.

\bibitem[{\citenamefont{Ruiz~Arriola et~al.}(2007)\citenamefont{Ruiz~Arriola,
  Calle~Cordon, and Pavon~Valderrama}}]{RuizArriola:2007wm}
\bibinfo{author}{\bibfnamefont{E.}~\bibnamefont{Ruiz~Arriola}},
  \bibinfo{author}{\bibfnamefont{A.}~\bibnamefont{Calle~Cordon}},
  \bibnamefont{and}
  \bibinfo{author}{\bibfnamefont{M.}~\bibnamefont{Pavon~Valderrama}}
  (\bibinfo{year}{2007}), \eprint{0710.2770}.

\bibitem[{\citenamefont{Epelbaum et~al.}(2002)\citenamefont{Epelbaum, Meissner,
  Gloeckle, and Elster}}]{Epelbaum:2001fm}
\bibinfo{author}{\bibfnamefont{E.}~\bibnamefont{Epelbaum}},
  \bibinfo{author}{\bibfnamefont{U.~G.} \bibnamefont{Meissner}},
  \bibinfo{author}{\bibfnamefont{W.}~\bibnamefont{Gloeckle}}, \bibnamefont{and}
  \bibinfo{author}{\bibfnamefont{C.}~\bibnamefont{Elster}},
  \bibinfo{journal}{Phys. Rev.} \textbf{\bibinfo{volume}{C65}},
  \bibinfo{pages}{044001} (\bibinfo{year}{2002}), \eprint{nucl-th/0106007}.

\bibitem[{\citenamefont{de~Swart et~al.}(1995)\citenamefont{de~Swart,
  Terheggen, and Stoks}}]{deSwart:1995ui}
\bibinfo{author}{\bibfnamefont{J.~J.} \bibnamefont{de~Swart}},
  \bibinfo{author}{\bibfnamefont{C.~P.~F.} \bibnamefont{Terheggen}},
  \bibnamefont{and} \bibinfo{author}{\bibfnamefont{V.~G.~J.}
  \bibnamefont{Stoks}} (\bibinfo{year}{1995}), \eprint{nucl-th/9509032}.

\bibitem[{\citenamefont{Brown and Machleidt}(1994)}]{Brown:1994pq}
\bibinfo{author}{\bibfnamefont{G.~E.} \bibnamefont{Brown}} \bibnamefont{and}
  \bibinfo{author}{\bibfnamefont{R.}~\bibnamefont{Machleidt}},
  \bibinfo{journal}{Phys. Rev.} \textbf{\bibinfo{volume}{C50}},
  \bibinfo{pages}{1731} (\bibinfo{year}{1994}).

\bibitem[{\citenamefont{Pavon~Valderrama and
  Arriola}(2005)}]{PavonValderrama:2005ku}
\bibinfo{author}{\bibfnamefont{M.}~\bibnamefont{Pavon~Valderrama}}
  \bibnamefont{and} \bibinfo{author}{\bibfnamefont{E.~R.}
  \bibnamefont{Arriola}}, \bibinfo{journal}{Phys. Rev.}
  \textbf{\bibinfo{volume}{C72}}, \bibinfo{pages}{044007}
  (\bibinfo{year}{2005}).

\bibitem[{\citenamefont{Dumbrajs et~al.}(1983)}]{Dumbrajs:1983jd}
\bibinfo{author}{\bibfnamefont{O.}~\bibnamefont{Dumbrajs}}
  \bibnamefont{et~al.}, \bibinfo{journal}{Nucl. Phys.}
  \textbf{\bibinfo{volume}{B216}}, \bibinfo{pages}{277} (\bibinfo{year}{1983}).

\bibitem[{\citenamefont{Kim and Zahed}(2009)}]{Kim:2008iy}
\bibinfo{author}{\bibfnamefont{K.-Y.} \bibnamefont{Kim}} \bibnamefont{and}
  \bibinfo{author}{\bibfnamefont{I.}~\bibnamefont{Zahed}},
  \bibinfo{journal}{JHEP} \textbf{\bibinfo{volume}{03}}, \bibinfo{pages}{131}
  (\bibinfo{year}{2009}), \eprint{0901.0012}.

\bibitem[{\citenamefont{Kim et~al.}(2009)\citenamefont{Kim, Lee, and
  Yi}}]{Kim:2009sr}
\bibinfo{author}{\bibfnamefont{Y.}~\bibnamefont{Kim}},
  \bibinfo{author}{\bibfnamefont{S.}~\bibnamefont{Lee}}, \bibnamefont{and}
  \bibinfo{author}{\bibfnamefont{P.}~\bibnamefont{Yi}}, \bibinfo{journal}{JHEP}
  \textbf{\bibinfo{volume}{04}}, \bibinfo{pages}{086} (\bibinfo{year}{2009}),
  \eprint{0902.4048}.

\bibitem[{\citenamefont{Hashimoto et~al.}(2009)\citenamefont{Hashimoto, Sakai,
  and Sugimoto}}]{Hashimoto:2009ys}
\bibinfo{author}{\bibfnamefont{K.}~\bibnamefont{Hashimoto}},
  \bibinfo{author}{\bibfnamefont{T.}~\bibnamefont{Sakai}}, \bibnamefont{and}
  \bibinfo{author}{\bibfnamefont{S.}~\bibnamefont{Sugimoto}}
  (\bibinfo{year}{2009}), \eprint{0901.4449}.

\bibitem[{\citenamefont{Feynman}(1988)}]{Feynman:1987hv}
\bibinfo{author}{\bibfnamefont{R.~P.} \bibnamefont{Feynman}},
  \emph{\bibinfo{title}{Variational Calculations in Quantum Field Theory}}
  (\bibinfo{publisher}{Proceedings International Workshop Wangerooge, F.R.
  Germany, 1-4 September 1987. Singapore, World Scientific Pub. Polley, L. D.
  E. L. Pottinger (Editors). pag 28-40.}, \bibinfo{year}{1988}).

\bibitem[{\citenamefont{Kaiser et~al.}(1997)\citenamefont{Kaiser, Brockmann,
  and Weise}}]{Kaiser:1997mw}
\bibinfo{author}{\bibfnamefont{N.}~\bibnamefont{Kaiser}},
  \bibinfo{author}{\bibfnamefont{R.}~\bibnamefont{Brockmann}},
  \bibnamefont{and} \bibinfo{author}{\bibfnamefont{W.}~\bibnamefont{Weise}},
  \bibinfo{journal}{Nucl. Phys.} \textbf{\bibinfo{volume}{A625}},
  \bibinfo{pages}{758} (\bibinfo{year}{1997}), \eprint{nucl-th/9706045}.

\bibitem[{\citenamefont{Kaiser et~al.}(1998)\citenamefont{Kaiser,
  Gerstendorfer, and Weise}}]{Kaiser:1998wa}
\bibinfo{author}{\bibfnamefont{N.}~\bibnamefont{Kaiser}},
  \bibinfo{author}{\bibfnamefont{S.}~\bibnamefont{Gerstendorfer}},
  \bibnamefont{and} \bibinfo{author}{\bibfnamefont{W.}~\bibnamefont{Weise}},
  \bibinfo{journal}{Nucl. Phys.} \textbf{\bibinfo{volume}{A637}},
  \bibinfo{pages}{395} (\bibinfo{year}{1998}), \eprint{nucl-th/9802071}.

\bibitem[{\citenamefont{Ademollo et~al.}(1969)\citenamefont{Ademollo,
  Veneziano, and Weinberg}}]{Ademollo:1969nx}
\bibinfo{author}{\bibfnamefont{M.}~\bibnamefont{Ademollo}},
  \bibinfo{author}{\bibfnamefont{G.}~\bibnamefont{Veneziano}},
  \bibnamefont{and} \bibinfo{author}{\bibfnamefont{S.}~\bibnamefont{Weinberg}},
  \bibinfo{journal}{Phys. Rev. Lett.} \textbf{\bibinfo{volume}{22}},
  \bibinfo{pages}{83} (\bibinfo{year}{1969}).

\bibitem[{\citenamefont{Kaplan and Savage}(1996)}]{Kaplan:1995yg}
\bibinfo{author}{\bibfnamefont{D.~B.} \bibnamefont{Kaplan}} \bibnamefont{and}
  \bibinfo{author}{\bibfnamefont{M.~J.} \bibnamefont{Savage}},
  \bibinfo{journal}{Phys. Lett.} \textbf{\bibinfo{volume}{B365}},
  \bibinfo{pages}{244} (\bibinfo{year}{1996}), \eprint{hep-ph/9509371}.

\bibitem[{\citenamefont{Savage}(1997)}]{Savage:1996tb}
\bibinfo{author}{\bibfnamefont{M.~J.} \bibnamefont{Savage}},
  \bibinfo{journal}{Phys. Rev.} \textbf{\bibinfo{volume}{C55}},
  \bibinfo{pages}{2185} (\bibinfo{year}{1997}), \eprint{nucl-th/9611022}.

\bibitem[{\citenamefont{Ericson et~al.}(2002)\citenamefont{Ericson, Loiseau,
  and Thomas}}]{Ericson:2000md}
\bibinfo{author}{\bibfnamefont{T.~E.~O.} \bibnamefont{Ericson}},
  \bibinfo{author}{\bibfnamefont{B.}~\bibnamefont{Loiseau}}, \bibnamefont{and}
  \bibinfo{author}{\bibfnamefont{A.~W.} \bibnamefont{Thomas}},
  \bibinfo{journal}{Phys. Rev.} \textbf{\bibinfo{volume}{C66}},
  \bibinfo{pages}{014005} (\bibinfo{year}{2002}), \eprint{hep-ph/0009312}.

\bibitem[{\citenamefont{Erkol et~al.}(2006{\natexlab{a}})\citenamefont{Erkol,
  Timmermans, Oka, and Rijken}}]{Erkol:2006eq}
\bibinfo{author}{\bibfnamefont{G.}~\bibnamefont{Erkol}},
  \bibinfo{author}{\bibfnamefont{R.~G.~E.} \bibnamefont{Timmermans}},
  \bibinfo{author}{\bibfnamefont{M.}~\bibnamefont{Oka}}, \bibnamefont{and}
  \bibinfo{author}{\bibfnamefont{T.~A.} \bibnamefont{Rijken}},
  \bibinfo{journal}{Phys. Rev.} \textbf{\bibinfo{volume}{C73}},
  \bibinfo{pages}{044009} (\bibinfo{year}{2006}{\natexlab{a}}),
  \eprint{nucl-th/0603058}.

\bibitem[{\citenamefont{Aliev and Savci}(2007)}]{Aliev:2006xm}
\bibinfo{author}{\bibfnamefont{T.~M.} \bibnamefont{Aliev}} \bibnamefont{and}
  \bibinfo{author}{\bibfnamefont{M.}~\bibnamefont{Savci}},
  \bibinfo{journal}{Phys. Rev.} \textbf{\bibinfo{volume}{D75}},
  \bibinfo{pages}{045006} (\bibinfo{year}{2007}), \eprint{hep-ph/0612144}.

\bibitem[{\citenamefont{Fernandez-Carames
  et~al.}(2008)\citenamefont{Fernandez-Carames, Gonzalez, and
  Valcarce}}]{FernandezCarames:2008en}
\bibinfo{author}{\bibfnamefont{M.~T.} \bibnamefont{Fernandez-Carames}},
  \bibinfo{author}{\bibfnamefont{P.}~\bibnamefont{Gonzalez}}, \bibnamefont{and}
  \bibinfo{author}{\bibfnamefont{A.}~\bibnamefont{Valcarce}},
  \bibinfo{journal}{Phys. Rev.} \textbf{\bibinfo{volume}{C77}},
  \bibinfo{pages}{054003} (\bibinfo{year}{2008}), \eprint{0804.4119}.

\bibitem[{\citenamefont{Hohler and Pietarinen}(1975)}]{Hohler:1974ht}
\bibinfo{author}{\bibfnamefont{G.}~\bibnamefont{Hohler}} \bibnamefont{and}
  \bibinfo{author}{\bibfnamefont{E.}~\bibnamefont{Pietarinen}},
  \bibinfo{journal}{Nucl. Phys.} \textbf{\bibinfo{volume}{B95}},
  \bibinfo{pages}{210} (\bibinfo{year}{1975}).

\bibitem[{\citenamefont{Grein}(1977)}]{Grein:1977id}
\bibinfo{author}{\bibfnamefont{W.}~\bibnamefont{Grein}},
  \bibinfo{journal}{Nucl. Phys.} \textbf{\bibinfo{volume}{B131}},
  \bibinfo{pages}{255} (\bibinfo{year}{1977}).

\bibitem[{\citenamefont{Mergell et~al.}(1996)\citenamefont{Mergell, Meissner,
  and Drechsel}}]{Mergell:1995bf}
\bibinfo{author}{\bibfnamefont{P.}~\bibnamefont{Mergell}},
  \bibinfo{author}{\bibfnamefont{U.~G.} \bibnamefont{Meissner}},
  \bibnamefont{and} \bibinfo{author}{\bibfnamefont{D.}~\bibnamefont{Drechsel}},
  \bibinfo{journal}{Nucl. Phys.} \textbf{\bibinfo{volume}{A596}},
  \bibinfo{pages}{367} (\bibinfo{year}{1996}), \eprint{hep-ph/9506375}.

\bibitem[{\citenamefont{Belushkin et~al.}(2007)\citenamefont{Belushkin, Hammer,
  and Meissner}}]{Belushkin:2006qa}
\bibinfo{author}{\bibfnamefont{M.~A.} \bibnamefont{Belushkin}},
  \bibinfo{author}{\bibfnamefont{H.~W.} \bibnamefont{Hammer}},
  \bibnamefont{and} \bibinfo{author}{\bibfnamefont{U.~G.}
  \bibnamefont{Meissner}}, \bibinfo{journal}{Phys. Rev.}
  \textbf{\bibinfo{volume}{C75}}, \bibinfo{pages}{035202}
  (\bibinfo{year}{2007}), \eprint{hep-ph/0608337}.

\bibitem[{\citenamefont{Erkol et~al.}(2006{\natexlab{b}})\citenamefont{Erkol,
  Timmermans, and Rijken}}]{Erkol:2006sa}
\bibinfo{author}{\bibfnamefont{G.}~\bibnamefont{Erkol}},
  \bibinfo{author}{\bibfnamefont{R.~G.~E.} \bibnamefont{Timmermans}},
  \bibnamefont{and} \bibinfo{author}{\bibfnamefont{T.~A.}
  \bibnamefont{Rijken}}, \bibinfo{journal}{Phys. Rev.}
  \textbf{\bibinfo{volume}{C74}}, \bibinfo{pages}{045201}
  (\bibinfo{year}{2006}{\natexlab{b}}).

\bibitem[{\citenamefont{Wang}(2007)}]{Wang:2007yt}
\bibinfo{author}{\bibfnamefont{Z.-G.} \bibnamefont{Wang}},
  \bibinfo{journal}{Phys. Rev.} \textbf{\bibinfo{volume}{D75}},
  \bibinfo{pages}{054020} (\bibinfo{year}{2007}), \eprint{hep-ph/0701176}.

\bibitem[{\citenamefont{Henneck}(1993)}]{Henneck:1993zz}
\bibinfo{author}{\bibfnamefont{R.}~\bibnamefont{Henneck}},
  \bibinfo{journal}{Phys. Rev.} \textbf{\bibinfo{volume}{C47}},
  \bibinfo{pages}{1859} (\bibinfo{year}{1993}).

\end{thebibliography}

\end{document}